\documentclass{aa}

\usepackage{bm} 

\usepackage[T1]{fontenc}
\usepackage{ae,aecompl}
\usepackage{graphicx}   
\usepackage{amsmath}    
\usepackage{amssymb}    

\usepackage[normalem]{ulem}

\usepackage[x11names]{xcolor}

\usepackage{mathtools}
\usepackage{natbib}
\bibpunct{(}{)}{;}{a}{}{,} 

\usepackage[breaklinks=true]{hyperref}

\hypersetup{
   colorlinks=true,
   linkcolor=blue,
   citecolor=blue,
   urlcolor=blue,
}

\usepackage{enumerate}

\newcommand{\mgas}{$M_\mathrm{g}$}
\newcommand{\fgas}{$F_\mathrm{g}$}
\newcommand{\mi}{$M_i$}
\newcommand{\taugas}{$\tau_g$}
\newcommand{\tausf}{$\tau_\mathrm{SF}$}
\newcommand{\mstar}{$M_\mathrm{star}$}
\newcommand{\logoh}{12$+$log(O/H)}
\newcommand{\lco}{$L_{\rm CO}^\prime$}
\newcommand{\mhi}{$M_{\rm HI}$}
\newcommand{\mhtwo}{$M_{\rm H2}$}
\newcommand{\aco}{$\alpha_{\rm CO}$}
\newcommand{\acosun}{$\alpha_{\rm CO}^{\odot}$}
\newcommand{\coone}{$^{12}$CO($1-0$)}
\newcommand{\cotwo}{$^{12}$CO($2-1$)}

\newcommand{\kms}{km\,s$^{-1}$}

\newcommand{\micron}{$\mu$m}

\newcommand{\msun}{M$_\odot$}

\newcommand{\htwo}{H$_2$}
\newcommand{\hi}{H{\sc i}}

\newcommand{\chisq}{$\chi^2$}

\newcommand{\tgal}{$t_{\rm gal}$}

\newcommand{\nii}{[N{\sc ii}]}

\newcommand{\epss}{$\epsilon_{\Large\star}$}

\newcommand{\zg}{$Z_g$}
\newcommand{\za}{$Z_a$}
\newcommand{\zw}{$Z_w$}
\newcommand{\mz}{$M_Z$}
\newcommand{\zetaw}{$\zeta_w$}
\newcommand{\zetaa}{$\zeta_a$}
\newcommand{\etaw}{$\eta_w$}
\newcommand{\etaa}{$\eta_a$}
\newcommand{\deta}{$\Delta\eta$}
\newcommand{\fent}{$f_\mathrm{ent}$}
\newcommand{\zej}{$Z_\mathrm{eject}$}
\newcommand{\vcirc}{$V_{\rm circ}$}

\newcommand{\ml}{$M/L$}
\newcommand\mug{$\mu_g$}
\newcommand\ye{y$_{\rm eff}$}

\newcommand{\lcounits}{${\rm K\,km\,s^{-1}\,pc^{2}}$}

\def\Mvir{\mbox{$M_{\rm vir}$}}

\newcommand{\papi}{Paper\,I}
\newcommand{\papii}{Paper\,II}
\defcitealias{Ginolfi+20}{Paper I}
\defcitealias{hunt20}{Paper II}

\begin{document}

\title{Scaling~relations~and~baryonic~cycling~in~local~star-forming~galaxies.\newline ~III.~Outflows,~effective~yields~and~metal~loading~factors}

\author{C.~Tortora\inst{\ref{inst1},\ref{inst2}}
    \and
    L.~K.~Hunt\inst{\ref{inst2}}
    \and
    M.~Ginolfi\inst{\ref{inst3},\ref{inst4}}
}

\institute{INAF -- Osservatorio Astronomico di Capodimonte, Salita
    Moiariello 16, 80131 - Napoli, Italy, \email{crescenzo.tortora@inaf.it}\label{inst1}
    \and
    INAF -- Osservatorio Astrofisico di Arcetri, Largo Enrico Fermi 5, I-50125
    Firenze, Italy\label{inst2}
    \and
    European Southern Observatory, Karl-Schwarzschild-Stra{\ss} e 2, 85748 Garching, Germany\label{inst3}
    \and
    Observatoire de Gen\'eve, Universit\'e de Gen\'eve, 51 Ch. des Maillettes, 1290 Versoix,
    Switzerland\label{inst4}
}

\date{Received XXX; accepted YYY}

\abstract{Gas accretion and stellar feedback processes link metal
content, star formation, and gas and stellar mass (and the
potential depth) in star-forming galaxies. Constraining this
hypersurface has been challenging because of the need for
measurements of \hi\ and \htwo\ gas masses spanning a broad
parameter space. A recent step forward has been achieved through
the ``Metallicity And Gas for Mass Assembly'' (MAGMA) sample of
local star-forming galaxies, which consists of
homogeneously-determined parameters and a significant quantity of
dwarf galaxies, with stellar masses as low as $\sim 10^5 -
10^{6}$\,\msun. Here, in the third paper of a series, we adopt a
``standard'' galactic chemical evolution model, with which we can
quantify stellar-driven outflows. In particular, we constrain the
difference between the mass-loading in accretion and outflows and
the wind metal-loading factor. The resulting model reproduces very
well the local mass-metallicity relation, and the observed trends
of metallicity with gas fraction. Although the difference in mass
loading between accreted and expelled gas is extremely difficult
to constrain, we find indications that, on average, the amount of
gas acquired through accretion is roughly the same as the gas lost
through bulk stellar outflows, a condition roughly corresponding
to a ``gas equilibrium'' scenario. In agreement with previous
work, the wind metal-loading factor shows a steep increase toward
lower mass and circular velocity, indicating that low-mass
galaxies are more efficient at expelling metals, thus shaping the
mass-metallicity relation. Effective yields are found to increase
with mass up to an inflection mass threshold, with a mild decline
at larger masses and circular velocities. A comparison of our
results for metal loading in outflows with the expectations for
their mass loading favors momentum-driven winds at low masses,
rather than energy-driven ones. }


\titlerunning{MAGMA $-$ outflows~and~metal~yields}
\authorrunning{C.~Tortora et al.}

\maketitle

\section{Introduction}\label{sec:introduction}

Star-forming galaxies evolve and assemble their mass by
accreting gas from the intergalactic medium (IGM),
converting it in stars, then expelling gas and metals, thus
changing the gas metallicity in the interstellar medium (ISM).
This baryon cycle is mainly regulated by internal
processes\footnote{We do not imply here that the environment does
not play a role in baryonic cycling, on the contrary; the idea is
that the environment can drive processes that take place within
the galaxy, so that ultimately they can be considered
``internal''.} \citep[e.g.,][]{wu17}, which are responsible for
metal and dust production, and the subsequent expulsion of
material. Outflows driven by stellar winds and supernovae (SNe)
explosions dominate in the lower-mass regime, while at higher
masses, outflows tend to be powered by feedback from
Active Galactic Nuclei 
\citep[AGN, e.g.,][]{Tremonti+04,dek_birn06,dalcanton07,Erb2008,Finlator2008,Zahid2014,deRossi+17_EAGLE,Lara-Lopez+19_yields}.

The efficiency of these processes is predicted to change with
galaxy properties, and in particular with the virial mass (\Mvir)
and stellar mass (\mstar) of the galaxies. These variations lead
to the most important relations among the main galaxy parameters
driving galaxy evolution, namely the
``main sequence'' correlation between \mstar\ and the star-formation rate 
\citep[SFR, e.g.,][the SFMS]{Brinchmann+04}; the mass-metallicity
relation \citep[e.g.,][the MZR]{Tremonti+04}; and the consequent
inverse correlation between metallicity and specific SFR
\citep[sSFR\,$\equiv$\,SFR/\mstar,
e.g.,][]{Mannucci+10,Lara-Lopez+10,hunt12,Hunt+16_I}. Baryonic
processes are thought to shape the SFMS and the MZR, in addition
to many other correlations among other stellar properties and dark
matter: the virial-to-stellar mass \citep[e.g.,][]{Moster+10} and
the baryonic Tully-Fisher \citep[e.g.,][]{Lelli+16_Tully-Fisher}
relations; the correlation between color and stellar population
gradients with mass \citep[e.g.,][]{Tortora+10CG}; and the
relation of the dark matter fraction and mass density slopes with
stellar mass \citep[e.g.,][]{Tortora+19_LTGs_DM_and_slopes}.

These scaling relations are not necessarily straight power laws,
but can also show curvatures, bending, breaks and U-shapes,
suggesting the existence of typical mass scales which regulate the
efficiency of the physical processes involved in baryonic cycling
\citep[e.g.,][]{dek_birn06}. \citet{Kannappan2013} identified two
transitions in galaxy morphology, gas fractions, and fueling
regimes. The first transition was considered a ``gas-richness''
threshold; galaxies below a transition mass of log(\mstar/\msun)
$\sim 9.5$ were found to be ``accretion-dominated'', characterized
by overwhelming gas accretion. The second \mstar\ transition
corresponds, instead, to a bimodality threshold, roughly
indicating a ``quenched'' regime at log(\mstar/\msun)\,$ > 10.5$,
populated by spheroid-dominated, gas-poor
galaxies. 

Both \mstar\ thresholds, and the mass regimes they delineate, are
really tracing gas content, and its intimate link with
star-formation activity \citep{Kannappan2013}. To further explore
this link, in the first paper of this series \citep[][hereafter
\papi]{Ginolfi+20}, we constructed a sample of galaxies in the
Local Universe with homogeneously-derived global measurements of
the main parameters that quantify galaxy evolution: namely \mstar,
SFR, metallicity O/H, and gas content including molecular gas mass
\mhtwo\ and atomic gas mass \mhi. The sample comprises $\sim$400
galaxies, and was dubbed ``Metallicity And Gas for Mass Assembly''
(MAGMA).

In a second paper \citep[][hereafter \papii]{hunt20},
we used MAGMA to explore how molecular and atomic gas properties vary as a function of \mstar\
and SFR.
The main focus of \papii\ was on feedback processes, and how they govern scaling relations
across the mass spectrum.
Essentially three processes play a role in regulating galaxy evolution\footnote{Here we are interested in
star-forming galaxies only, and do not discuss the effects of AGN feedback on massive galaxies.}:
\textit{(P1)}~preventive feedback, the (lack of) availability of cold baryons from the host halo;
\textit{(P2)}~inefficiency of the star-formation process, the conversion of the available gas into stars; and
\textit{(P3)}~ejective feedback, the production of energy and momentum, and the expulsion of material including gas, dust, and metals.
In \papii, we investigated two of the three main feedback mechanisms behind
the inefficiency of low-mass halos to form stars \citep[e.g.,][]{behroozi13,moster13,graziani15,graziani17}:
namely the availability of cold gas to form stars \textit{(P1)}, and the inefficiency of the
star-formation process itself \textit{(P2)}.

Here, in the third paper of the series, we examine the remaining
mechanism, \textit{(P3)}, the efficiency of metal enrichment and
ejective feedback. Understanding the processes that shape galaxy
scaling relations with metallicity has proved to be a challenging
task. It has been historically important to compare observations
with the ``closed-box'' model, assuming that the total baryon
content in the galaxy is fixed and the gas and stellar content are
only determined by star formation and the stellar yields
\citep[e.g.,][]{pagel75}. However, it is widely recognized that
galaxies do not evolve as closed boxes
\citep[e.g.,][]{Tremonti+04}, and more complex models have
considered inflows and outflows of material from the galactic
ecosystem
\citep[e.g.,][]{larson72,pagel75,tinsley80,pagel81,edmunds84,edmunds90,Tremonti+04,dalcanton07,
Erb2008,Finlator2008,Recchi+08,Peeples_Shankar11,Dayal2013,Zahid2014,peng14}.
More recently, hydrodynamical simulations have tested these
predictions
\citep[e.g.,][]{ma16,deRossi+17_EAGLE,Torrey+19_Illustris}.

Our approach here is mainly empirical, requiring a sample with
measurements of \mstar, SFR, O/H, and gas content, homogeneously
determined over a wide range of stellar mass and gas fractions;
MAGMA is such a sample. In this paper, Sect.\ \ref{sec:sample}
describes briefly the MAGMA sample and how it was assembled. Then,
Sect.\ \ref{sec:formalism} presents the theoretical formalism for
metallicity evolution and the scaling of metallicity with \mstar,
SFR, and gas content. The computation of outflow parameters and
metal-loading factors for MAGMA galaxies is described in Sect.\
\ref{sec:loading}, and their impact on shaping the
mass-metallicity relation in Sect.\ \ref{sec:shaping}. We recast
our results in the context of the effective yields in Sect.\
\ref{sec:yields}, and discuss the physical implications for wind
metallicity and mass loading of the winds in Sect.\
\ref{sec:feedback}. Our conclusions are drawn in Sect.\
\ref{sec:conclusions}.

\section{MAGMA sample}\label{sec:sample}

\citetalias{Ginolfi+20} presented the MAGMA sample, comprising 392
local galaxies having homogeneously derived global quantities
including \mstar, SFR, metallicities [\logoh], and gas masses
(both atomic, \mhi, and molecular, \mhtwo, and total
\mgas\,=\,\mhi\,$+$\,\mhtwo). Molecular gas masses were obtained
by measurements of the CO luminosity, \lco. We assembled this
sample by combining a variety of previous surveys at $z \sim 0$
with new observations of CO in low-mass galaxies. Only galaxies
with robust detections of physical parameters were retained, that
is to say we required clear detections of CO and \hi. Galaxies
were eliminated if they showed optical signatures of an AGN in the
parent samples \citep[obtained from line-ratio diagnostics,
similar to][]{Baldwin1981}. We also eliminated potentially
\hi-deficient galaxies
by applying a cut based on \hi-deficiency measurements, 
where available \citep[e.g.,][]{Boselli2009,Boselli2014a}.
The MAGMA sample spans more than 5 orders of magnitude in \mstar, SFR,
and \mgas, and a factor of $\sim 50$ in metallicity [$Z$, \logoh].

Because of potential systematics that could perturb our results,
in \citetalias{Ginolfi+20} we homogenized the sample by
recalculating both \mstar\ and SFR in a uniform way. To calculate
SFR, we adopted the hybrid formulations of \citet{Leroy+19}, based
on combinations of GALEX and WISE luminosities. New \mstar\
measurements were computed using $3.4$\,\micron\ (or IRAC
3.6\,\micron\ when not available) luminosities from fluxes in the
ALLWISE Source Catalogue \citep{Wright2010} and a
luminosity-dependent \ml\ from \citet{Hunt+19}; the continuum
contribution from free-free emission was subtracted from these
luminosities before the \mstar\ computation. Both SFR and \mstar\
assume a \citet{Chabrier2003} IMF.

For metallicity, \logoh, we adopted either ``direct''
calibrations, based on electron temperatures
\citep[e.g.,][]{izotov07}, or the \nii-based strong-line
calibration by \citet[][PP04N2]{Pettini2004}. When PP04N2-derived
metallicities were not available, they were converted from the
original calibration to PP04N2 according to
\citet{Kewley_Ellison08}. PP04N2 is the metallicity
calibration closest to the direct one
\citep[e.g.,][]{Andrews_Martini2013,Hunt+16_I}, which makes it
appropriate for a sample like MAGMA that spans a wide range of
\mstar\ and SFR.

The molecular gas mass was determined from \lco\ based mainly on
\coone; only in a few cases did we have to convert from \cotwo\ to
\coone\ using ``standard'' prescriptions
\citep[e.g.,][]{leroy09,schruba12}. The conversion factor \aco\
from CO luminosity to molecular gas mass was calculated according
to the piecewise relation presented in \papii, with \aco$\propto
Z^{1.55}$, for \logoh$< 8.69$ and
\acosun\,$=\,3.2$\,\msun\,(\lcounits)$^{-1}$ \citep[see
also][]{Saintonge+11_COLDGASS_I}. The impact of a different
metallicity dependence is discussed in \papii. Here we do not
include a factor of 1.36 to account for helium. For more details
about the parent surveys, as well as parameter determinations and
calibrations, quality checks, or the derivation of scaling
relations, we refer the reader to \citetalias{Ginolfi+20} and
\citetalias{hunt20}.

\section{Metallicity evolution and scaling relations: formalism}\label{sec:formalism}

Much effort has been spent over the last fifty years to understand
the theoretical framework behind the baryon exchange cycle and
metallicity evolution in galaxies
\citep[e.g.,][]{larson72,pagel75,tinsley80,pagel81,edmunds84,edmunds90,Tremonti+04,dalcanton07,
Erb2008,Finlator2008,Recchi+08,spitoni10,Peeples_Shankar11,Dayal2013,Zahid2014,peng14}.
It has been fairly well established that bulk outflows and
galactic winds driven by star formation are necessary to explain
the MZR and other scaling relations involving
metallicity. Here we follow these pioneering works, and explore
the behavior of the MZR, SFR, and gas content in the MAGMA
galaxies. MAGMA provides the important constraints of gas
measurements together with SFR and metallicity, thus enabling,
for the first time, a quantification of the mass and metallicity
loading in stellar outflows, and their role in shaping the
MZR in the Local Universe.

\subsection{Galactic chemical evolution}\label{sec:gce}

In what follows, we rely heavily on the formulations for galactic chemical evolution by \citet{pagel09}.
We first make the assumption of instantaneous recycling, namely that the chemical enrichment provided
by stellar evolution, nucleosynthesis, and expulsion of metals into the ISM take place ``instantaneously''
relative to the timescales of overall galaxy evolution.

There are two main quantities that this assumption makes
time-invariant: the first is the return fraction $R$, the mass
fraction of a single stellar generation that is returned to the
ISM; the second is the (true) stellar yield, $y$, defined as the
mass of an element newly produced and expelled by a generation of
stars relative to the mass that remains locked up in long-lived
stars and compact remnants. The metal mass produced relative to
the initial mass of a single stellar generation is then
$q\,=\,\alpha\,y$ where $\alpha\,=\,1 - R$, (or ``lock-up''
fraction), with the implicit assumption that stellar lifetimes can
be neglected so that $R$ is a net mass return fraction \citep[see
e.g.,][]{Vincenzo+16_yields}.

These parameters are governed by the IMF, and in particular the
mass fraction of a stellar generation above a certain limit, since
these are the main contributors to metal production. We further
assume that metals are homogeneously mixed at all times, and that
initally there are no stars and no metals, only the initial gas
\mgas$(t=0)$ with mass \mi.

There are four main physical quantities involved in and constrained by chemical evolution:
the mass in stars, \mstar;
the star-formation rate SFR, here referred to as $\psi$;
the mass in gas, \mgas;
and the mass in metals, \mz, or the metal abundance in the gas, \zg\,=\,\mz/\mgas.
The mass of stars that have been born up to time $t$ can be written as:
\begin{equation}
S(t)\,=\,\int_0^t\ \psi(t^\prime)\ dt^\prime\quad .
\label{eqn:mstar_all}
\end{equation}
Consequently, the mass \mstar\ remaining in stars and long-lived remnants is:
\begin{equation}
M_{\Large\star}(t)\,=\,\alpha\ S(t)\quad .
\label{eqn:mstar}
\end{equation}
The total baryonic mass in a galactic system is the sum of gas and stars:
\begin{equation}
M_{\rm tot}(t)\,=\,M_{\Large\star}(t)\ + \ M_{\rm g}(t)\,=\,M_i - M_w(t) + M_a(t) \quad ,
\label{eqn:mtot}
\end{equation}
where $M_w$ is the mass expelled in bulk winds and outflows, and
$M_a$ is the mass of newly-accreted gas. This is essentially a
mass continuity equation that defines the mass conservation
between infall, outflows, and star formation. Thus, the time
derivative of the gas mass can be written as:
\begin{equation}
\dot{M}_\mathrm{g}\,=\,\dot{M}_a - \dot{M}_w - \dot{M}_{\Large\star}\quad .
\label{eqn:mgtimederiv}
\end{equation}

The time variation of the mass of metals \mz\ in the ISM gas depends on $\psi$, $R$, the stellar yield $q$,
$\dot{M}_a$, and $\dot{M}_w$ as follows:
\begin{equation}
\dot{M}_Z\,=\,Z_g\,R\,\psi + q\,\psi - Z_g\,\psi + Z_a\,\dot{M}_a
- Z_w\,\dot{M}_w \quad , \label{eqn:mztimederiv}
\end{equation}
where \za\ and \zw\ are the metallicities of the accreted and
outflow gas, respectively. The first two terms on the right-hand
side of Eq. \eqref{eqn:mztimederiv} correspond to the addition of
metals of ejecta from stars; the third term to the loss to the ISM
from astration; and the remaining terms to the metals gained from
accretion and lost in bulk stellar outflows from winds and SNe.

Equation \eqref{eqn:mztimederiv} can be simplified by two common assumptions:
we take $\dot{M}_a$ and $\dot{M}_w$ to be directly proportional to $\psi$
\citep[e.g.,][]{Erb2008,Peeples_Shankar11,Dayal2013,Lilly2013}:
\begin{eqnarray}
\eta_a & = & \dot{M}_a / \psi\quad, \nonumber \\
\eta_w & = & \dot{M}_w / \psi\quad,
\label{eqn:massloading}
\end{eqnarray}
where we have introduced the mass-loading factors \etaa\ and
\etaw. It is fairly well established that the latter relation is
true, namely that the mass in outflows is proportional to SFR, at
least to its integral over time
\citep[e.g.,][]{Muratov+15,christensen16}. In contrast, the former
proportionality is more subject to speculation, although evidence
is mounting that gas accretion powers star formation
\citep[e.g.,][]{fraternali12,sanchez14,tacchella16}. In any case,
this assumption provides the noteable advantage of analytical
simplicity; we plan to investigate more completely its
ramifications in a future paper. Following
\citet{Peeples_Shankar11}, we also introduce the metal-loading
factors \zetaa\ and \zetaw\ relating \zg, \za, \etaa, \zw, and
\etaw:
\begin{equation}
\zeta_a\,\equiv\, \left( \frac{Z_a}{Z_g}
\right)\,\frac{\dot{M}_a}{\psi}\,=\,\left( \frac{Z_a}{Z_g} \right)
\eta_a, \label{eqn:zetaa_def}
\end{equation}
and
\begin{equation}
\zeta_w\,\equiv\, \left( \frac{Z_w}{Z_g}
\right)\,\frac{\dot{M}_w}{\psi}\,=\,\left( \frac{Z_w} {Z_g}
\right) \eta_w\quad. \label{eqn:zetaw_def}
\end{equation}
\noindent \zetaw\ and \zetaa\ essentially describe the efficiency
with which gas accretion and star-formation-driven outflows change
metal content within the galaxy. Thus, we can rewrite
Eqs.\,\eqref{eqn:mgtimederiv} and \eqref{eqn:mztimederiv} as:
\begin{equation}
\dot{M}_\mathrm{g}\,=\,\psi\,(\eta_a - \eta_w - \alpha)\quad,
\label{eqn:mgtimederiv_again}
\end{equation}
and
\begin{equation}
\dot{M}_Z\,=\,\psi\,[q - \alpha\,Z_g + Z_g\,(\zeta_a -
\zeta_w)]\quad. \label{eqn:mztimederiv_again}
\end{equation}

Our observations do not directly constrain \mz, but rather \zg, so
we need to reformulate Eq.\,\eqref{eqn:mztimederiv_again}:
\begin{equation}
\dot{M}_Z\,=\,\frac{d}{dt}\left( M_g\,Z_g \right)\,=\,M_g\,\dot{Z}_g + Z_g\,\dot{M}_g
\end{equation}
in order to solve for $\dot{Z}_g$:
\begin{equation}
\dot{Z}_g\,=\,\frac{\psi}{M_g}\ [q + Z_g\,(\zeta_a - \zeta_w -
\eta_a + \eta_w)]\quad, \label{eqn:zgtimederiv}
\end{equation}
where we have used Eq.\,\eqref{eqn:mgtimederiv_again} for
$\dot{M}_g$ and Eq.\,\eqref{eqn:mztimederiv_again} for $\dot{M}_Z$.

The necessary ingredients are now in place to derive the solution for the dependence of \zg\ on the remaining
parameters.
First, we know that SFR is related to gas mass by:
\begin{equation}
\psi\,=\,\epsilon_{\Large\star}\,M_g
\label{eqn:sfr}
\end{equation}
where \epss\ is the inverse depletion time or time-scale for star
formation. Equations \eqref{eqn:mgtimederiv_again} and
\eqref{eqn:sfr} are coupled, so must be solved simultaneously.
Assuming that \epss\ does not vary with time, we can thus
integrate Eq.\,\eqref{eqn:mgtimederiv_again},
to obtain the following solution for the gas mass:
\begin{equation}
M_{g}(t) = M_i\ e^{t \, \epsilon_{\Large\star} \, (\eta_a -
\eta_w -\alpha)}\quad. \label{eqn:mg_sol}
\end{equation}
Substituting this solution into
Eq.\,\eqref{eqn:zgtimederiv} and integrating, we obtain:
\begin{equation}
Z_g\,=\,\left( \frac{q}{\zeta_w - \zeta_a + \eta_a - \eta_w}
\right)\, \left[ 1 - \left( \frac{M_g}{M_i} \right)^{\frac{\zeta_a
- \zeta_w - \eta_a + \eta_w}{\eta_a - \eta_w - \alpha}} \right]
\quad, \label{eqn:zg}
\end{equation}
where it can be seen that the time dependence is encapsulated
in the evolution of the gas \citep[\mgas/\mi, e.g.,][]{Recchi+08}.
Equation\,\eqref{eqn:zg} is essentially telling us that galaxies
evolve along a hypersurface
constrained by \mstar, \mgas, gas metallicity, and accretion and
wind mass and metal loadings.
The galaxy's position on this
hypersurface can change with time as gas is accreted, stars are
formed, and mass and metals are ejected in winds.
However, at each time, the relation constraining these fundamental variables is
approximately invariant.

The ratio of the system gas mass \mgas\ and the initial gas mass
\mi\ can be written in terms of the baryonic gas mass fraction,
\mug\,=\,$M_g/(M_g + M_{\Large\star})$:
\begin{equation}
\frac{M_g}{M_i}\,=\,\frac{\mu_g}{1 + (\mu_g - 1)\,\left( \frac{\eta_a - \eta_w}{\alpha} \right)}
\label{eqn:mgmi}
\end{equation}
where we have assumed that the total baryonic mass is described by
Eq.\,\eqref{eqn:mtot}. This is again an expression of mass
conservation, meaning that no gas has been expelled altogether
from the galaxy, but rather has been recycled into stars as
described in Eq.\,\eqref{eqn:sfr}.

Similar solutions for \zg\ were first formulated by
\citet{edmunds84}, and later by \citet{Recchi+08},
\citet{Erb2008}, \citet{spitoni10}, \citet{Peeples_Shankar11},
\citet{Dayal2013}, and many others. These earlier works mostly
assumed that \zw/\zg\,=\,1 in a so-called homogeneous wind
\citep[e.g.,][]{pagel09}, leading to \zetaw\,=\,\etaw. This
assumption simplifies Eq.\,\eqref{eqn:zg} placing the dependence
of \etaw\ only in the exponent of \mgas/\mi. However, as has been
shown previously \citep[e.g.,][]{Peeples_Shankar11} and as we show
below for MAGMA, this is almost certainly an over-simplified
assumption because the metallicity of the outflowing material \zw\
is not generally the same as the ISM gas, \zg, especially at low
masses. We further discuss this simplification and its
implications later on in the paper (see also Appendix
\ref{app:hom_wind}).

\subsection{Limitations of the model} \label{sec:limitations}

First, Eq. \eqref{eqn:zg} is highly sensitive to the value of \etaa$-$\etaw$-\alpha$ in the denominator of
the exponent of \mgas/\mi.
For convenience in what follows, we define:
\begin{equation}
\Delta \eta = \eta_a - \eta_w
\label{eqn:deta}
\end{equation}
When \deta$\,=\,\alpha$, the exponent for \mgas/\mi\ in Eq.
\eqref{eqn:zg} becomes infinite. This equality,
\deta\,$\approx\,\alpha$,  is exactly what would be expected for
the ``equilbrium scenario'' for gas and metallicity evolution, in
which gas accretion is balanced by star formation and winds
\citep[e.g.,][]{Dave+12,Mitra+15}, such that
$\dot{M}_\mathrm{g}\,\approx\,0$. From Eq.
\eqref{eqn:mgtimederiv_again}, this implies that either
$\psi\,\approx\,0$ which is not physically relevant here, or
\deta\,$\approx\,\alpha$. In other words, for ``equilibrium'', the
difference between the accretion \etaa\ and wind \etaw\
coefficients is balanced by the lock-up fraction $\alpha$.
However, this formalism gives for the equilibrium scenario a
solution for \zg\ that is apparently
ill-conditioned\footnote{Although, taking the numerator of the
exponent of \mgas/\mi\ in Eq. \eqref{eqn:zg} as a constant, gives:
$$\lim_{a \to 0} \frac{y\,(1-x^{a/b})}{a} \to -\frac{y}{b}\,\ln(x)$$
which is roughly the closed-box solution with no gas inflow or
outflow (see Sect. \ref{sec:yields}).}. When \deta\,=\,$\alpha$,
Eq. \eqref{eqn:mg_sol} [and Eq. \eqref{eqn:mgmi}] are irrelevant,
as \mgas\,$=$\,\mi.

More generally, the trend of gas growth in Eq. \eqref{eqn:mg_sol}
depends on the sign of the exponent \deta\,$-\,\alpha$.
This exponent is a critical component of Eq. \eqref{eqn:zg} because of the sensitivity to the sign of the
exponent of \mgas/\mi.
Where $\dot{M}_\mathrm{g}\,>\,0$, with \deta\,$-\,\alpha$ slightly positive, a finite \zg\ solution is obtained
only when this exponent is negative, and conversely,
if \deta\,$-\,\alpha$ is slightly negative, then this exponent should be positive.

Equation \eqref{eqn:zg} is also sensitive to the gas fraction \mug\ through
Eq. \eqref{eqn:mgmi}.
As also noticed by \citet{Recchi+08}, this approach leads to constraints on the possible
values of gas fraction \mug.
Depending on the value of \deta\,$-\,\alpha$,
very small \mug\ leads to infinite solutions for \zg.

In any case, despite the complexity of possible solutions,
Eq. \eqref{eqn:zg} is a useful tool to assess the interplay of gas and
metallicity in galaxies.
In the next section, we explore this by letting the data guide the possible solutions.

\bigskip
\section{Mass and metal loading in MAGMA}\label{sec:loading}

MAGMA combines homogeneous measurements of \mstar, SFR, \zg, and
\mgas; so, for the first time, we can infer from observations the
mass loading factors, \etaa\ and \etaw, as well as the
metal-loading factor \zetaw. To do this, we first make another
common assumption, namely that the accreted gas is pristine
\citep[\za\,=\,0, $\zeta_{a}=0$,
e.g.,][]{Erb2008,Peeples_Shankar11,Dayal2013,Lilly2013,Creasey2015}.
Although this assumption may not be exactly true for low-redshift
galaxy populations, there is considerable observational evidence
that the gas accreted by local galaxies is significantly metal
poor \citep[e.g.,][]{tosi82,sanchez14,fraternali17}.

To apply the formalism described above to the observations,
we convert the MAGMA oxygen abundance \logoh\ to \zg\ according
to:
\begin{equation}
\log (Z_g)\,=\,\log(\mathrm{O/H}) + 0.956\quad ,
\label{eqn:zgconv}
\end{equation}
where the underlying assumption is that oxygen measures the enrichment of a primordial mix
of gas comprising approximately 75\% hydrogen by mass
\citep[e.g.,][]{Peeples_Shankar11,chisholm18}.
We adopt the yields and return fractions for the \citet{Chabrier2003} IMF
from \citet{Vincenzo+16_yields} using the \citet{romano10} stellar yields:
$(y,R)\,=\,(0.037,0.455)$.
We have purposely not considered a conversion of O/H to total metal abundance \zg,
because of the inherent uncertainties in this conversion
\citep[e.g.,][]{Vincenzo+16_yields}; thus our value for $y$ is the
oxygen yield $y_{\rm O}$ from \citet{Vincenzo+16_yields}.

From Eq.\,\eqref{eqn:zg}, and fixing \zetaa\ to 0, because of our
assumption that inflowing gas is pristine, it can be seen that
there are two variables that shape the \zg--\mstar--SFR--\mgas\
hypersurface: \deta\ and \zetaw. With this formulation, it is
impossible to determine separately the accretion and outflow
mass-loading factors, \etaa\ and \etaw. Physically, this means
that the important driving factor is the difference \deta, that is
to say the dominance (or not) of gas accretion over bulk outflows.
Below we describe the results from a Bayesian approach for
inferring mass- and metallicity-loading factors in MAGMA.

\begin{figure*}[!t]
\begin{center}
\hbox{
\includegraphics[angle=0,width=0.33\linewidth]{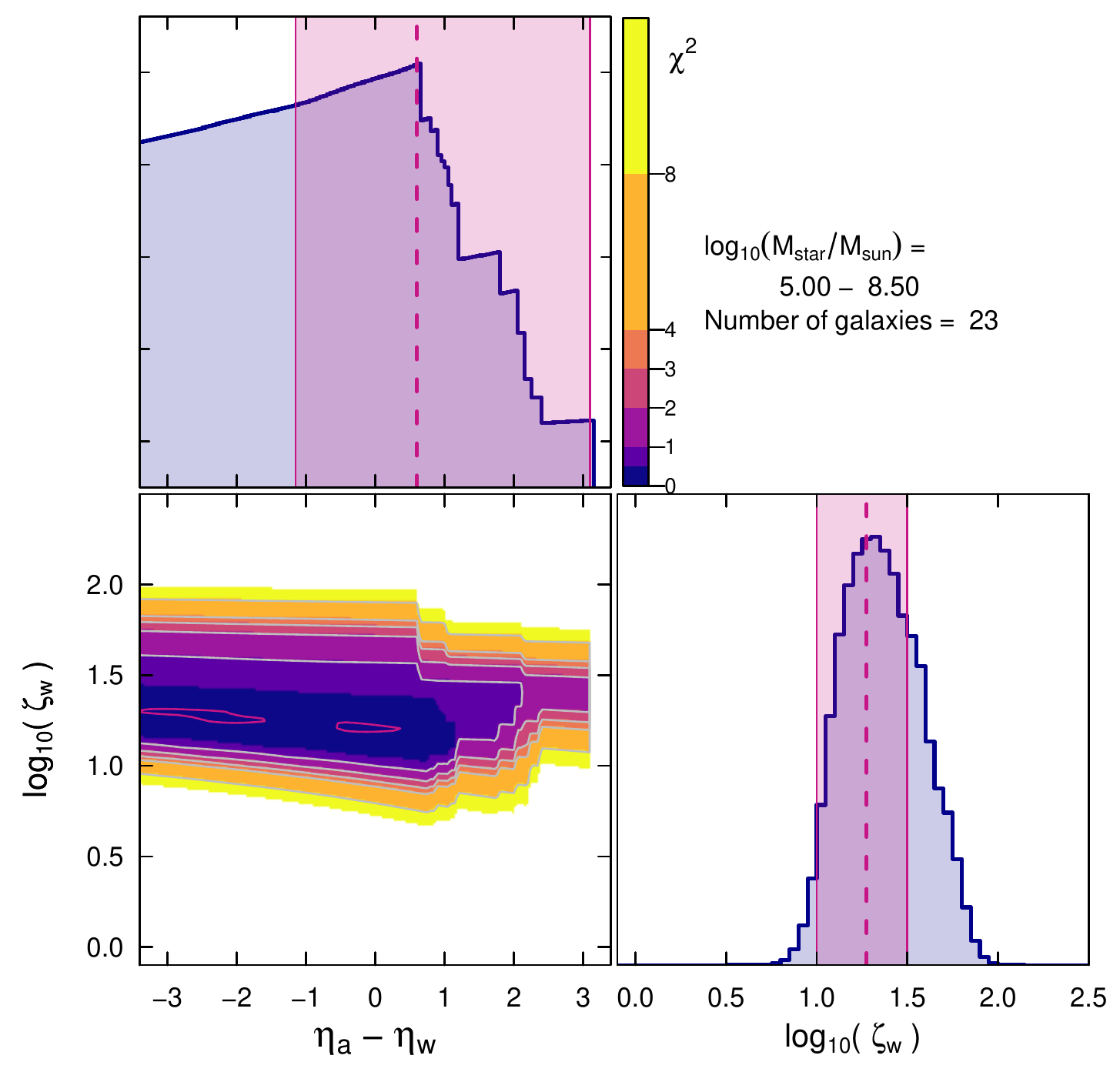}
\includegraphics[angle=0,width=0.33\linewidth]{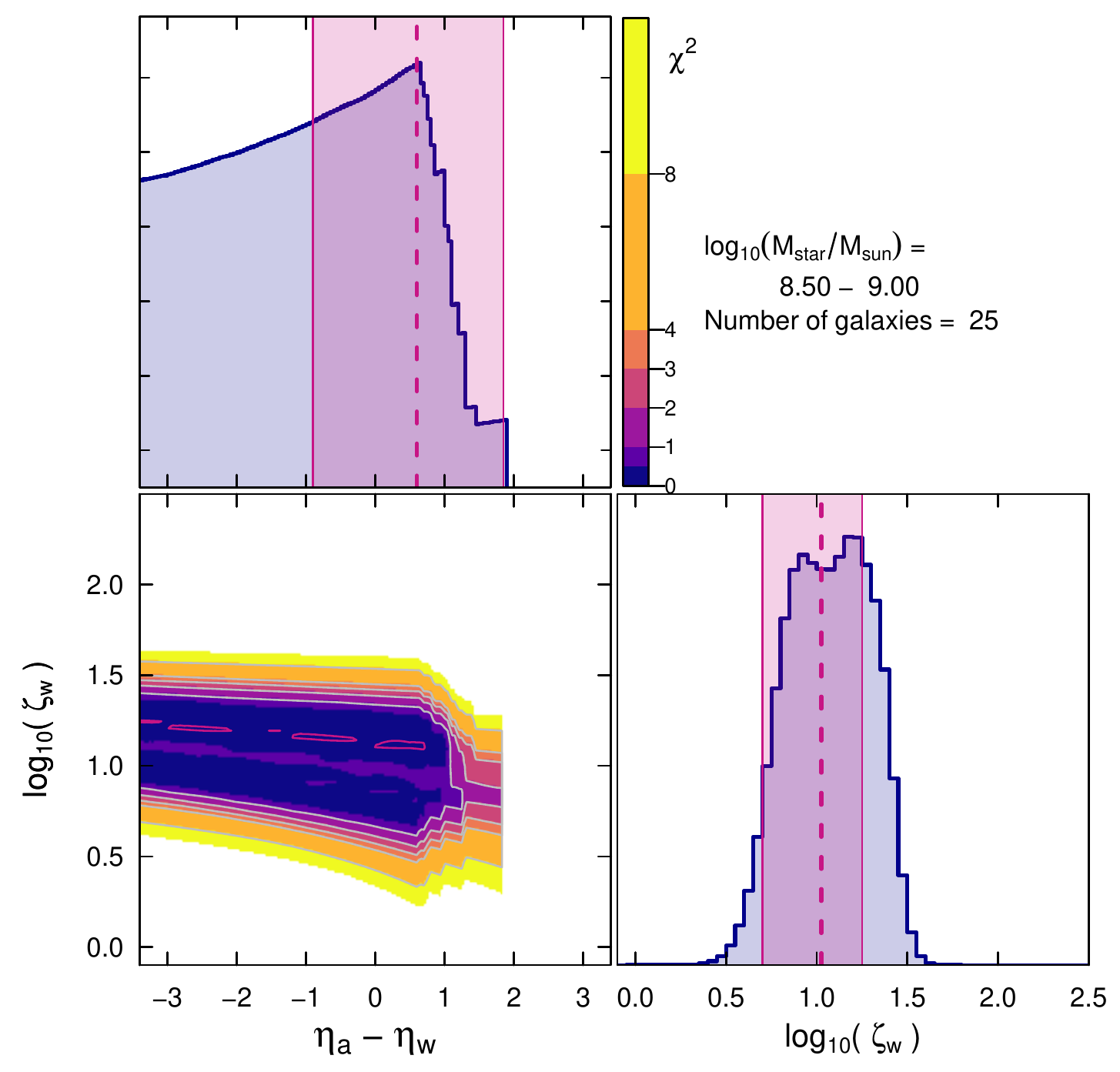}
\includegraphics[angle=0,width=0.33\linewidth]{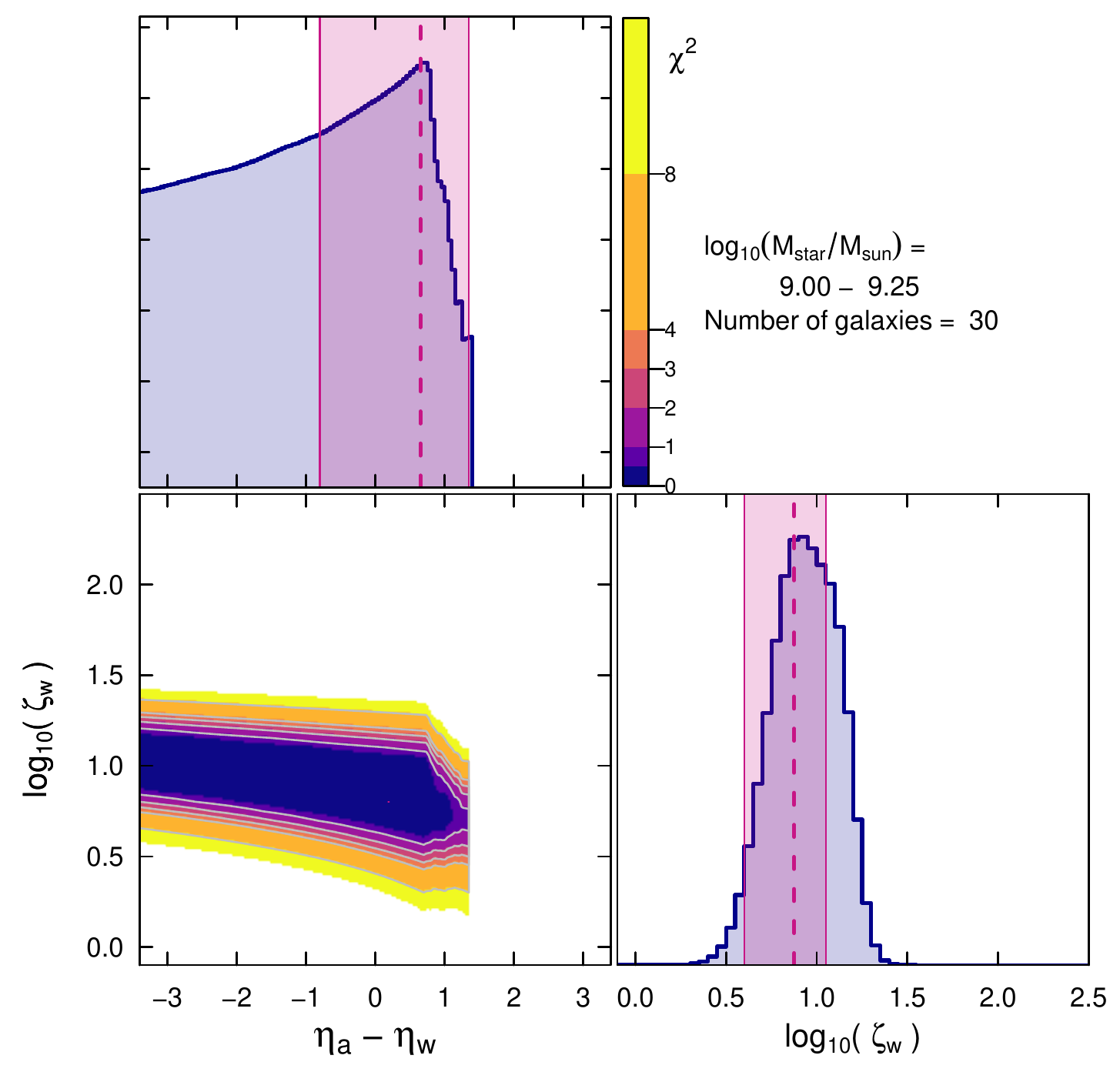}
}
\vspace{\baselineskip}
\hbox{
\includegraphics[angle=0,width=0.33\linewidth]{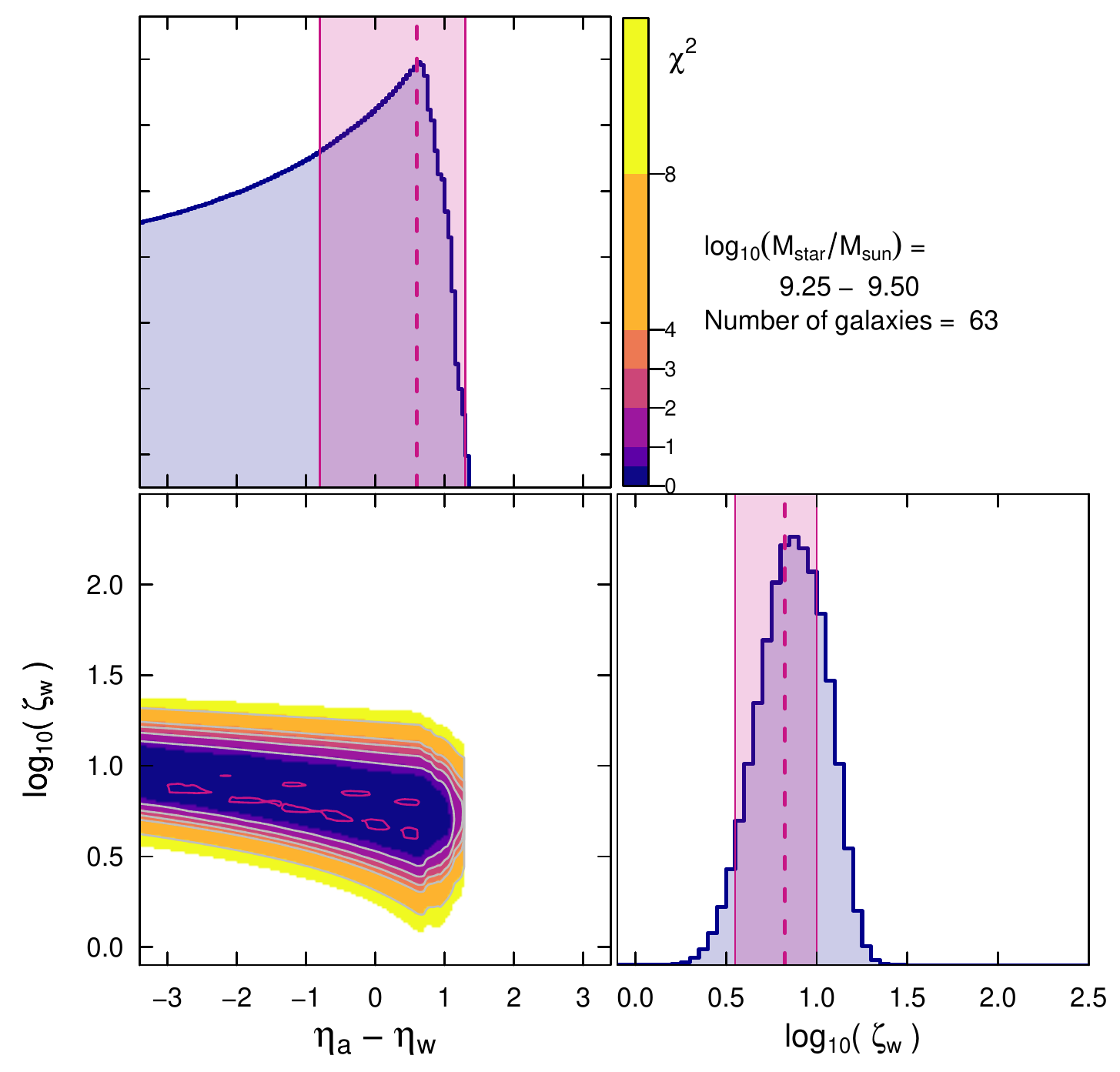}
\includegraphics[angle=0,width=0.33\linewidth]{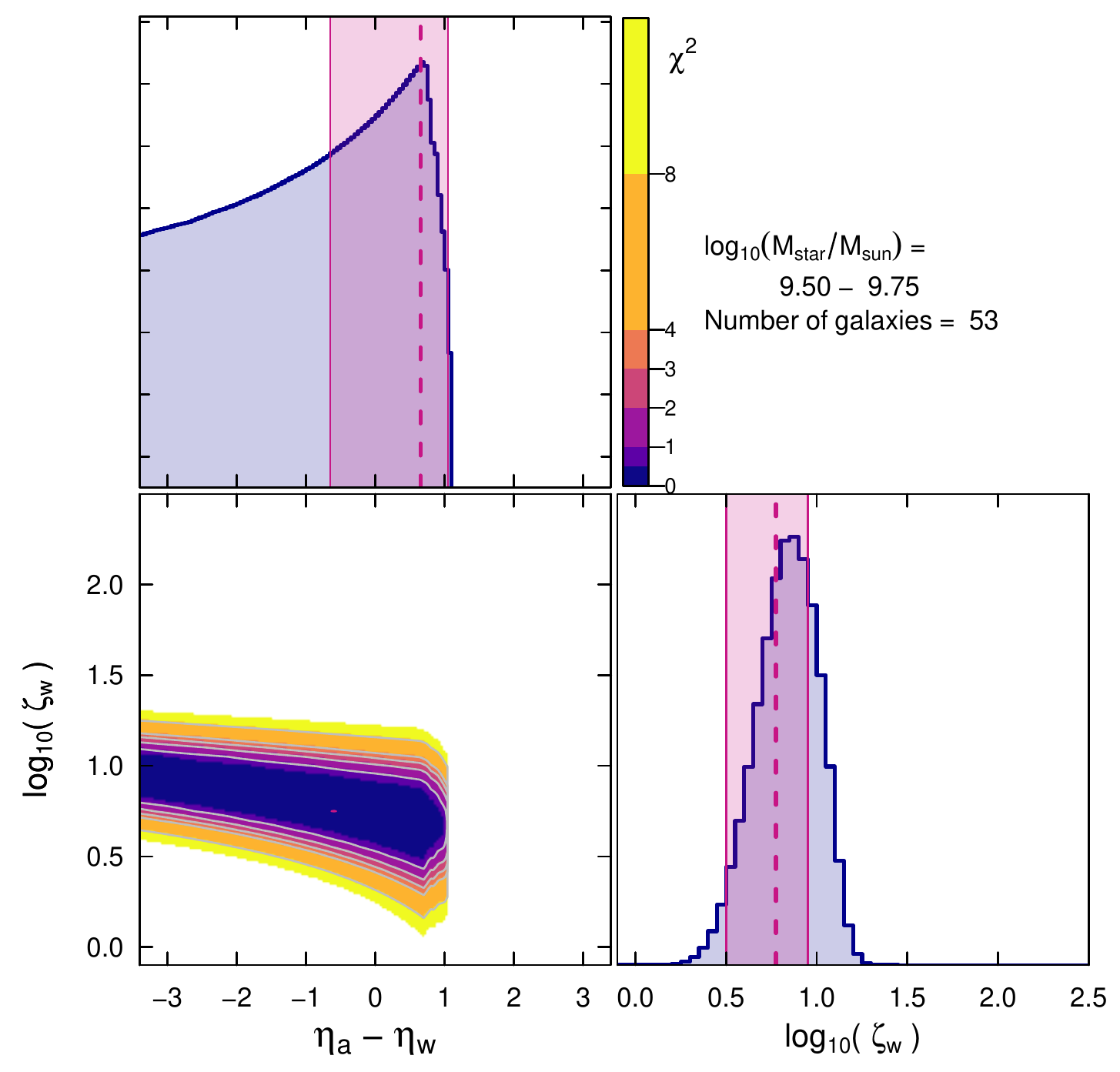}
\includegraphics[angle=0,width=0.33\linewidth]{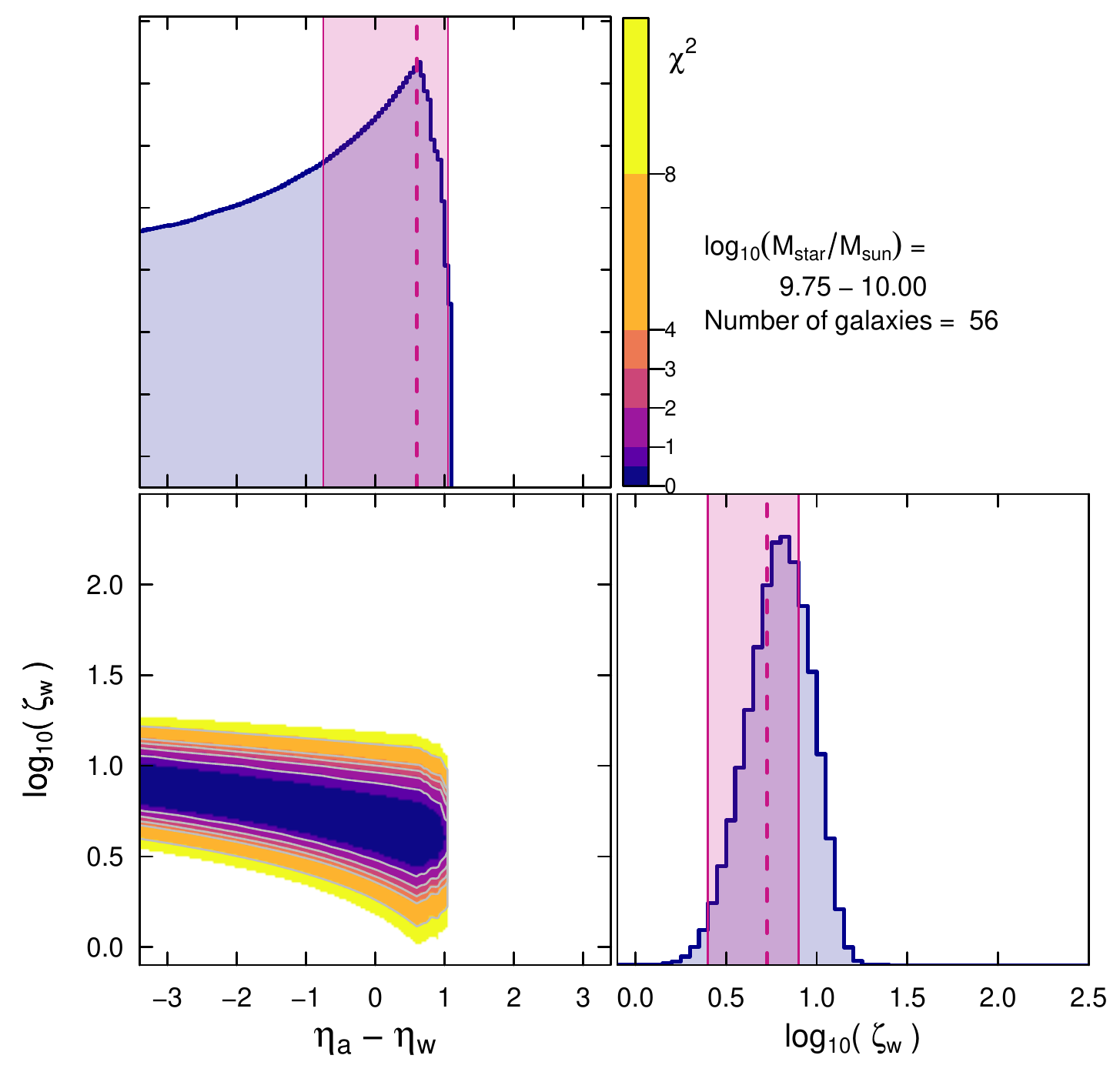}
}
\vspace{\baselineskip}
\hbox{
\includegraphics[angle=0,width=0.33\linewidth]{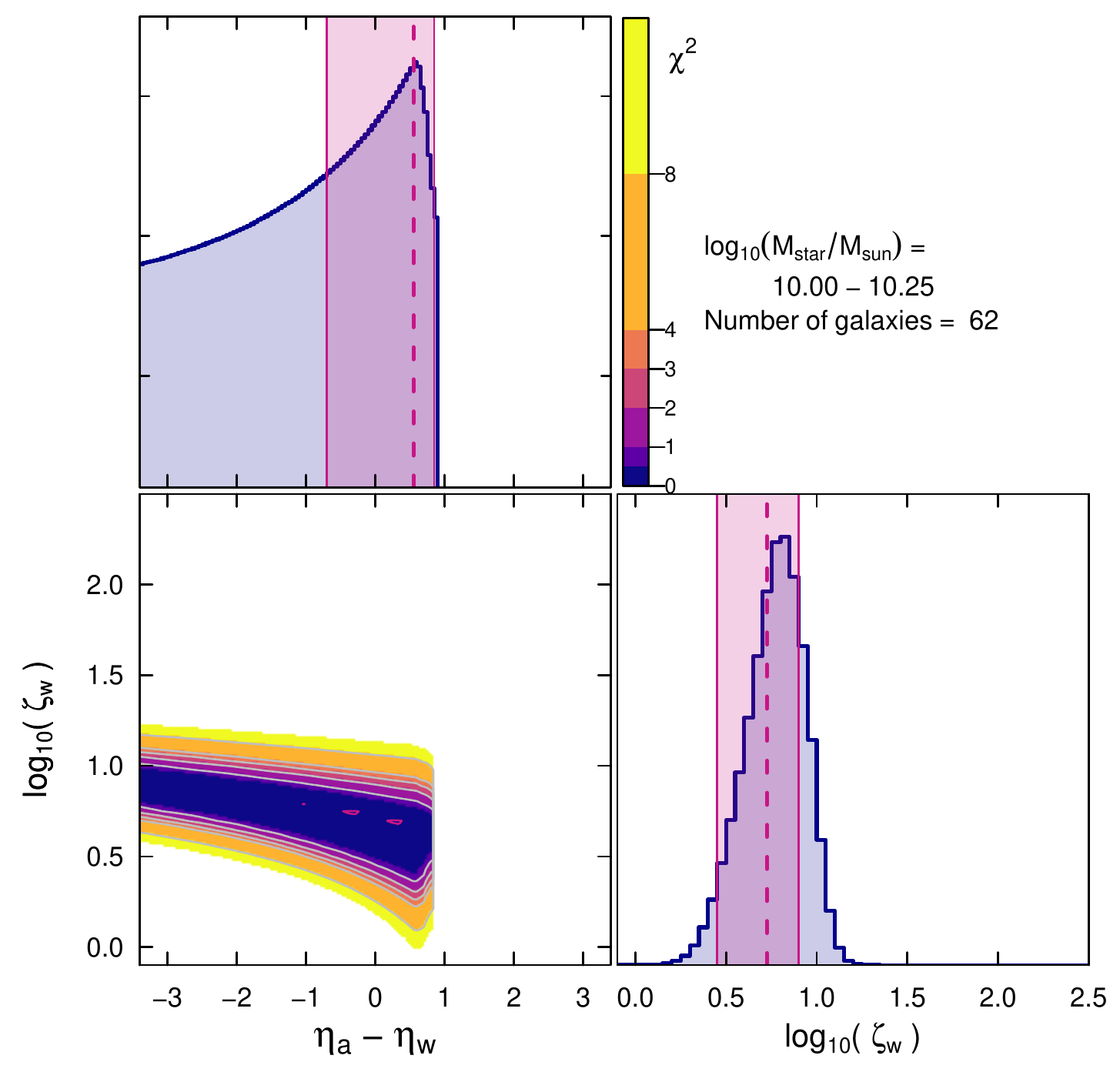}
\includegraphics[angle=0,width=0.33\linewidth]{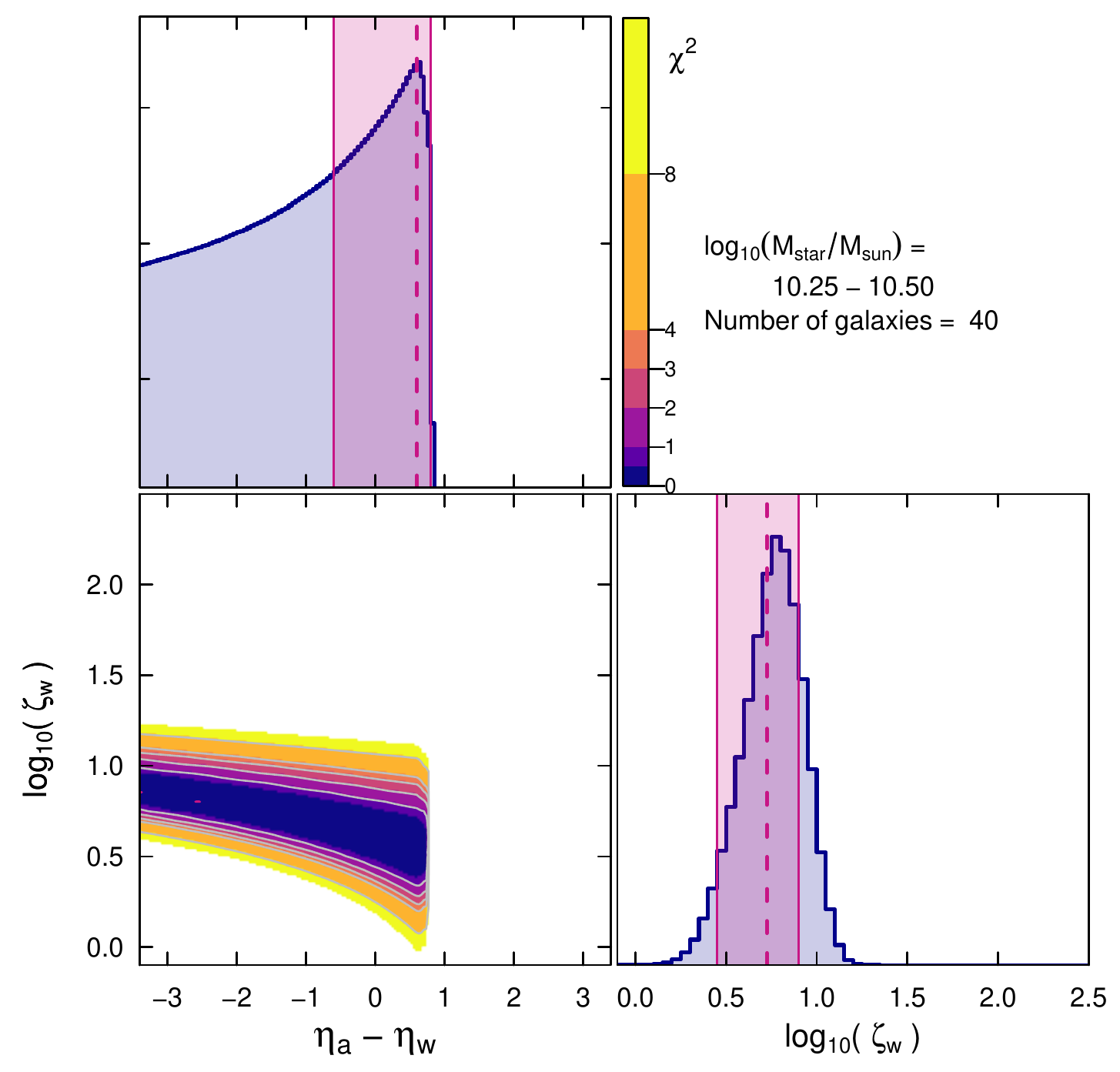}
\includegraphics[angle=0,width=0.33\linewidth]{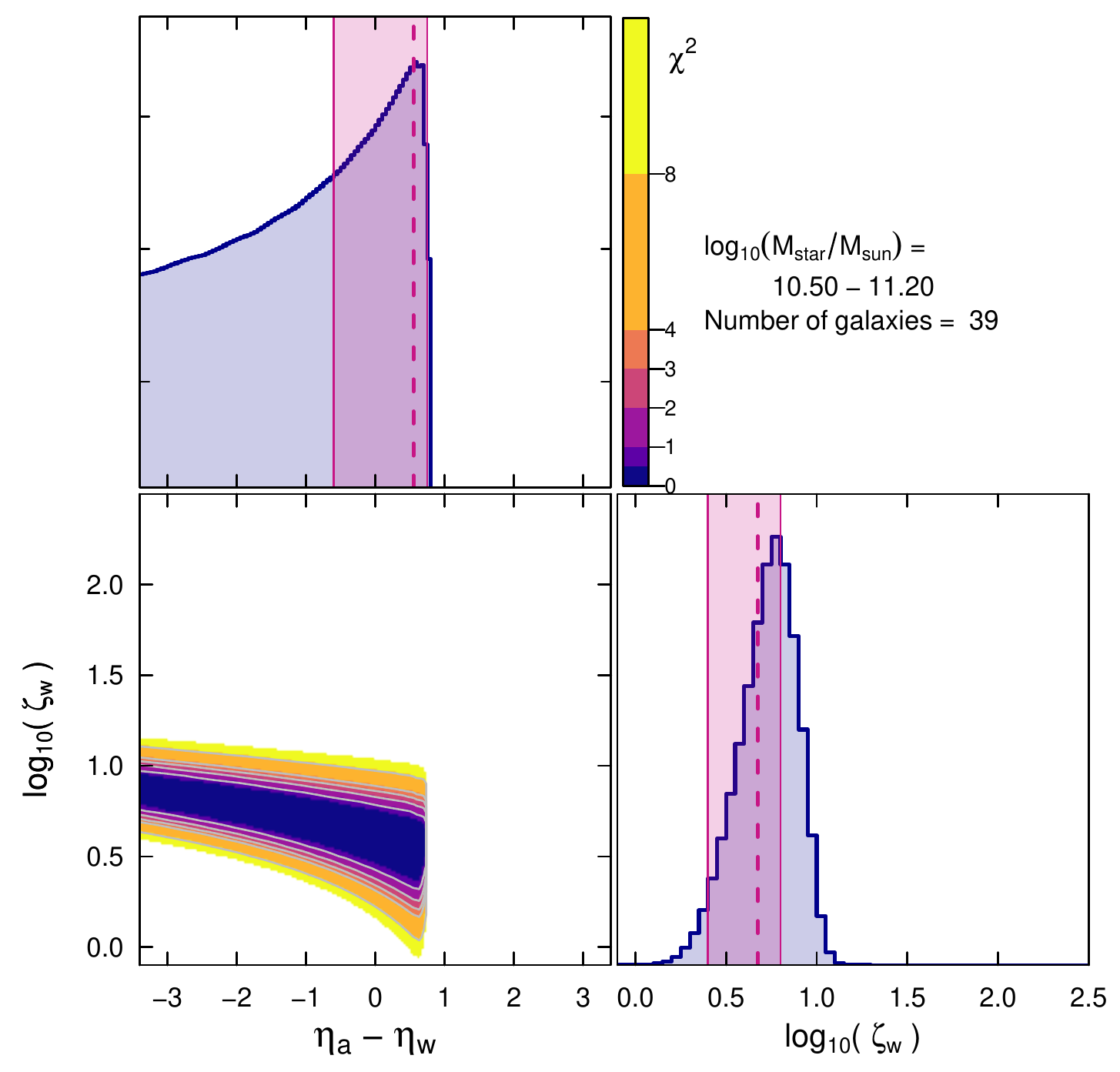}
}
\end{center}
\vspace{-1.2\baselineskip} \caption{Corner plots of \chisq\
surface as a function of the model parameters (\deta, \zetaw) for
MAGMA binned by \mstar.
The violet contours correspond to the minimum \chisq\
value. The top and right panels of each corner plot report the
probability density distributions for the marginalized parameters;
confidence intervals ($\pm 1\sigma$) are shown as violet-tinted
shaded rectangular regions, and the MLE (PDF median) is shown by a
vertical dashed line.
\label{fig:bayes} }
\end{figure*}

\subsection{Bayesian approach}
\label{sec:bayes}

For each galaxy with four observables (\mstar, SFR, \zg, \mgas),
we need to determine two parameters, \deta\ and \zetaw. However,
only two observables enter into Eq.\,\eqref{eqn:zg}: \zg\ and
\mug. Thus, the problem is severely underdetermined. To overcome
this limitation, we have divided MAGMA into \mstar\ bins, and
assumed that \deta\ and \zetaw\ are the same for each mass bin.
The \mstar\ bins were defined to guarantee a sufficient number of
galaxies in each bin, but also to adequately emphasize the
lowest-\mstar\ MAGMA galaxies.

To find the best-fit combination of (\deta, \zetaw) for each \mstar\ bin,
we first calculate $\chi^2$ value for each of N galaxies
in the bin\footnote{This is not particularly computationally tedious because
we have a few tens of galaxies within each mass bin, and only two parameters to determine.}:
\begin{equation}
\chi_{i,j}^{2} = \left(\frac{Z_g^{\mathrm{obs}(j)} -
Z_g^{\mathrm{mod}(i)}}{\sigma} \right)^{2}\quad , \label{eqn:chi2}
\end{equation}
where \zg$^{\mathrm{obs}(j)}$ for the $j^\mathrm{th}$ galaxy is
given by Eq.\,\eqref{eqn:zgconv} and \zg$^{\mathrm{mod}(i)}$ by
Eqs. \eqref{eqn:zg} and \eqref{eqn:mgmi} for the $i^\mathrm{th}$
parameter pair (\deta, \zetaw); we have assumed a constant error
$\sigma\,=\,0.1$, similar to the deviation of the MZR fit
discussed in \papi. \chisq\ is computed over the grid of \deta\
and \zetaw\ parameters, in order to obtain $N$ values [indexed by
$j$ in Eq. \eqref{eqn:chi2}] of \chisq\ for each parameter pair
(\deta, \zetaw).

Since only one value of \chisq\ can be associated with each unique parameter pair,
a decision must be made for the assignment of this single value for the $N$ galaxies in the mass bin.
This can sometimes be accomplished through stacking \citep[e.g.,][]{Belfiore+19_bathtub},
but we cannot do that here, because
of the way \zg$^\mathrm{mod}$ is defined [Eqs. \eqref{eqn:zg}, \eqref{eqn:mgmi}].
We experimented with several methods; the approach that best captures the sensitivity
of the \chisq\ distribution to the variation in the parameters (\deta, \zetaw)
is to assign the single \chisq\ value
to the mean of the lowest quartile in the \chisq\ distribution of the $N$ galaxies in the mass bin.
This is a rather arbitrary choice but the numbers of galaxies in each \mstar\ bin
are sufficient to sample in a meaningful way the galaxies with the best fits.
Thus for each mass bin, assigning this single \chisq\ value to each parameter pair,
we obtain the distribution of \chisq\ values for that bin.

We then apply the standard Bayesian formulation:
\begin{equation}\label{eqn:bayes}
P({\bm\theta}|{\bm D}) \propto P({\bm \theta})
P({\bm D}|{\bm \theta})
\end{equation}
to derive the full posterior probability distribution $P({\bm
\theta}|{\bm D})$ of the parameter vector ${\bm
\theta}$\,=\,(\deta, \zetaw), given the data vector ${\bm
D}$\,=\,(\zg, \mug). This posterior is proportional to the product
of the prior $P({\bm \theta)}$ on all model parameters (the
probability of a given model being obtained without knowledge of
the data), and the \emph{likelihood} $P({\bm D}|{\bm \theta})$
that the data are compatible with a model generated by a
particular set of parameters. The data are assumed to be
characterized by Gaussian uncertainties, so the likelihood of a
given model is proportional to $\exp(-\chi^2/2)$. We assume
uniform priors: \hbox{\deta\,$\in$\,[-3.4,3.4]} in linearly
sampled steps, and $\log_{10}$(\zetaw)\,$\in$\,[-0.1,2.5] in
logarithmic steps of 0.05\,dex.

The best-fit parameter vector (\deta, \zetaw) is evaluated by
constructing the probability density functions (PDFs), weighting
each model with the likelihood $\exp(\chi^2/2)$, and normalizing
to ensure a total probability of unity. Specifically, this was
achieved by marginalization over other parameters:
\begin{equation}
\label{eqn:marginalize}
P({\bm\theta}|{\bm D})\,= \int \,P({\bm \theta},Y|{\bm D})\,dY
\end{equation}
where $Y$ is either \deta\ or \zetaw, excluding the parameter of
interest (\zetaw\ or \deta, respectively).

The results for the different mass bins are shown in Fig.  \ref{fig:bayes}.
The \chisq\ surfaces show the relation between \deta\ and \zetaw.
For each \mstar\ bin, the PDF median is shown as a vertical dashed
line, and the 1$\sigma$ confidence intervals (CIs) by a shaded
violet region. We have taken the maximum-likelihood estimate (MLE)
to be the median of the PDF
for \zetaw\ and the mode for \deta.
This choice is discussed more fully below.

Figure \ref{fig:bayes} shows that \deta\ is degenerate; low values
of \chisq\ can be achieved for a broad range of \deta. However,
strikingly, the spread of allowed values for \deta\ at low \mstar\
(in the two upper left-most panels in Fig. \ref{fig:bayes}) is
much broader than for higher stellar masses. In the higher \mstar\
bins, the allowed distribution of \deta\ is truncated to
increasingly small values of \deta. This implies that \etaa, with
equal probability, can be significantly larger than \etaw\ at low
masses, as also shown by the PDFs in each panel.
Another clear result that emerges from Fig. \ref{fig:bayes} is that higher \zetaw\ occur 
at lower \mstar.

Figure \ref{fig:bayes} also shows that while the PDF of \zetaw\ shows a form similar to a Gaussian
distribution, and is thus fairly well constrained, the PDF for \deta\ is not.
It peaks at a given value, but then falls off toward lower values gradually, leading
to the conclusion that the best-fit value may depend on the interval over which the
uniform prior is sampled.
To investigate this, we have performed the Bayesian calculations using two different intervals
in \deta:
\hbox{\deta\,$\in$\,[-10,3.4]} and
\hbox{\deta\,$\in$\,[0,3.4]},
again in linearly sampled steps.
These cornerplots are shown in Appendix \ref{app:bayesian},
and the best-fit values for \deta\ and \zetaw\ for \hbox{\deta\,$\in$\,[0,3.4]} are given in the last two columns
of Table \ref{tab:loading_data}.

\begin{table}[!h]
\caption{\deta\ and \zetaw\ obtained in bins of mass for Bayesian
method$^{a}$.} \label{tab:loading_data}
\begin{tabular}{ccc@{\extracolsep{6pt}}cc@{\extracolsep{3pt}}c}
\hline
\\
&& \multicolumn{2}{c}{\deta\,$\in$\,[-3.4,3.4]}
& \multicolumn{2}{c}{\deta\,$\in$\,[0,3.4]} \\
\cline{3-4} \cline{5-6}
\\
log(\mstar/\msun)\,bin & N & \deta\ & $\log_{10}$\zetaw\ & \deta\ & $\log_{10}$\zetaw\ \\ 
\\
\hline
\\
5--8.5       & 23 &  0.60        &  1.275      &     0.60       &   1.275\\
8.5--9       & 25 &  0.60        &  1.025      &     0.60       &   0.925\\
9--9.25      & 31 &  0.65        &  0.875      &     0.65       &   0.775\\
9.25--9.5    & 63 &  0.60        &  0.825      &     0.60       &   0.725\\
9.5--9.75    & 53 &  0.65        &  0.775      &     0.65       &   0.675\\
9.75--10     & 56 &  0.60        &  0.725      &     0.60       &   0.625\\
10--10.25    & 62 &  0.55        &  0.725      &     0.55       &   0.575\\
10.25--10.50 & 40 &  0.60        &  0.725      &     0.60       &   0.575\\
10.5--11.20  & 39 &  0.55        &  0.675      &     0.55       &   0.525\\
\\
\hline
\\
\end{tabular}
$^{a}$~The mode of the PDF is given for \deta, and the median for \zetaw.
\end{table}

\begin{figure}[!t]
\begin{flushright}
\includegraphics[width=0.971\linewidth]{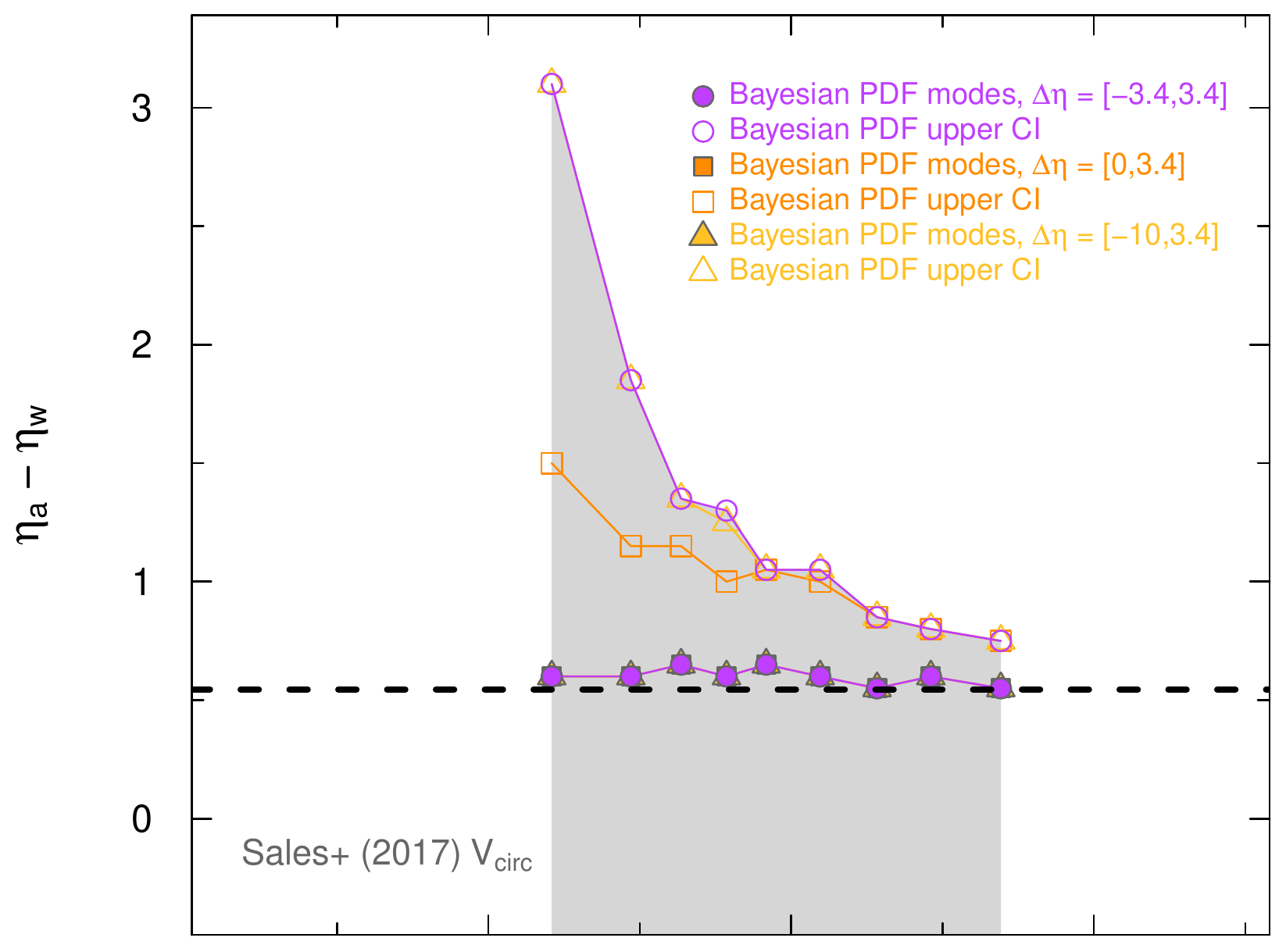} \\
\vspace{-0.5\baselineskip}
\includegraphics[width=0.98\linewidth]{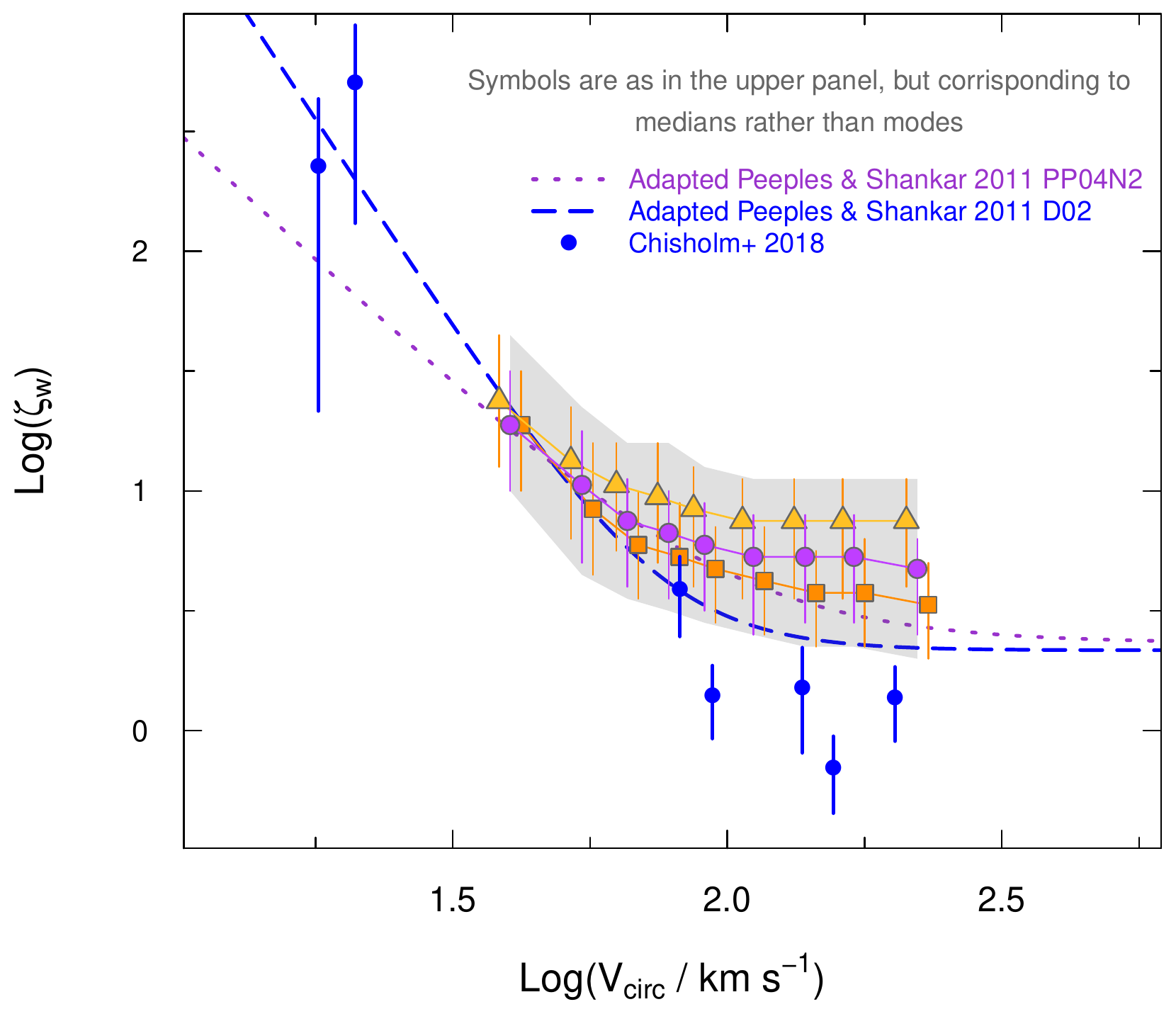}
\end{flushright}
\caption{Difference of mass-loading factors \deta\ (top panel) and
metal-loading factors \zetaw\ (bottom) as a function of virial
velocities obtained from the BTFR from \citet{Sales2017}. In both
panels, filled symbols show the Bayesian results, and in the top
panel, open ones illustrate the upper bound of the PDF $1\sigma$
confidence intervals (CI). Different symbol types correspond to
the interval of \deta\ in the Bayesian priors, as described in the
text; the symbols for \deta\,$\in$\,[0,3.4] and
\deta\,$\in$\,[-10,3.4] are (arbitrarily) offset along the
abscissa for better visibility. The gray regions give the
$\pm\,1\sigma$ uncertainties in the best-fit values. Also shown is
the ``equilibrium'' asymptote, namely \deta\,=\,$\alpha$ (see
Sect. \ref{sec:gce}). In the top panel, the two estimates of
\deta\ from different priors are exactly coincident. In the bottom
panel, the formulations by \citet{Peeples_Shankar11} for two O/H
calibrations, redone here for our value of $y$, are given by
(purple) short-dashed and (blue) long-dashed curves for the PP04N2
and D02 calibrations, respectively. Also shown are the
metal-loading factors computed by \citet{chisholm18} from COS
data. The text gives more details.} \label{fig:loadings_vcirc}
\end{figure}

Results are shown graphically in Fig. \ref{fig:loadings_vcirc},
where \deta\ is plotted against \vcirc\ in the upper panel and
\zetaw\ vs. \vcirc\ in the lower one. Because the outflow
strengths probably depend on the depth of the potential well, we
assess them in terms of the virial velocities, \vcirc, as proposed
by \citet{Peeples_Shankar11}. We have calculated virial velocities
in several ways, but ultimately adopt $V_\mathrm{out}$, velocities
from the baryonic Tully-Fisher relation (BTFR) predicted by
APOSTLE/EAGLE simulations from \citet{Sales2017}, focused on the
low-mass regime. Other alternatives are also viable, such as the
power-law BTFR from \citet{Lelli+16_SPARC}, or velocities obtained
from the \Mvir--\mstar\ relation via halo abundance matching
\citep{Moster+10}. However, at low \mstar, and low baryonic
masses, the inferred virial velocities change significantly
according to whether the relation is a straight power law as in
\citet[e.g.,][]{Lelli+16_SPARC} or inflected at low masses as in
\citet{Sales2017}. The simple power-law relation gives \vcirc\
values that are apparently too low relative to the rest of the
sample, while the \vcirc\ values by \citet{Sales2017} are more
consistent. Moreover, because the low-mass end of our sample is
somewhat extreme, there is also some doubt about the applicability
of the usual halo-abundance matching techniques. In any case, our
conclusions are independent of the formulation used to translate
the mass scale to the kinematic one.

As shown in Fig. \ref{fig:loadings_vcirc}, the different intervals
used for the Bayesian priors give fairly consistent results both
for \deta\ and for \zetaw. The best-fit Bayesian value for \deta\
is $\approx\,\alpha$ for all the prior intervals sampled, and
there is a clear trend for \zetaw\ to increase toward lower
\vcirc\ (or stellar mass, as we show later). Over the mass range
sampled by MAGMA, \zetaw\ is almost a factor of ten higher at
lower \mstar\ (\vcirc) relative to the most massive galaxies.

Also shown in the bottom panel of Fig. \ref{fig:loadings_vcirc}
are the estimates of \zetaw\ obtained by \cite{chisholm18} with
Cosmic Origins Spectrograph (COS) observations of ultraviolet
absorption lines. By analyzing the underlying stellar continuum,
and using the observed line profiles to constrain optical depth
and covering fraction, \citet{chisholm18} were able to quantify
the mass outflow rate, $\dot{M}_w$. Then, through photoionization
modeling, they obtain column densities of each element that are
then used to constrain the metallicity of the outflow \zw. As
shown in \citet{chisholm18}, their results are well fit by the
analytical predictions of \citet{Peeples_Shankar11}, with yields
in the range 0.007--0.038, which agree with our reference value of
$y\,=\,0.037$.

\subsection{Comparison with a closed-form approximation}
\label{sec:closedform}

Figure \ref{fig:loadings_vcirc} also illustrates the
predictions of \citet{Peeples_Shankar11} using their formulation
for the MZR \citep[taken from][and converted to a
\citealt{Chabrier2003} IMF]{Kewley_Ellison08}.
\citet{Peeples_Shankar11} inferred \zetaw\ from the behavior of
\mgas\ and \zg\ with \mstar, using known derivatives derived
empirically; we discuss this approach more fully below. Figure
\ref{fig:loadings_vcirc} shows their formulation relating \zetaw\
and \vcirc:
\begin{equation}
\zeta_w \,=\, \left( \frac{V_0}{V_{\rm circ}} \right)^b + \zeta_0 \quad ,
\label{eqn:zetawmodel}
\end{equation}
where, for consistency, we have recomputed their fits using our
higher adopted yield (for $y_{\rm O}$) of 0.037 taken from
\citet{Vincenzo+16_yields} for a \citet{Chabrier2003} IMF. As
shown in Fig. \ref{fig:loadings_vcirc}, the PP04N2 results are in
good agreement with MAGMA, while the D02 calibration
\citep{Denicolo+02} is slightly discrepant, and more consistent
with \citet{chisholm18}, despite the higher yield. This
illustrates how the computed \zetaw\ values depend on the adopted
yields, and underscores the necessity of a self-consistent
framework. We have performed our entire metallicity analysis with
lower (and higher) yields than the ones used here, and the results
do not change.

\citet{Peeples_Shankar11} assessed \deta\ and \zetaw\
based on known derivatives calculated from observed scaling relations
of \mgas\ and \zg\ with \mstar. In some sense, this is
complementary to our Bayesian approach, but may also skew results
as we show below. Following \citet{pagel09}, the time dependence
in Eq.\,\eqref{eqn:mgtimederiv_again} can be eliminated by noting
that $\dot{M}_{\Large\star}\,=\,\alpha\,\psi$. Thus, by changing
variables we obtain:

\begin{equation}
\frac{dM_g}{dM_{\Large\star}}\,=\,\frac{\dot{M}_g}{\dot{M}_{\Large\star}}\,=\,\frac{\dot{M}_a - \dot{M}_w}{\alpha\,\psi} - 1
\,=\,\left( \frac{\eta_a - \eta_w}{\alpha} \right) - 1
\quad .
\label{eqn:mgmassderiv}
\end{equation}
The same can be done for Eq.\,\eqref{eqn:mztimederiv_again}, and
eliminating the time dependence as above gives:
\begin{equation}
\frac{dM_Z}{dM_{\Large\star}}\,=\,y + Z_g\,\left( \frac{\zeta_a -
\zeta_w}{\alpha} - 1 \right) \quad . \label{eqn:mzmassderiv}
\end{equation}

Our observations do not directly constrain \mz, but rather \zg, so we need to
formulate $dZ_g/dM_{\Large\star}$ as follows:
\begin{equation}
\frac{dM_Z}{dM_{\Large\star}}\,=\,Z_g\,\frac{dM_g}{dM_{\Large\star}}
+ M_g\,\frac{dZ_g}{dM_{\Large\star}}\quad .
\end{equation}
The left-hand side is given by Eq.\,\eqref{eqn:mzmassderiv}, and
$dM_g/dM_{\Large\star}$ by Eq.\,\eqref{eqn:mgmassderiv}, so we can
solve for $dZ_g/dM_{\Large\star}$:
\begin{equation}
\frac{dZ_g}{dM_{\Large\star}}\,=\,\frac{1}{\alpha\,M_g}\,\left[ q + Z_g\,(\zeta_a - \zeta_w - \eta_a + \eta_w) \right]\quad .
\label{eqn:zgmassderiv}
\end{equation}
As they should be, Eqs.\,\eqref{eqn:zgmassderiv} and
\eqref{eqn:mgmassderiv} are related to Eqs.\,\eqref{eqn:zgtimederiv}
and \eqref{eqn:mgtimederiv_again} by the simple derivative,
$\dot{M}_{\Large\star}\,=\,\alpha\,\psi$.
Eqs.\,\eqref{eqn:zgmassderiv} and \eqref{eqn:mgmassderiv} are
slightly different from the ones derived by
\citet{Peeples_Shankar11}, because of the different
assumptions they made for the stellar enrichment given by
Eq.\,\eqref{eqn:mztimederiv}.

For a given \mgas\,vs.\,\mstar\ relation, we know $dM_g/d$\mstar,
the left side of Eq.\,\eqref{eqn:mgmassderiv}; for a given MZR, we
know $dZ_g/d$\mstar, the left side of
Eq.\,\eqref{eqn:zgmassderiv}. Thus, in the spirit of
\citet{Peeples_Shankar11}, \deta\,=\,$\eta_a - \eta_w$ and \zetaw\
can be directly constrained observationally. Inverting Eqs.
\eqref{eqn:mgmassderiv} and \eqref{eqn:zgmassderiv} gives \deta\
and \zetaw\ explicitly:
\begin{equation}
\eta_a - \eta_w\,=\,\alpha\,\left( F_g\ \frac{d\log M_g}{d\log M_{\Large\star}} + 1 \right) \quad ,
\label{eqn:deltaeta}
\end{equation}
\begin{eqnarray}
\zeta_w&=&\frac{q}{Z_g} - \alpha\,\left[ F_g \left( \frac{d\log Z_g}{d\log M_{\Large\star}} +
\frac{d\log M_g}{d\log M_{\Large\star}} \right) + 1 \right] \nonumber \\
&=&\frac{q}{Z_g} - (\eta_a - \eta_w) - \alpha\,F_g\,\frac{d\log Z_g}{d\log M_{\Large\star}} \quad .
\label{eqn:zetaw}
\end{eqnarray}
In Eq.\,\eqref{eqn:zetaw}, we have defined
\fgas\,$\equiv$\,\mgas/\mstar. \citet{Peeples_Shankar11} were the
first to derive such a formulation for \zetaw; as mentioned in
Sect. \ref{sec:formalism}, the formalism here differs slightly
from theirs because of the way they dealt with metal production in
stellar ejecta.

The problem with Eq. \eqref{eqn:deltaeta} is that for positive
$d\log\,M_g/d\log$\,\mstar, \deta\ is by definition $>$ 0, because
\fgas\ is $\geq\,0$. In \citetalias{hunt20}, the robust
least-squares power-law relation\footnote{For all statistical
analysis, we rely on the {\it R} statistical package: R Core Team
(2018), R: A language and environment for statistical
  computing, R Foundation for Statistical Computing, Vienna, Austria
  (\url{https://www.R-project.org/}).}
between \mgas\ and \mstar\ is found to be:
\begin{equation}
\log(M_{\rm g})\,=\,(0.72\,\pm\,0.02)\ \log(M_{\Large\star}) + (2.51\,\pm\,0.22)\quad ,
\label{eqn:mgasmstar}
\end{equation}
thus with $d\log\,M_g/d\log$\,\mstar\,$\approx\,$0.7,
roughly consistent with earlier works \citep[e.g.,][]{leroy08}.
Because of this positive derivative, to infer \deta\ by
relying on Eq. \eqref{eqn:deltaeta} and $d\log\,M_g/d\log$\,\mstar\
would give results that are inherently skewed toward higher values of \deta.
Qualitatively, we would expect that as a galaxy evolves, the stellar mass continues to grow,
while the gas mass diminishes; such a scenario would give $d\log\,M_g/d\log$\,\mstar\,$< 0$,
and possibly \deta\,$< \alpha$.
Consequently, from a physical point of view,
it may be inaccurate or unrealistic to invoke the sample's overall $d\log\,M_g/d\log$\,\mstar\ to
calculate \deta.
What we require is an estimate of \deta\ for individual galaxies which may or may not be related
to the statistical properties of the parent sample at z$\approx$0.

\begin{figure*}[!t]
\includegraphics[width=0.48\linewidth]{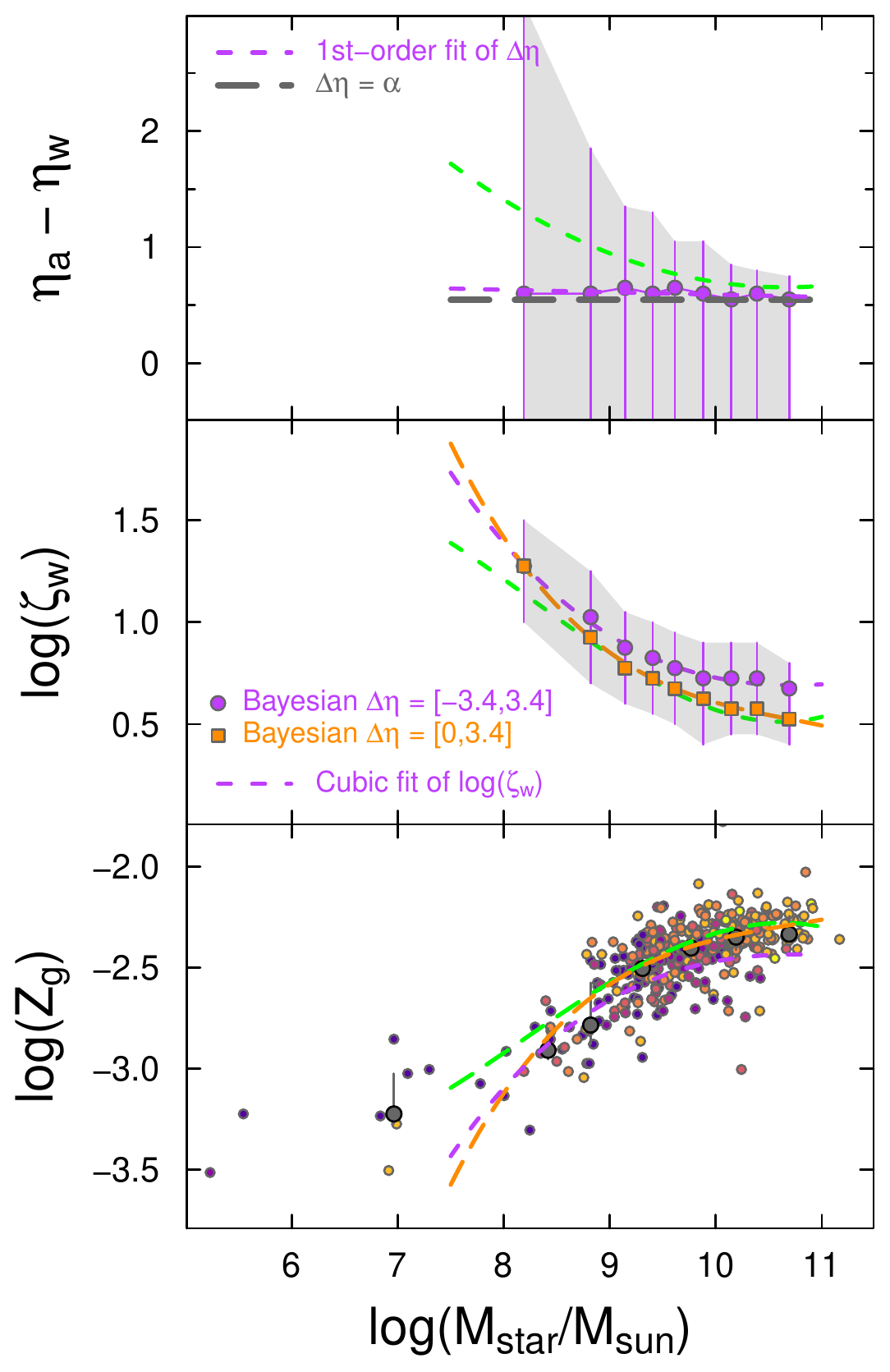}
\hspace{0.04\linewidth}
\includegraphics[width=0.48\linewidth]{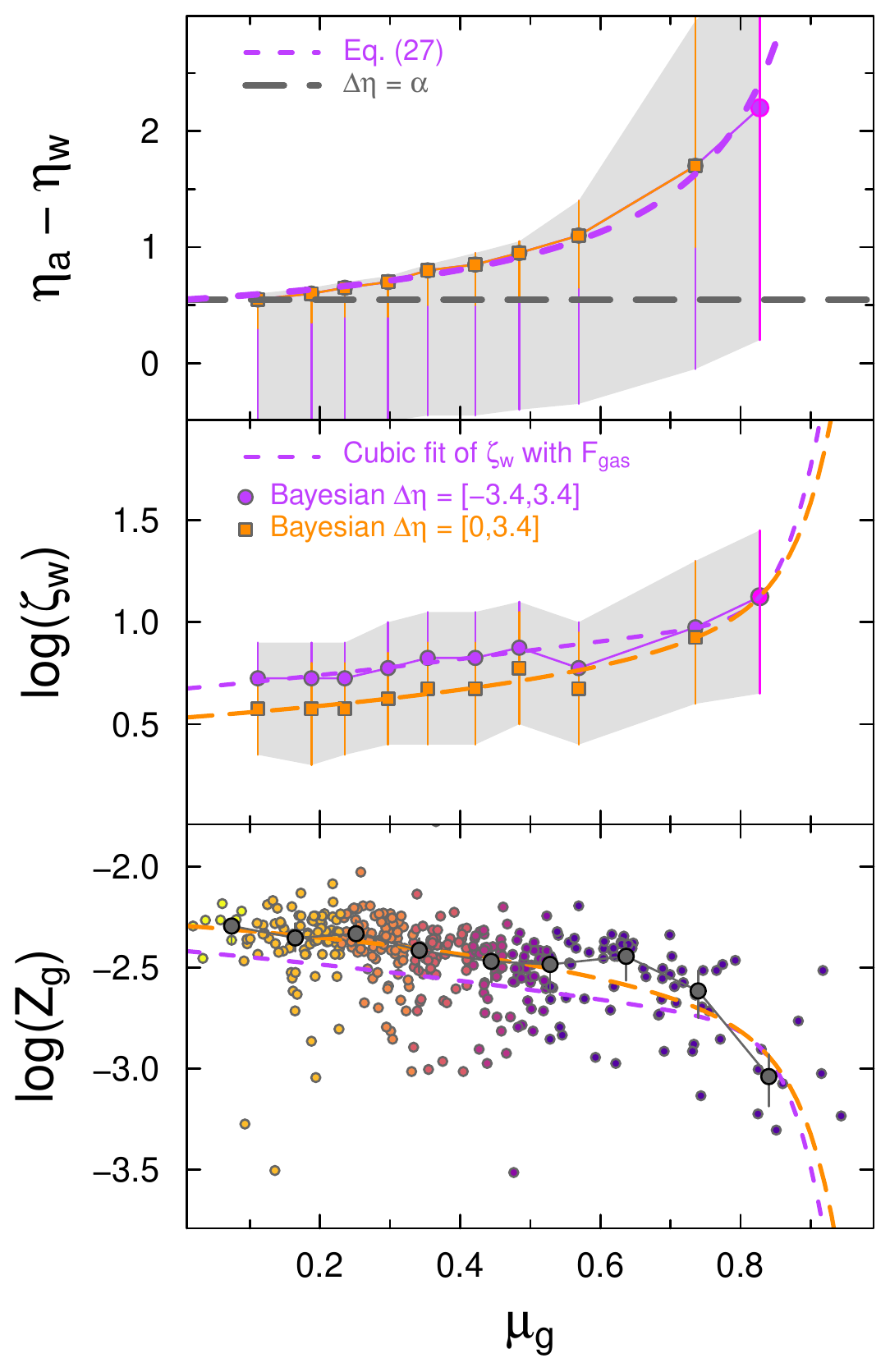}\\
\vspace{-0.8\baselineskip} \caption{Difference of mass-loading
factors \deta\ and metal-loading factors \zetaw\ inferred from the
Bayesian analysis, and observed \zg\ in the bottom panel, as a
function of \mstar\ on the left, and as a function of \mug\ on the
right. In the two upper panels, the best-fit Bayesian values are
shown as filled symbols (circles or squares according to the
\deta\ prior), with error bars corresponding to the $\pm\,1\sigma$
excursion, which is also illustrated by light-gray shaded regions.
The rightmost point (magenta circle) in the top and middle right
panels corresponds to the ``extra'' high-\mug\ bin shown in Fig.
\ref{fig:bayes_highmug}. In the top panels, a horizontal dashed
line illustrates the ``equilibrium solution'' with
\deta\,=\,$\alpha$. In the top left panel, the (purple) dashed
curve is a linear power-law fit of \deta\ to log(\mstar), while
the dashed curve in the top right panel shows Eq.
\eqref{eqn:deltaeta}, with
$d$\,log\,\mgas/$d$\,log\mstar\,=\,0.72, as found in
\citetalias{hunt20}. The fits to the curves for \zetaw\ in the
middle panels are cubic: log(\zetaw) as a function of log(\mstar)
in the left panel, and \zetaw\ as a function of \fgas\ in the
right. In the lower panels, observed MAGMA \zg\ are shown as small
filled circles, color coded by gas fraction, while the \zg\
medians and 1$\sigma$ deviations are shown by large (black) filled
circles and error bars. The dashed curves in the lower panels
correspond to the calculation of \zg\ based on the fits in the
upper panels (as a function of \mug\ on the left, and log(\mstar)
on the right): the long-dashed curves (dark orange) correspond to
the \deta$\geq$0 priors, while the short-dashed ones (purple) to
the ``symmetric'' priors. The green curves in the left panels are
described in the text, obtained by fitting individual galaxy
\deta\ and \zetaw\ calculated using Eqns. \eqref{eqn:deltaeta} and
\eqref{eqn:zetaw}. } \label{fig:loadings}
\end{figure*}

In any case, Fig. \ref{fig:loadings_vcirc} illustrates that
\zetaw\ is not particularly sensitive to \deta. The predictions of
\zetaw\ from \citet{Peeples_Shankar11} using
$d\log\,M_g/d\log$\,\mstar\ are fairly consistent with those we
find with the Bayesian approach, even though the former requires
\deta\,$>$\,0 and the latter suggests that \deta\,$\approx$\,0.

\section{Shaping the mass-metallicity relation with outflows}\label{sec:shaping}

In the previous Section, we estimated \deta\ and \zetaw\
through a Bayesian approach. Here we parameterize \deta\ and
\zetaw\ as a function of both \mstar\ and \mug\ to predict
quantitatively how mass loading \deta\ and metal loading \zetaw\
shape the MZR.

\subsection{Stellar mass, metallicity, and outflows} \label{sec:mzr_outflows}

First, we incorporate \mstar\ as the defining variable.
Figure \ref{fig:loadings} (left panel) shows the trends of
Bayesian-inferred \deta\ and \zetaw\ with observed \zg\ as a
function of \mstar\ (in logarithmic space). The bottom left panel
shows the predictions for \zg\ that are inferred from the
functional forms we have found for \deta\ and \zetaw\ (see Table
\ref{tab:loading_fits}). The middle left panel shows \zetaw\
together with the robust cubic best-fit
polynomials
with coefficients as given in Table \ref{tab:loading_fits}. The cubic
polynomial has no physical significance; rather our objective is
to parameterize \deta\ and \zetaw\ relative to \mstar\ in order to
compute \zg.

The slight differences ($\sim$0.15\,dex for \mstar$\ga
10^9$\,\msun, long-dashed orange- vs. short-dashed purple curves)
in \zetaw\ according to the \deta\ prior are evident in the middle
left panel in Fig. \ref{fig:loadings} and the lower one. When the
constraint on the prior \deta$>$0, the Bayesian inference of
\zetaw\ gives lower values, and is a better approximation of
observed \zg\ (orange long-dashed curve). On the other hand, when
the \deta\ priors  include negative values \zetaw\ is larger (see
also Fig. \ref{fig:loadings_vcirc}), but \zg\ is underestimated
(purple short-dashed curve). \textit{Since the curves shown in the
lower panel are not fits to \zg, but rather the parameterization
of \zg\ as a function of \deta\ and \zetaw, this may be telling us
that positive values over negative \deta\ are preferred, at least
in the current evolutionary states of the MAGMA galaxies.}

The green curves in the top two left panels of Fig.
\ref{fig:loadings} illustrate the polynomial fits (quadratic for
\deta\ and cubic for $\log$\,\zetaw) that would be obtained had we
calculated \deta\ and \zetaw\ from Eqns. \eqref{eqn:deltaeta} and
\eqref{eqn:zetaw}, respectively,
using $d\log Z_g / d\log M_{\Large\star}$ and $d\log M_g/d\log M_{\Large\star}$ derived
from the MAGMA scaling relations.
The corresponding green curve
in the lower left panel is the approximation of \zg\ that would be
obtained from these fits. It is clear that for
$\log$\,\mstar/\msun\,$\ga$\,8.5, \zg\ is equally well estimated
by the priors of \deta\,$\geq$\,0 and the fits to individual
parameters computed with  Eqns. \eqref{eqn:deltaeta} and
\eqref{eqn:zetaw}. Wind metal loading \zetaw\ estimated with the
symmetric prior of \deta\ falls short ($\sim$\,0.15\,dex) of
observed \zg\ in this mass range. In this mass range, Fig.
\ref{fig:loadings} also shows that \deta\ is unimportant; the main
factor that shapes \zg\ is \zetaw.

On the other hand, the fit to \deta\ in the top panel
[green curve, Eq. \eqref{eqn:deltaeta}] is significantly rising
toward lower \mstar, unlike the Bayesian values for \deta\ that,
with both sets of priors, is roughly constant $\approx\,\alpha$.
The influence of \deta\ at low \mstar\
($\log$\,\mstar/\msun\,$\la$\,8.5) is clearly shown by the green
curve in the lower left panel of Fig. \ref{fig:loadings}. The
fall-off of \zg\ toward low \mstar\ is less severe than would be
predicted by a flat \deta, and is more consistent with the
observations.

\begin{table}[!h]
\caption{Robust polynomial fits to Bayesian estimates of \deta\
and \zetaw$^\mathrm{a}$} \label{tab:loading_fits}
\resizebox{\linewidth}{!}{
\begin{tabular}{ccccc}
\hline
\\
\multicolumn{1}{c}{Dependent} & \multicolumn{1}{c}{$a_0$} &
\multicolumn{1}{c}{$a_1$} & \multicolumn{1}{c}{$a_2$} &
\multicolumn{1}{c}{$a_3$} \\
\multicolumn{1}{c}{variable} \\
\\
\hline
\\
\multicolumn{5}{c}{$x\,=\,\log$\,\mstar} \\
\\
\deta               & $0.80\,\pm\,0.14$ & $-0.02\,\pm\,0.01$ & $-$ & $-$ \\
$\log$\,\zetaw$^\mathrm{c}$ & $22.518\,\pm\,16.4$ & $-5.4453\,\pm\,5.24$ & $0.4462\,\pm\,0.56$ &  $-0.0120\,\pm\,0.02$ \\
$\log$\,\zetaw$^\mathrm{d}$ & $39.750\pm\,1.9$ & $-10.7839\,\pm\,0.61$ & $0.9972\,\pm\,0.06$ &  $-0.0310\,\pm\,0.002$ \\
\\
\\
\hline
\\
\multicolumn{5}{c}{$x\,=\,$\fgas} \\
\\
\deta $^\mathrm{b}$ & $\alpha$ & $\alpha$\,$d\log\,M_g/d\log$\,\mstar  & $-$ & $-$ \\
\zetaw$^\mathrm{c}$ & $4.699\,\pm\,0.54$ & $3.4871\,\pm\,1.39$ & $-1.0556\,\pm\,0.76$ & $0.1468\,\pm\,0.11$ \\
\zetaw$^\mathrm{d}$ & $3.394\,\pm\,0.34$ & $2.0313\,\pm\,0.87$ & $-0.2162\,\pm\,0.48$ & $0.0469\,\pm\,0.07$ \\
\\
\hline
\end{tabular}
}
\begin{flushleft}
{\footnotesize
$^\mathrm{a}$~$d(x)\,=\,a_0 + a_1\,x + a_2\,x^2 +
a_3\,x^3$, where $d$ corresponds to the dependent variable, either
\deta, \zetaw, or $\log$\,\zetaw\ as defined in the table body, and $x$ is the
independent variable, either \fgas\ or $\log$\,\mstar.
The coefficients are given to several digits because of the high degree of the
polynomials; with only two digits, as would be consistent with the large uncertainties,
the resulting values of the dependent variable would be incorrect. \\
$^\mathrm{b}$This is not a fit, but rather Eq. \eqref{eqn:deltaeta}.\\
$^\mathrm{c}$Symmetric prior on \deta\ (Figs. \ref{fig:bayes} and \ref{fig:bayes_fbary}). \\
$^\mathrm{d}$Prior \deta\,$\geq 0$ (Fig. \ref{fig:bayes_0}).}
\end{flushleft}
\end{table}

\begin{figure}[!t]
\centering
\includegraphics[width=0.98\linewidth]{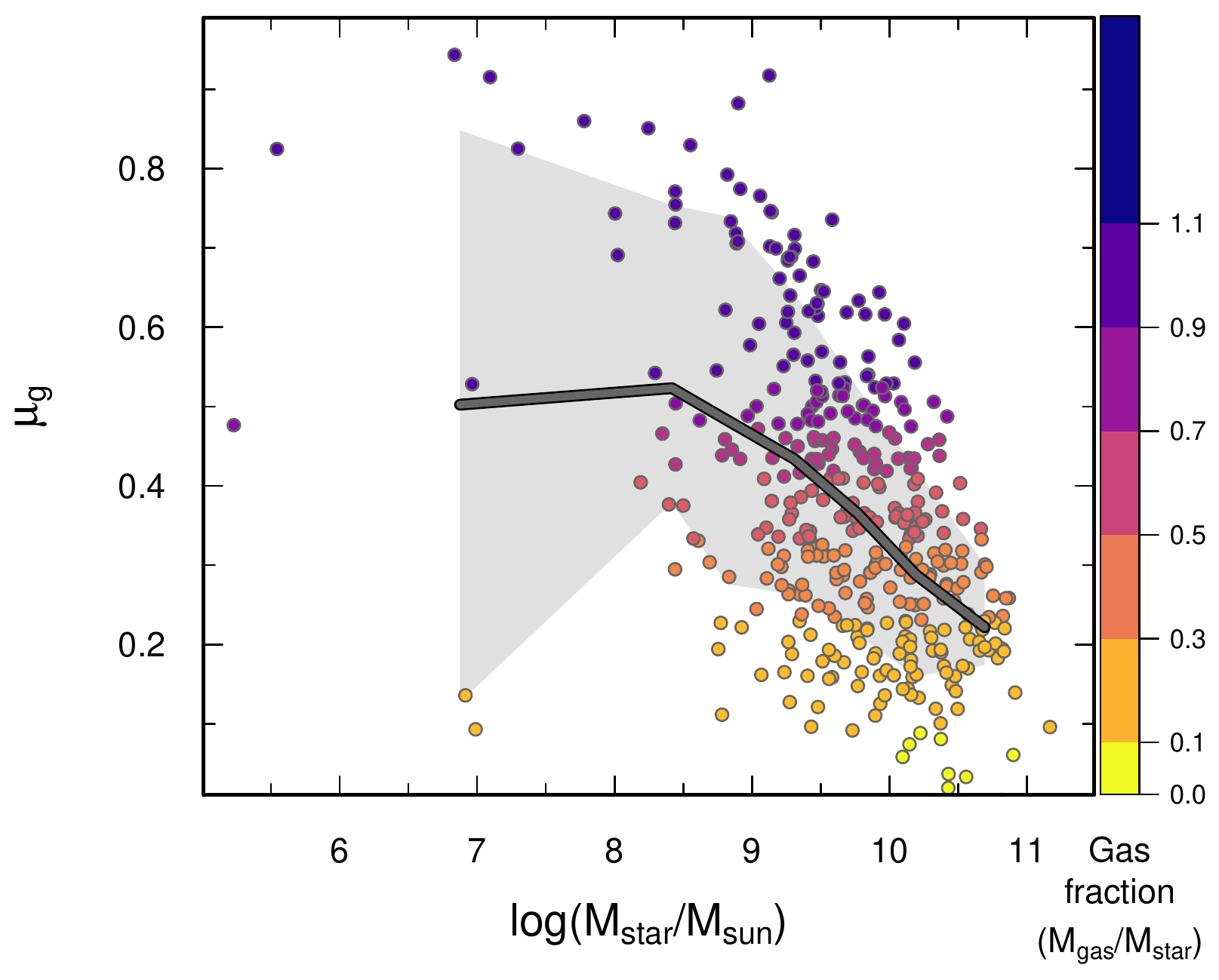}
\caption{Baryonic gas fraction \mug\ plotted against (log) \mstar\
for MAGMA galaxies. Galaxies are coded by gas fraction, \fgas, as
shown in the vertical color bar. MAGMA medians are shown as a
heavy gray line, with $\pm 1\sigma$ excursions as the gray region.
As noted in the text, the two low-mass galaxies with low \mug\ are Sextans\,A and WLM,
with probable substantial underestimates of the total gas content.
} \label{fig:mug}
\end{figure}

\subsection{Gas fraction, metallicity, and outflows}\label{sec:mug_outflows}

In \citetalias{hunt20} [see also Eq. \eqref{eqn:mgasmstar}], we
found that gas content \fgas\ grows with decreasing stellar mass
\citep[see also
e.g.,][]{Haynes2011,Huang2012,Boselli2014b,Saintonge2017,Catinella2018}.
Figure \ref{fig:mug} shows this comparison explicitly for baryonic
gas fraction \mug\ and \mstar\ in MAGMA galaxies. There is a clear
trend for higher \mug\ as \mstar\ decreases, but in the very
lowest \mstar\ regime, below log(\mstar/\msun)\,$\la$\,8.5, the
scatter is large. In MAGMA, most galaxies in this regime have
\mug\,$\ga$\,0.5, but there are two exceptions, Sextans\,A and
WLM, at very low \mug\ despite low global stellar mass. This
apparent inconsistency probably arises from underestimates of the
total gas content corresponding to the localized regions where CO
was detected
\citep[e.g.,][]{Shi2015,Shi2016,Elmegreen2013,rubio15}, and
indicates the inherent difficulties of probing the gas, stellar,
and metal contents in these extreme dwarf systems. Although the
median trend for less-massive galaxies to be more gas rich is
evident down to \mstar$\sim 3\times 10^8$\,\msun, the large
overall scatter makes it virtually impossible to infer trends of
\deta\ and \zetaw\ with \mug\ from their trends with \mstar.

\begin{figure*}[!t]
\begin{center}
\hbox{
\includegraphics[angle=0,width=0.33\linewidth]{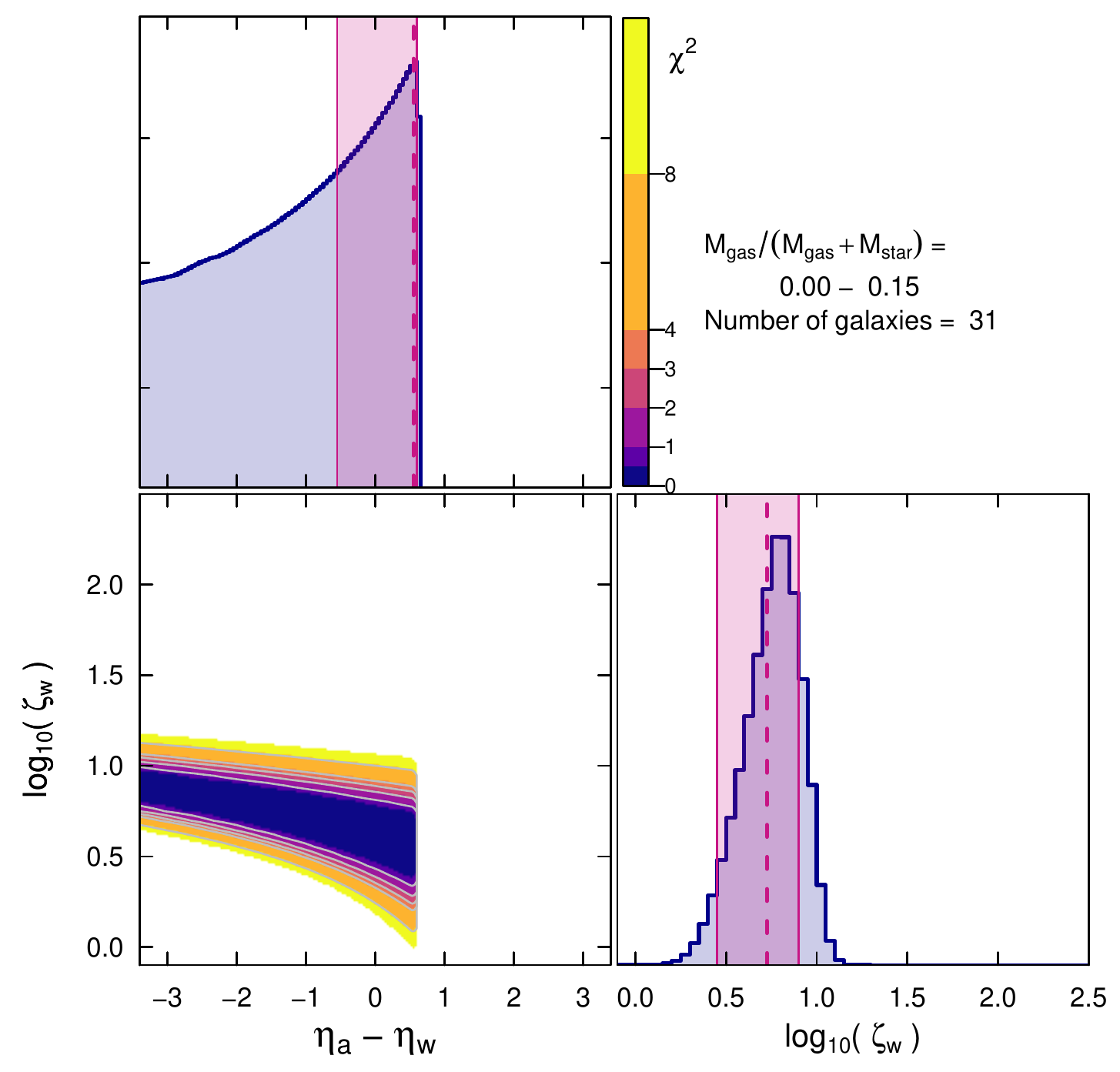}
\includegraphics[angle=0,width=0.33\linewidth]{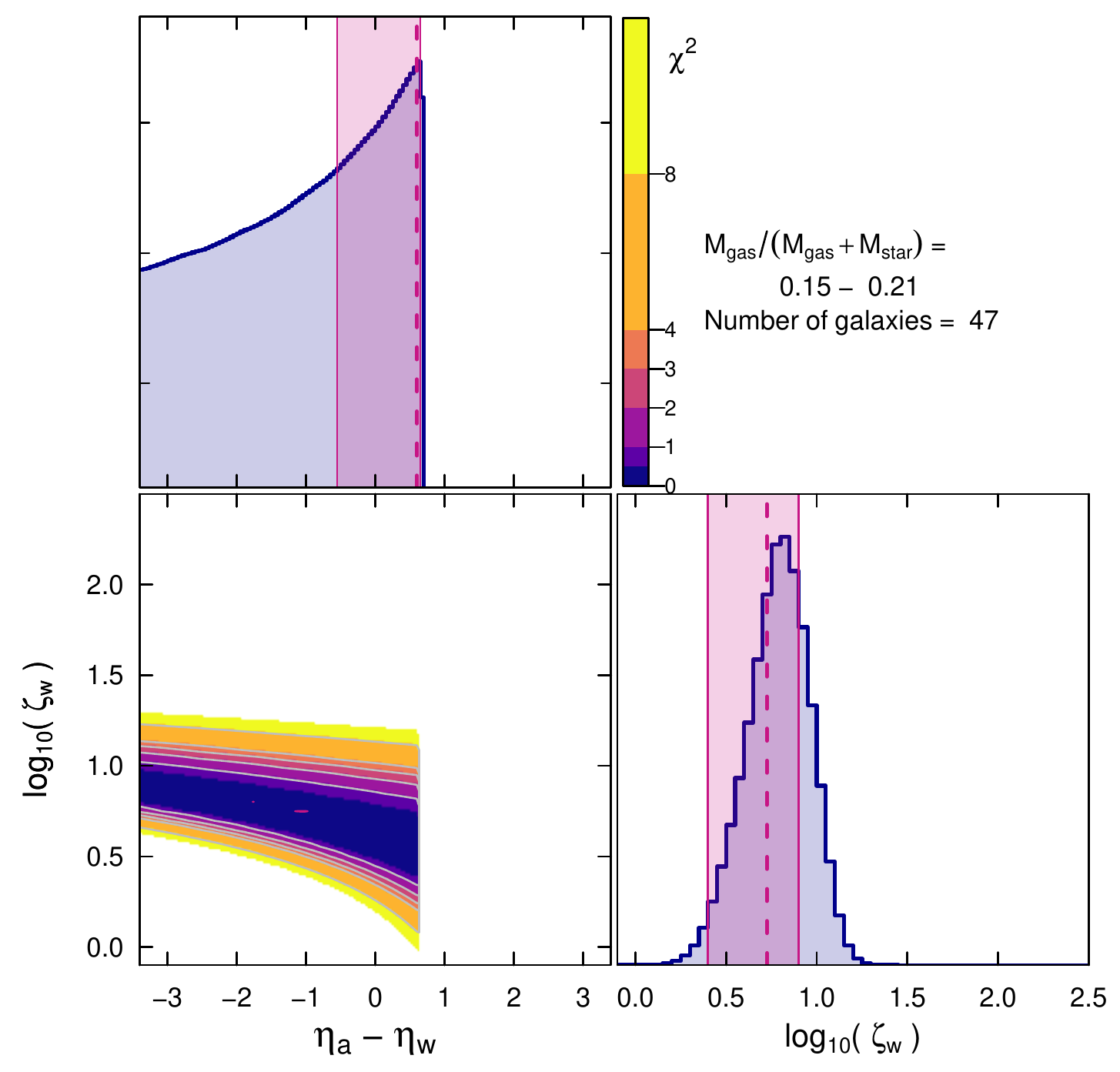}
\includegraphics[angle=0,width=0.33\linewidth]{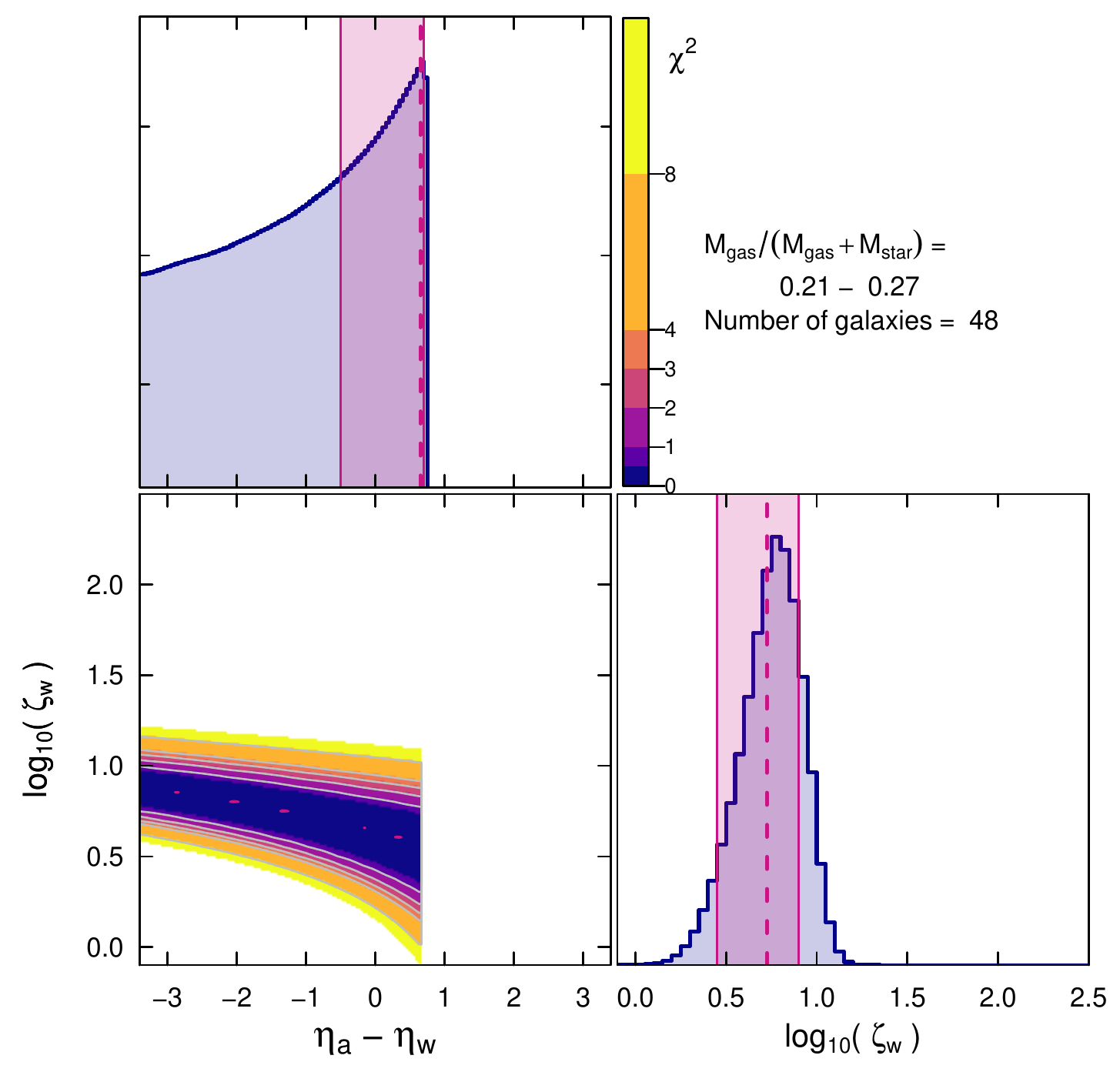}
}
\vspace{\baselineskip}
\hbox{
\includegraphics[angle=0,width=0.33\linewidth]{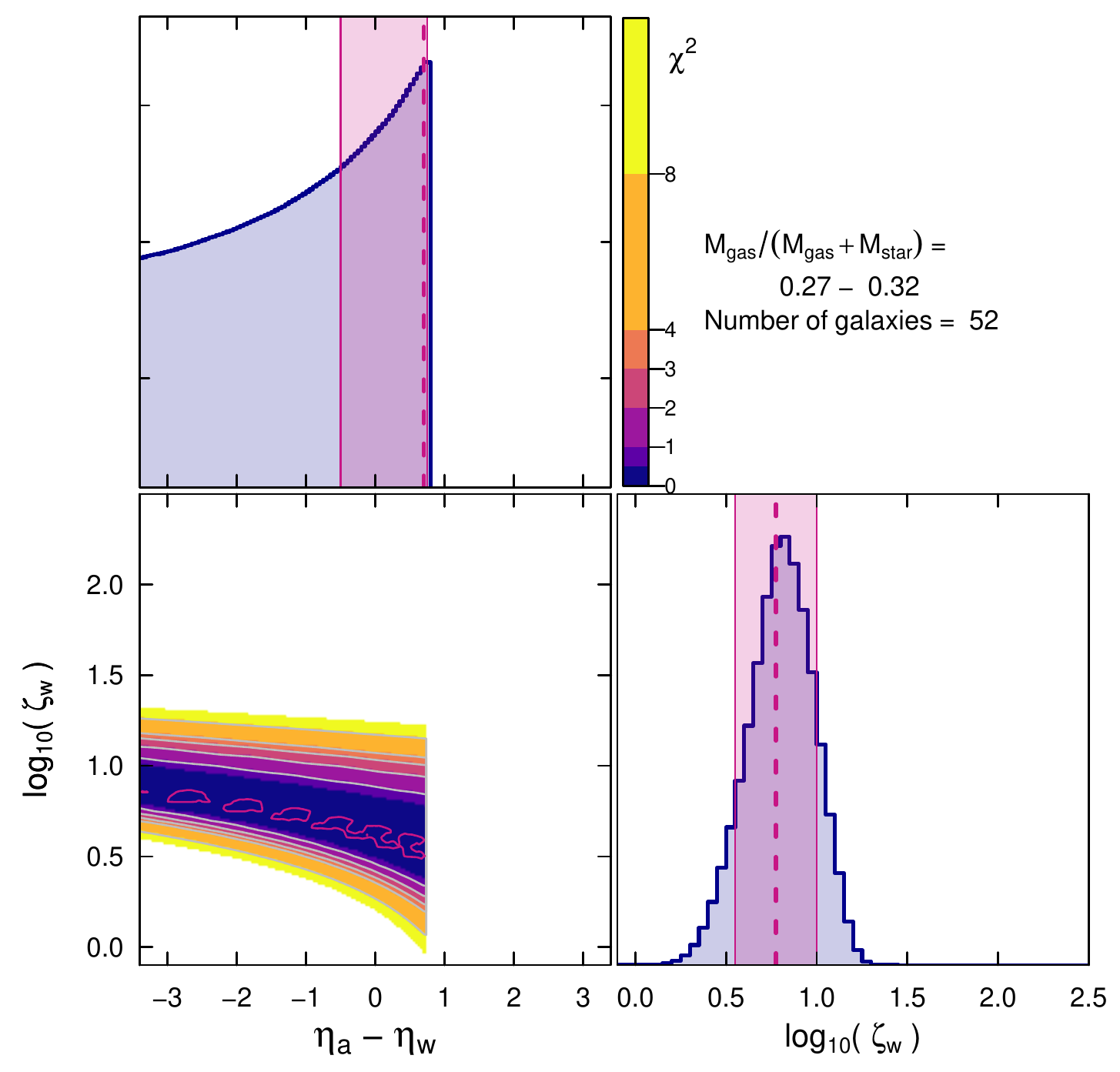}
\includegraphics[angle=0,width=0.33\linewidth]{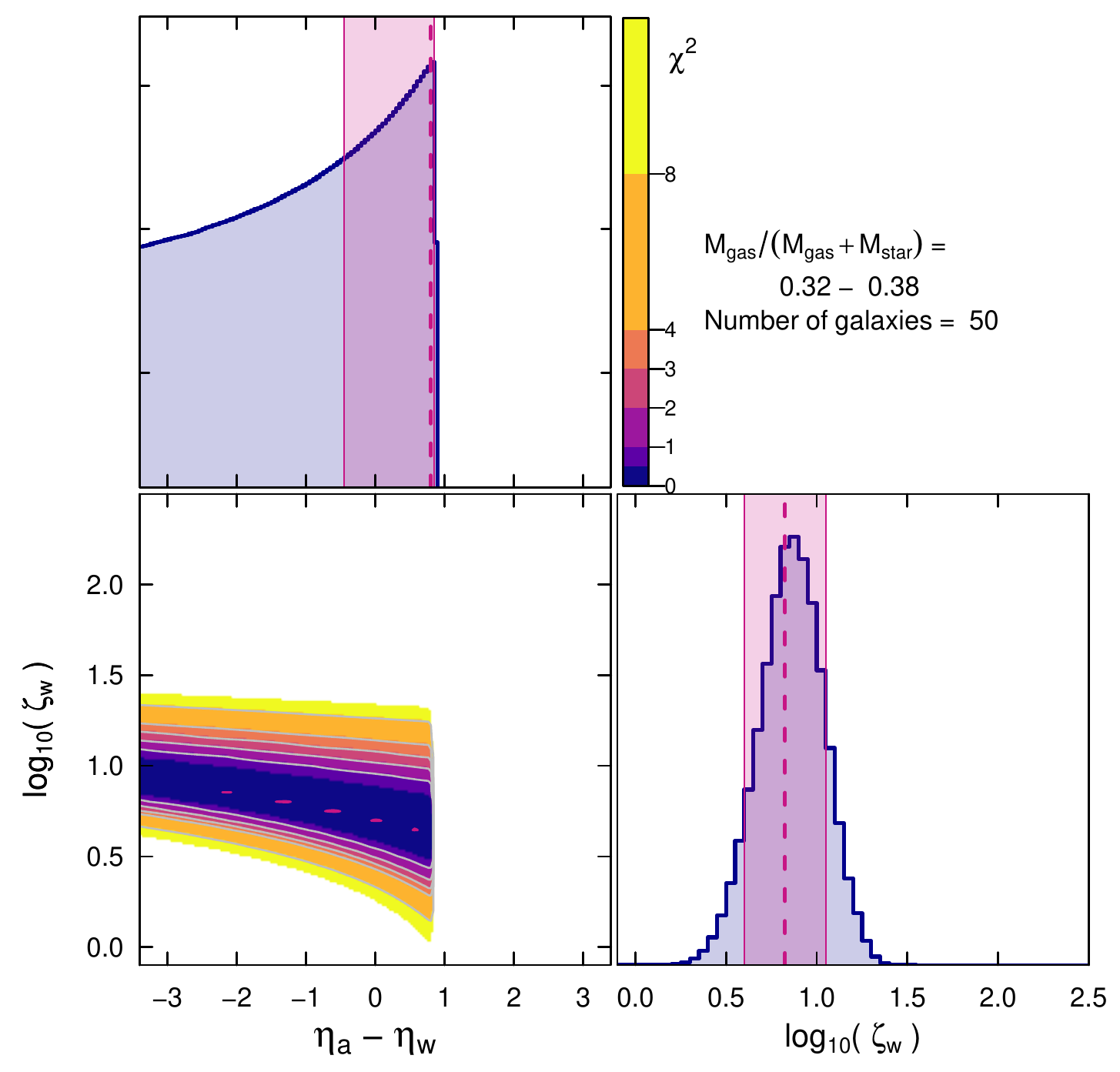}
\includegraphics[angle=0,width=0.33\linewidth]{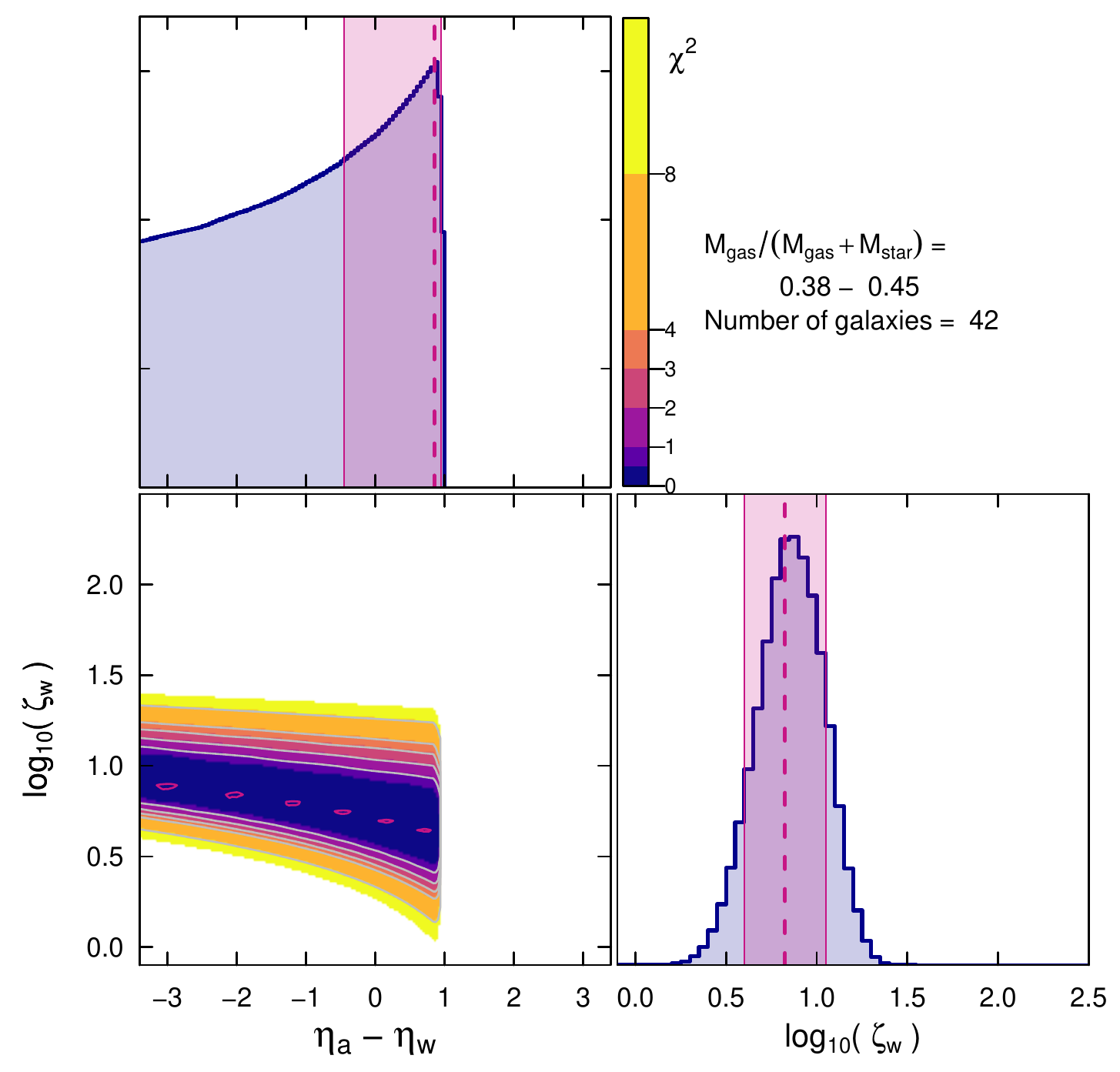}
}
\vspace{\baselineskip}
\hbox{
\includegraphics[angle=0,width=0.33\linewidth]{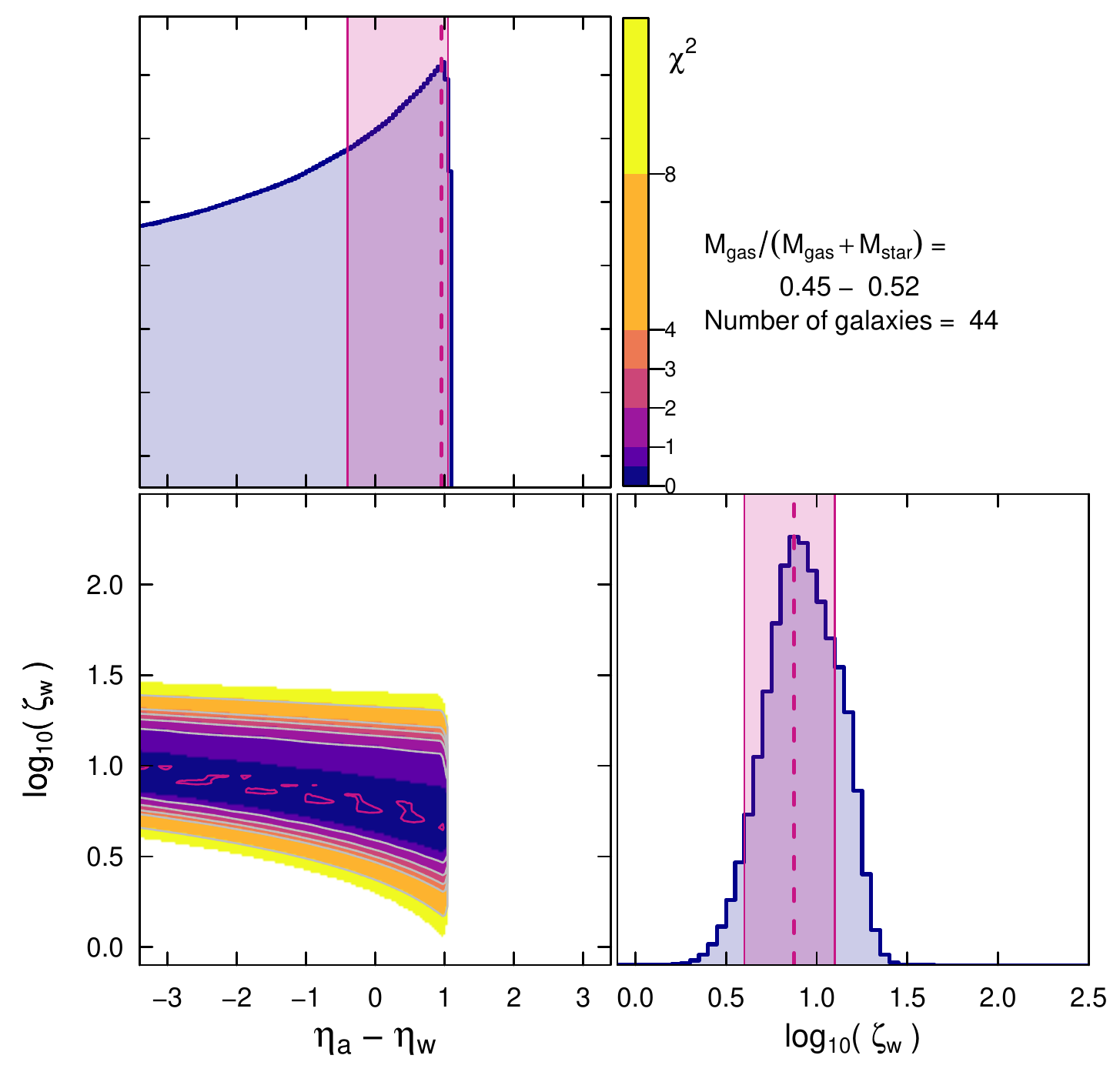}
\includegraphics[angle=0,width=0.33\linewidth]{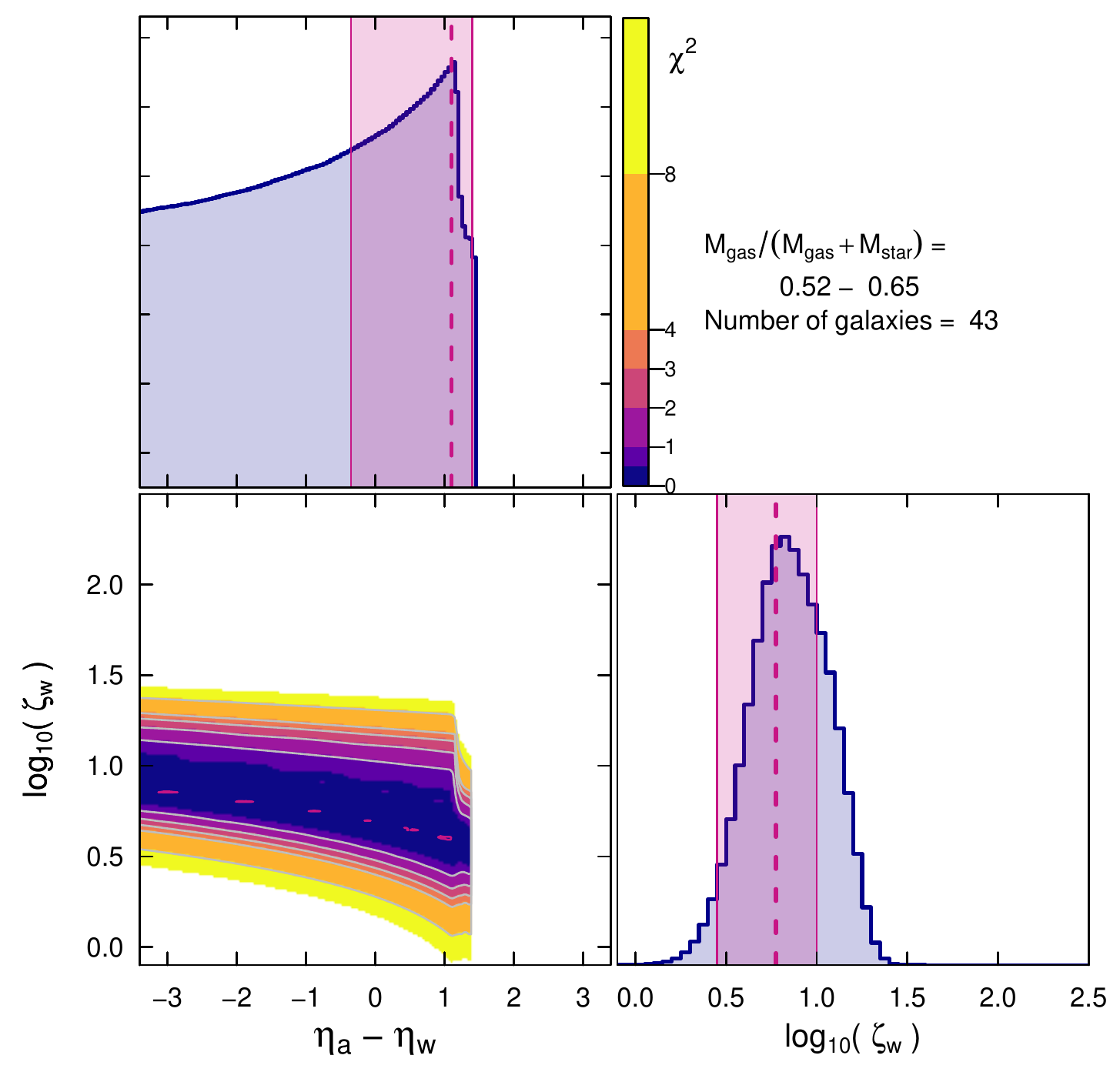}
\includegraphics[angle=0,width=0.33\linewidth]{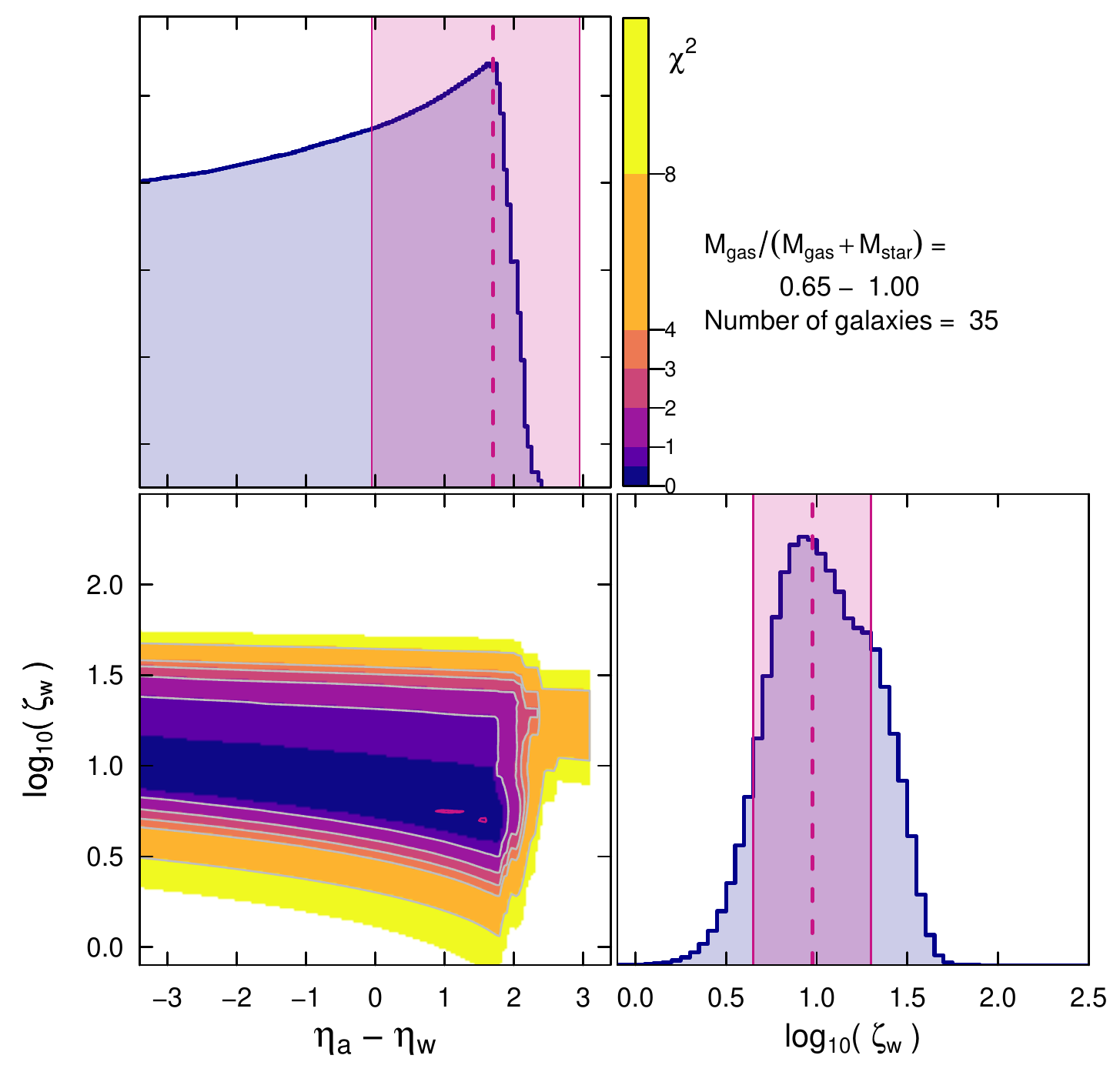}
}
\end{center}
\vspace{-1.2\baselineskip}
\caption{Corner plots of \chisq\
surface as a function of the model parameters (\deta, \zetaw) for
MAGMA binned by \mug. The violet contours correspond to the minimum \chisq\
value. The top and right panels of each corner plot report the
probability density distributions for the marginalized parameters;
confidence intervals ($\pm 1\sigma$) are shown as violet-tinted
shaded rectangular regions, and the MLE (PDF mode for \deta, median for \zetaw) is shown by a
vertical dashed line.
\label{fig:bayes_fbary}
}
\end{figure*}

Thus, to assess the dependence of \deta\ and \zetaw\
on \mug, we cannot use the analysis of Sect. \ref{sec:bayes}.
Consequently, we repeated the Bayesian estimation by binning in \mug, rather than in \mstar.
As before, bin boundaries were chosen such that a similar number of galaxies is
found in each of the nine bins; the priors for \deta\ and \zetaw\ are as before
(see Sect. \ref{sec:bayes}).
The results for the symmetric prior on \deta\,$\in$\,[-3.4,3.4] are shown in Fig. \ref{fig:bayes_fbary},
where the layout of the figure is the same as in Fig. \ref{fig:bayes}.
Here there is a trend for larger \deta\ excursions at high \mug\ (see lower panels),
similar to what we found for low \mstar.
There is also a tendency for the distribution of \zetaw\ to be broader
at high \mug, that is also reflected in the broader range of \deta.

\begin{figure}[!h]
\begin{center}
\includegraphics[angle=0,width=0.98\linewidth]{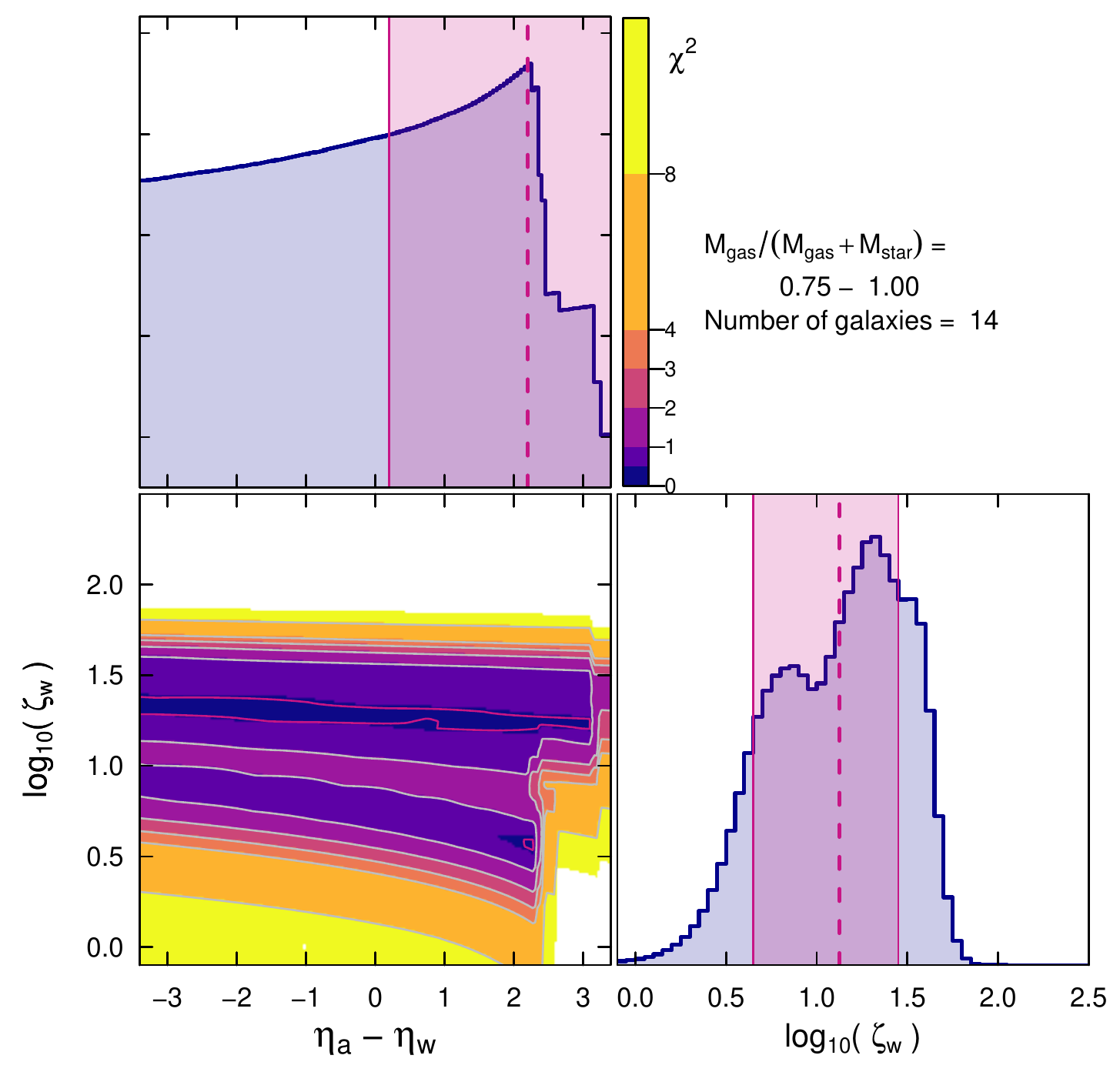}
\end{center}
\vspace{-1.2\baselineskip}
\caption{Corner plot of \chisq\
surface as a function of the model parameters (\deta, \zetaw) for
the highest \mug\ bin MAGMA, here with a higher boundary than in Fig. \ref{fig:bayes_fbary}: \mug$>$0.75.
As in previous figures,
the violet contours correspond to the minimum \chisq\
value, and the top and right panels report the
probability density distributions for the marginalized parameters.
Confidence intervals ($\pm 1\sigma$) are shown as violet-tinted
shaded rectangular regions, and the MLE (PDF mode for \deta, median for \zetaw) is shown by a
vertical dashed line.
\label{fig:bayes_highmug}
}
\end{figure}

At high \mug, there is some evidence for a bimodal distribution of
\zetaw, as shown in Fig. \ref{fig:bayes_highmug}, where we have
raised the boundary of the highest \mug\ bin. In this \mug\ bin,
the rightmost one plotted in Fig.  \ref{fig:bayes_fbary}, 
the best-fit (mode) \deta\ is higher than
any. Nevertheless, as can be seen in the lower-right panel, higher
values of \zetaw\ are possible for virtually every \deta\ sampled,
so that the dependence of \zetaw\ on \deta\ is apparently
degenerate, as we concluded in Sect. \ref{sec:bayes}.

To parameterize \deta\ and \zetaw, here we adopt \fgas\ as the
independent variable because it is statistically better behaved
than \mug, as a result of the simpler dependence on \mstar.
However, it is more convenient to compare quantities to \mug\ as
shown in the right panel of Fig. \ref{fig:loadings} where Bayesian
estimates for \deta, \zetaw, and observed \zg\ are plotted as a
function of \mug. The curve in the top right panel is not a fit,
but rather the prediction of Eq. \eqref{eqn:deltaeta} using
$d\log\,M_g/d\log$\,\mstar\,=\,0.72 from Eq.
\eqref{eqn:mgasmstar}. The middle right panel shows \zetaw\
together with the robust cubic best-fit polynomial with
coefficients as given in Table \ref{tab:loading_fits}. Metallicity
vs.  \mug\ is given in the bottom right panel, where the
long-dashed curve corresponds not to a fit, but rather to the
prediction of Eq.\,\eqref{eqn:zg} with the functional forms for
\zetaw\ (cubic polynomial) and \deta. The difference in \zetaw\
for the two different priors for \deta\ in the Bayesian analysis
are analogous to those for \mstar\ (left panel Fig.
\ref{fig:loadings}). At low \mug, \zg\ is better approximated by
\zetaw\ inferred from \deta$> 0$ relative to \zetaw\ from the
symmetric prior.

The influences of \zetaw\ and \deta\ are clearly seen in the trend
for \zg\ with \mug. The steep increase of \zetaw\ and \deta\
toward high \mug\ is reflected in the dramatic reduction of
metallicity toward high \mug. The shape of the MAGMA \zg\ curve
relative to \mug\ resembles that found by \citet[][see her Fig.
3]{Erb2008}, for $z \sim 2$ galaxies, but with some important
differences. The first is that \citet{Erb2008} assume that
\zw/\zg\,=\,1, that is to say that \zetaw\,=\,\etaw. Appendix
\ref{app:hom_wind} presents the trivial derivation of this result.
Under this assumption, \zg\ in Eq.\,\eqref{eqn:zg} is essentially
governed by \etaa, with \etaw\ entering only into the $M_g/M_i$
exponent (see also Appendix \ref{app:hom_wind}). Consequently,
there is less variation at low \mug, with the prediction by
\citet{Erb2008} assuming a constant asymptotic value for \mug\ as
high as $\sim\,0.35$. By considering the change in \zw\ relative
to \zg, as we do here, \zg\ never reaches a truly asymptotic
value, although there is flattening toward lower \mug.

The second difference is the assumption for $R$ (thus $\alpha$) in the metallicity treatment of \citet{Erb2008}.
She uses the same metallicity calibration as adopted here (PP04N2), but assumes that $\alpha$,
the fraction of gas retained in stars, is unity, thus neglecting the gas that is expelled in SNe and stellar winds.
This would imply a difference $-0.34$\,dex in her and our absolute metallicity scale.
An additional factor is the different yield adopted here ($y\,=\,0.037$) relative to $y\,=\,0.019$
used by \citet{Erb2008}.
Considering these two numerical differences, the absolute scale in \citet{Erb2008} would
be reduced by $\sim 0.6$\,dex.
Such a difference is consistent with the galaxies shown by \citet{Erb2008}, considering that
her sample is at $z \sim 2$, while MAGMA is at $z \sim 0$.
Thus, as observed, lower \zg\ ($\sim -2.7$\,dex) at low gas fractions would be expected for her sample relative to ours
\citep[e.g.,][]{Erb+06,Maiolino+08}.

\subsection{Considerations}
\label{sec:considerations}

Interpretation of our results is not altogether straightforward.
On the one hand, the Bayesian analysis shows that \deta\ as a function of \mstar\ tends
toward $\alpha$ which would indicate a condition of ``gas equilibrium''
\cite[e.g.,][]{Dave+12,Mitra+15}.
On the other, \deta\ as a function of \mug\ does not show this behavior, but rather resembles what
would be predicted by Eq. \eqref{eqn:deltaeta}.
This distinction stems from the highly imperfect correlation of \mug\ with \mstar\
as shown in Fig. \ref{fig:mug}.
Figures  \ref{fig:bayes} and \ref{fig:bayes_fbary}
also illustrate that a wide range of \deta\ is possible (\deta$< 0$, \deta$< \alpha$, \deta$\geq \alpha$),
but that this only weakly affects \zetaw.

Our analysis also shows that within a $1\sigma$ range,
\deta\ can be $< \alpha$, over all mass ranges,
but possibly more importantly,
\deta\ can be significantly $> \alpha$ toward low \mstar\ and high \mug\ (see Fig. \ref{fig:loadings}).
We take this to imply that at low \mstar\ and high \mug\ there is more possibility
for increased overall gas accretion (\etaa) relative to wind outflows (\etaw).
This could be related to the stochastic nature of star-formation histories in low-mass
galaxies \citep[e.g.,][]{weisz12,madau14reversal,krumholz15,emami19},
and to the availability of gas.

Figure \ref{fig:loadings} shows that \deta\,$\approx \alpha$
almost independently of \mstar, thus implying ``gas equilibrium''
at all masses; however, \deta\ increases as a function of \mug\
roughly as predicted by Eq. \eqref{eqn:deltaeta}. Because we would
expect that younger, less evolved, galaxies contain more gas, thus
higher \mug, this would seem to suggest that \deta\ could depend
on the evolutionary phase of a galaxy. Since we cannot know the
star-formation histories (SFHs) of the individual galaxies, this
suggestion is difficult to test. However, we can fix a simple SFH,
such as a falling or rising exponential, and experiment with what
such a ``toy model'' would predict for $d\log\,M_g/d\log$\,\mstar\
as a function of time. If we consider that the timescale of star
formation \tausf\ is tightly linked to the gas depletion time
\taugas\ (\taugas\,=\,\epss$^{-1}$), we can explicitly calculate
the growth of \mstar\ and \mgas\ using Eqs. \eqref{eqn:mstar} and
\eqref{eqn:mg_sol}, respectively.

These toy-model calculations are shown in Appendix \ref{app:toymodel}.
As expected, $d\log\,M_g/d\log$\,\mstar\ depends clearly on \deta;
both positive and negative values of $d\log\,M_g/d\log$\,\mstar\
are possible depending on the SFH, the age of the galaxy \tgal,
the timescale for star formation \tausf, and most importantly on \deta.
The value of \deta, relative to $\alpha$ in the exponent of Eq. \eqref{eqn:mg_sol},
determines whether gas mass is increasing or diminishing with time.
Since stellar mass is monotonically increasing, the value of \deta\ sets
the value of $d\log\,M_g/d\log$\,\mstar.
Nevertheless, our results show that \deta\ is almost impossible to constrain,
demonstrating behavior that varies according to whether the independent variable is
\mstar\ or gas content \mug.

As discussed in Sect. \ref{sec:limitations}, when \deta\,=\,$\alpha$, as for the gas-equilibrium scenario,
Eq. \eqref{eqn:zg} is numerically intractable
because of the zero denominator in the exponent for \mgas/\mi.
This same condition gives $\dot{M}_g$\,=\,0 and \mgas\,=\,\mi\ [see Eq. \eqref{eqn:mgmi}].
Because of the relatively long total-gas depletion times, $\sim$\,a few Gyr,
even without a true equilibrium solution,
essentially \mgas/\mi\ remains unity throughout most of the lifetime of the
galaxy [see Eq. \eqref{eqn:mg_sol}].

An important aspect of our formalism is that, in essence, through
Eq. \eqref{eqn:sfr}, we assume that the SFH corresponds also to the time evolution of gas mass.
Moreover, because we assume that total system mass is conserved and that the time derivative of accreted
gas mass $\dot{M}_a$ is proportional to SFR through \etaa, it is
possible that the formalism itself requires a rough equality between
\deta\ and $\alpha$, the lock-up fraction.

Despite the degeneracy of \deta\ and the inability of our
formalism to determine unambiguously its behavior, \zetaw\ is
fairly well determined (see Figs. \ref{fig:loadings_vcirc} and
\ref{fig:loadings}). After discussing effective yields in Sect.
\ref{sec:yields}, in the remainder of the paper, we focus on
\zetaw, and what can be learned from our results about metal- and
mass-loading in the stellar-driven outflows of star-forming
galaxies.

\section{Effective yields}\label{sec:yields}

Above we have quantified a scenario for the relation between \zg\
and gas content in galaxies that requires significant gas
accretion and stellar-driven outflows. In this section, we recast
our results in the context of the effective yield, as measured
against the predictions from a different, ``closed-box'',
paradigm. A galaxy evolving as a closed box obeys a simple
relationship between the metallicity and the gas mass fraction
\citep{Searle_Sargent72,pagel75}:
\begin{equation}
\label{eqn:Z_closed_box}
Z_{\rm g} = y \, \ln(1/ \mu_{\rm g})
\end{equation}
where \mug\ is the baryonic gas mass fraction and $y$ is defined
as the ``true'' yield (Sect. \ref{sec:gce}). As gas is converted
into stars, metal content grows and the gas mass diminishes. If
the galaxy evolves as a closed box, then the yield $y$ in
Eq.\,\eqref{eqn:Z_closed_box} should be equal to the
nucleosynthetic yield. However, galaxies do not evolve as closed
systems, so we expect the overall metal yield to differ from the
nucleosynthetic value. The effective yield \ye\ can be defined by
inverting Eq.\,\eqref{eqn:Z_closed_box}
\citep[e.g.,][]{Lequeux+79}:
\begin{equation}\label{eqn:yeff}
y_{\rm eff} = \frac{Z_{\rm g}}{\ln(1/ \mu_{\rm g})} \quad .
\end{equation}
The effective yield is expected to be lower than the nucleosynthetic value $y$
because metals in galaxies are lost from the system through outflows,
and the metal content of the ISM gas will be diluted with
infall of metal-poor gas \citep[e.g.,][]{edmunds90}.

\begin{figure}[!h]
\centering
\includegraphics[width=0.98\linewidth]{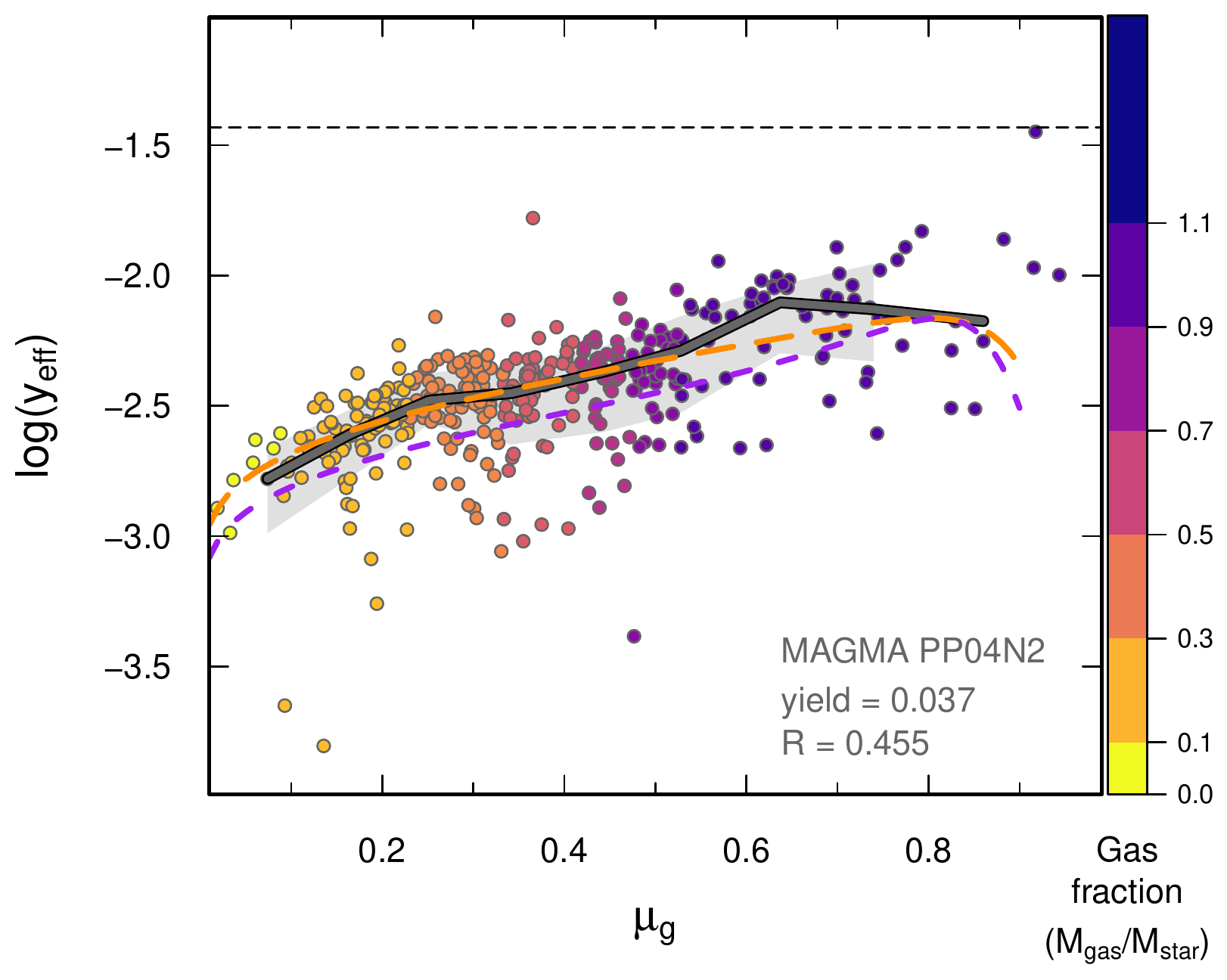}
\caption{Effective yield as a function of \mug. Galaxies are coded
by gas fraction, \fgas, as shown in the vertical color bar. MAGMA
medians are shown as a heavy gray line, with $\pm 1\sigma$
excursions as the surrounding gray regions. The two dashed curves
correspond to the predictions from the \deta\ and \zetaw\
parameterizations (see also Fig. \ref{fig:loadings}), with the
symmetric \deta\ priors given as a short-dashed (purple) curve,
and the \deta\,$\geq$\,0 priors as a long-dashed (orange) one. The
horizontal dashed line corresponds to our adopted true yield
$y\,=\,0.037$ (see text).} \label{fig:yeff_gasfraction}
\end{figure}

\begin{figure*}[!t]
\centering
\includegraphics[width=0.48\linewidth]{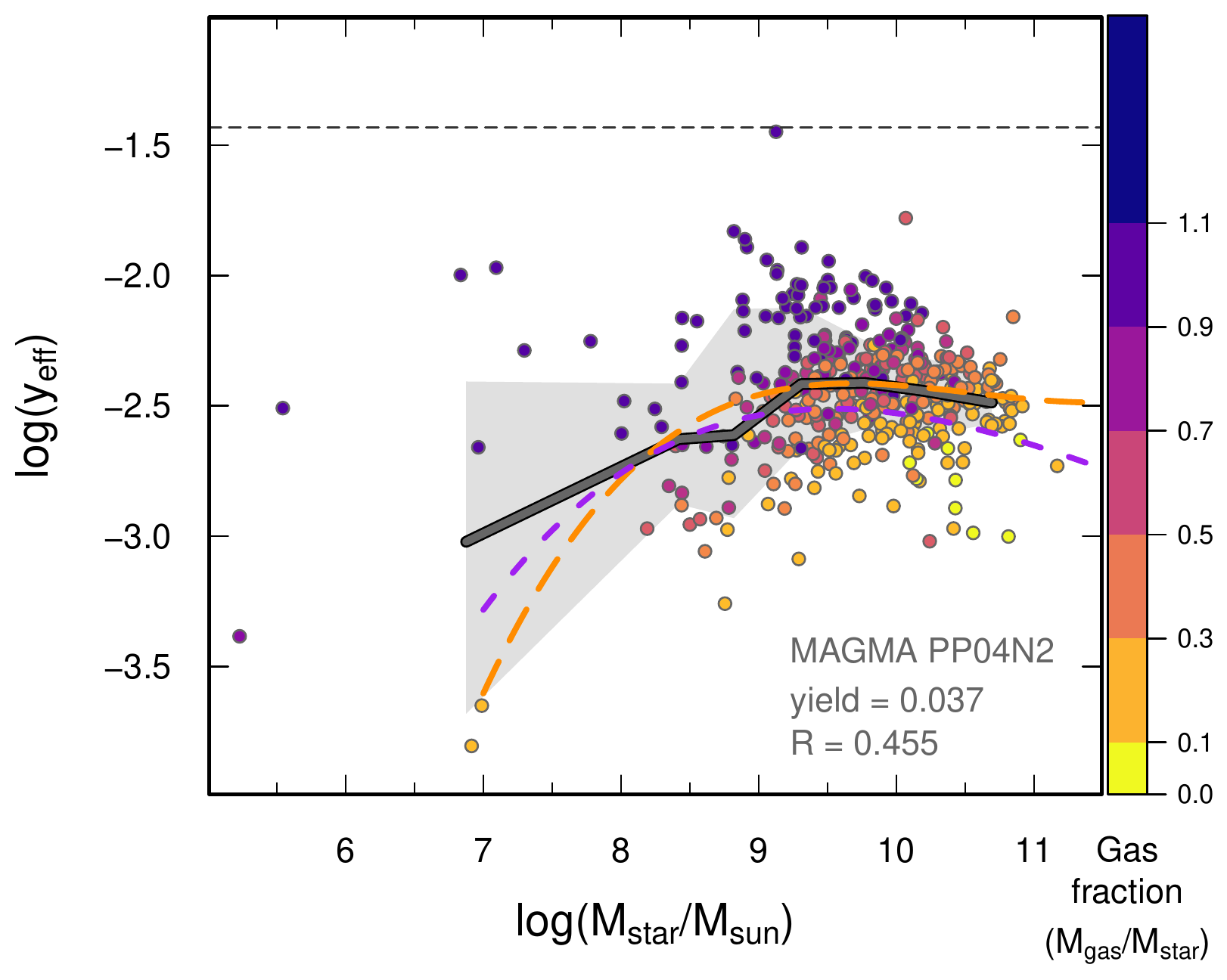}
\hspace{0.01\linewidth}
\includegraphics[width=0.48\linewidth]{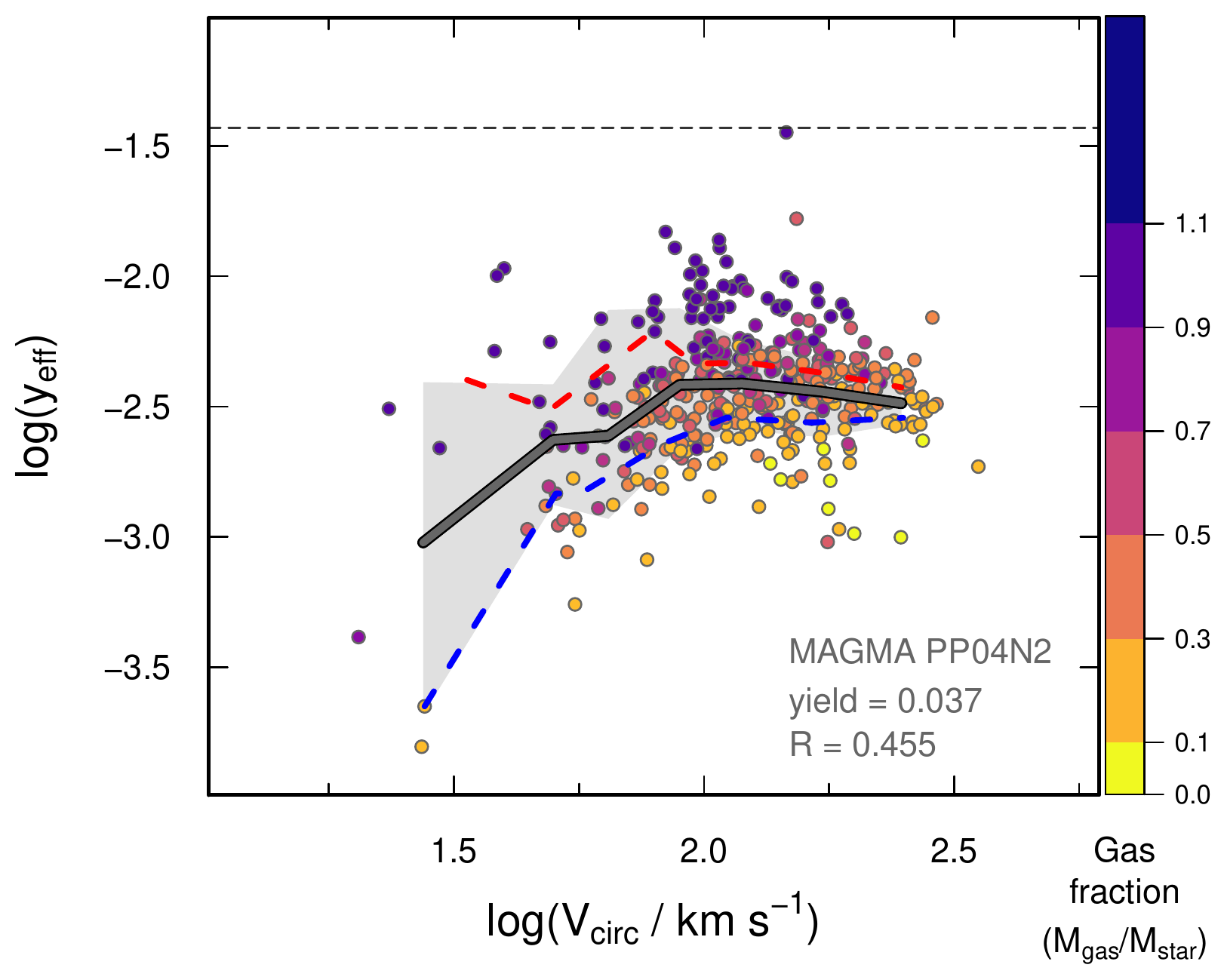}
\caption{Effective yield as a function of \mstar\ (left panel) and
\vcirc\ (right). Galaxies are coded by gas fraction, \fgas, as
shown in the vertical color bar. MAGMA medians are shown as a
heavy gray line, with $\pm 1\sigma$ excursions as the surrounding
gray regions. Also shown in the left panel as two dashed curves
are the predictions from the \deta\ and \zetaw\ parameterizations
(see also Fig. \ref{fig:loadings}), with the symmetric \deta\
priors given as a short-dashed (purple) curve, and the
\deta\,$\geq$\,0 priors as a long-dashed (orange) one. The right
panel illustrates galaxies with high star-formation efficiency
(short gas depletion times) with a blue dashed curve, and those
with low efficiency
(long depletion times) as a red one. 
The horizontal short-dashed line corresponds to the yield of 0.037 adopted in this paper. 
See text
for more details.} \label{fig:yeff}
\end{figure*}

The effective yields \ye\ for MAGMA, computed using
Eq.\,\eqref{eqn:yeff}, are shown in Fig.
\ref{fig:yeff_gasfraction}, plotted against baryonic gas mass
fraction \mug. The model prediction from Eq.\,\eqref{eqn:zg}
through our parameterizations of \deta\ and \zetaw\ is shown as
two curves, corresponding to the two different \deta\ priors. For
reference, our adopted value of $y$ is shown as a horizontal
short-dashed line. As also evident in previous figures, the
results from the \deta\,$\geq$\,0 prior (long-dashed orange curve)
are in slightly better agreement with the data. The inflection of
\ye\ at high \mug\ is due mainly to the strong downward curvature
of \zg\ as seen in Fig. \ref{fig:loadings}. Moreover, \ye\ is
always (with a single possible exception at high \mug, NGC\,4731)
significantly smaller than the true yield $y\,=\,0.037$ adopted
here. \citet{edmunds90} has shown that this behavior for \ye\
would be expected if metals were lost through stellar- and
SNe-driven bulk outflows, or diluted by accretion of metal-poor
gas. MAGMA shows a strong confirmation of this scenario. It is
clear from Fig. \ref{fig:yeff_gasfraction} that our parameterized
model is a good approximation to the data (see also Fig.
\ref{fig:loadings}).

Models that do not take into account the metal-enriched bulk
outflows at low \mstar, namely assuming that \zw\,$=$\,\zg,
predict that at high \mug, \ye\ would approach the nucleosynthetic
value $y$ \citep[e.g.,][]{Erb2008}. This can be seen by expanding
Eq.\,\eqref{eqn:yeff} in a Taylor's series
\citep[see][]{dalcanton07,filho13} so that for high \mug, we have:
\begin{equation}
y_{\rm eff} \approx \frac{M_Z}{M_{\Large\star}} \quad .
\label{eqn:yeffapprox}
\end{equation}
Thus, in very gas-rich galaxies, the effective yield would be
approximately independent of gas fraction. As pointed out by
\citet{dalcanton07}, metal-poor accreted gas will lower the
metallicity of the system, but it will also augment \mug, thus
maintaining the system in a pseudo-closed-box equilibrium. The
linear dependence of \ye\ on metal mass \mz, as shown by
Eq.\,\eqref{eqn:yeffapprox}, also means that metal-enriched
outflows are very effective at lowering the effective yield in
galaxies with high gas-mass fractions \citep[see][]{dalcanton07}.
Indeed, Fig. \ref{fig:yeff_gasfraction} shows that even at high
\mug, \ye\ is significantly lower than the true yield $y$.

It is important to remember that Eq.\,\eqref{eqn:yeff} measures \zg\
in the ionized gas, but the gas content \mug\ is derived from
neutral gas (\hi\ $+$ \htwo). There is some evidence that the
metal contents of these two gas phases are different; in  UV
absorption studies of metal-poor dwarf galaxies
\citep[e.g.,][]{thuan05,lebouteiller13}, the metallicity of the
neutral gas in these systems was always found to be lower than in
the ionized gas. A similar conclusion was reached by
\citet{filho13} and \citet{thuan16} by comparing \ye\ with $y$,
and attributing the difference to the lower metallicity in the
neutral gas.

Figure \ref{fig:yeff} plots the effective yield \ye\ against
stellar mass \mstar\ (left panel) and circular velocity, \vcirc\
(right). Even though there is much scatter at low \mstar\ and
\vcirc\ (\mstar\,$\la 3 \times 10^{8}$\,\msun\ and \vcirc\ $\la
50$\,\kms), \ye\ is increasing with mass and velocity up to $3
\times 10^{9}$\,\msun\ and $\sim 100$\,\kms, in qualitative
agreement with similar trends reported in the literature
\citep[e.g.,][]{Garnett02_yields,Pilyugin+04,Tremonti+04,dalcanton07,Ekta_Chengalur10}.
For more massive galaxies, there is an inflection in the relations
with a mild decline at masses and velocities larger than $3 \times
10^{9}$\,\msun\ and $\sim 100$\,\kms.

The right panel of Fig. \ref{fig:yeff} also shows the MAGMA sample
divided into two regimes in gas depletion times, \taugas, defined
as \mgas/SFR (see \papii); galaxies with long \taugas\
(\taugas\,$>\,4 \times 10^9$\,yr) are shown as a dashed red curve,
and short times \taugas\,$<\,4 \times 10^9$\,yr as a dashed blue
one. The gas depletion time \taugas\ is the inverse of the
so-called star-formation efficiency (SFE, see \papii), that also
enters into Eq. \eqref{eqn:sfr}, so short \taugas\ would
correspond to higher SFE, and vice versa. There can be as much as
a factor of ten in \ye\ between these two categories of galaxies,
with a mean difference of $\sim\,2-3$ in \ye. At fixed mass,
larger yields correspond to longer \taugas\ or lower SFE. In these
galaxies with longer \taugas, the peak in \ye\ at \mstar\,$\sim 3
\times 10^{9}$\,\msun\ seems to disappear, while there is a steady
decline of \ye\ toward lower \mstar\ for shorter depletion times
\citep[see also][]{Lara-Lopez+19_yields}. Longer \taugas\ are also
associated with higher gas fractions (see color coding in Fig.
\ref{fig:yeff}); thus Eq.\,\eqref{eqn:yeffapprox} may be telling
us that the higher effective yields are simply a function of the
enhanced impact of metal-enriched outflows in gas-rich galaxies.

\section{Characterizing stellar feedback through winds}\label{sec:feedback}

By adopting the formalism for galactic chemical evolution
formulated by \citet{pagel09}, we have quantified with MAGMA the
metal-weighted mass loading factor \zetaw\ (=\,\etaw\,\zw/\zg),
and with somewhat less success, the difference in mass loading of
gas accretion and outflows, \deta. However, we are interested in
the individual mass-loading factor relative to the SFR, \etaw, and
the metal content of the winds, \zw. In this section, we explore
the constraints for mass loading \etaw\ and \zw\ provided by our
results.

Assuming that \zw\,=\,\zg, as done in much previous work
\citep[e.g.,][]{Erb2008,Finlator2008,Dayal2013,Hunt+16_I,Belfiore+19_bathtub},
implies that virtually all of the material in the outflow is ISM
gas. We note that \zg\ could be only a lower limit to the
metallicity of the wind \citep[e.g.,][]{Peeples_Shankar11},
because the metals produced by SNe would be expected to enrich the
gas above that of the galaxy \citep[e.g.,][]{vader86}. Thus, to
estimate \zw, we also need to consider the fraction of gas that
has been entrained in the wind, \fent, and the metallicity of the
SNe ejecta, \zej\ \citep[see][]{dalcanton07,Peeples_Shankar11}:
\begin{eqnarray}
Z_w & = & f_\mathrm{ent}\,Z_g + (1 - f_\mathrm{ent})\,Z_\mathrm{eject} \nonumber\\
\frac{Z_w}{Z_g} & = & f_\mathrm{ent} + (1 - f_\mathrm{ent})\,\frac{Z_\mathrm{eject}}{Z_g}\quad .
\label{eqn:fent}
\end{eqnarray}

\begin{figure*}[!ht]
\centering
\includegraphics[width=0.48\linewidth]{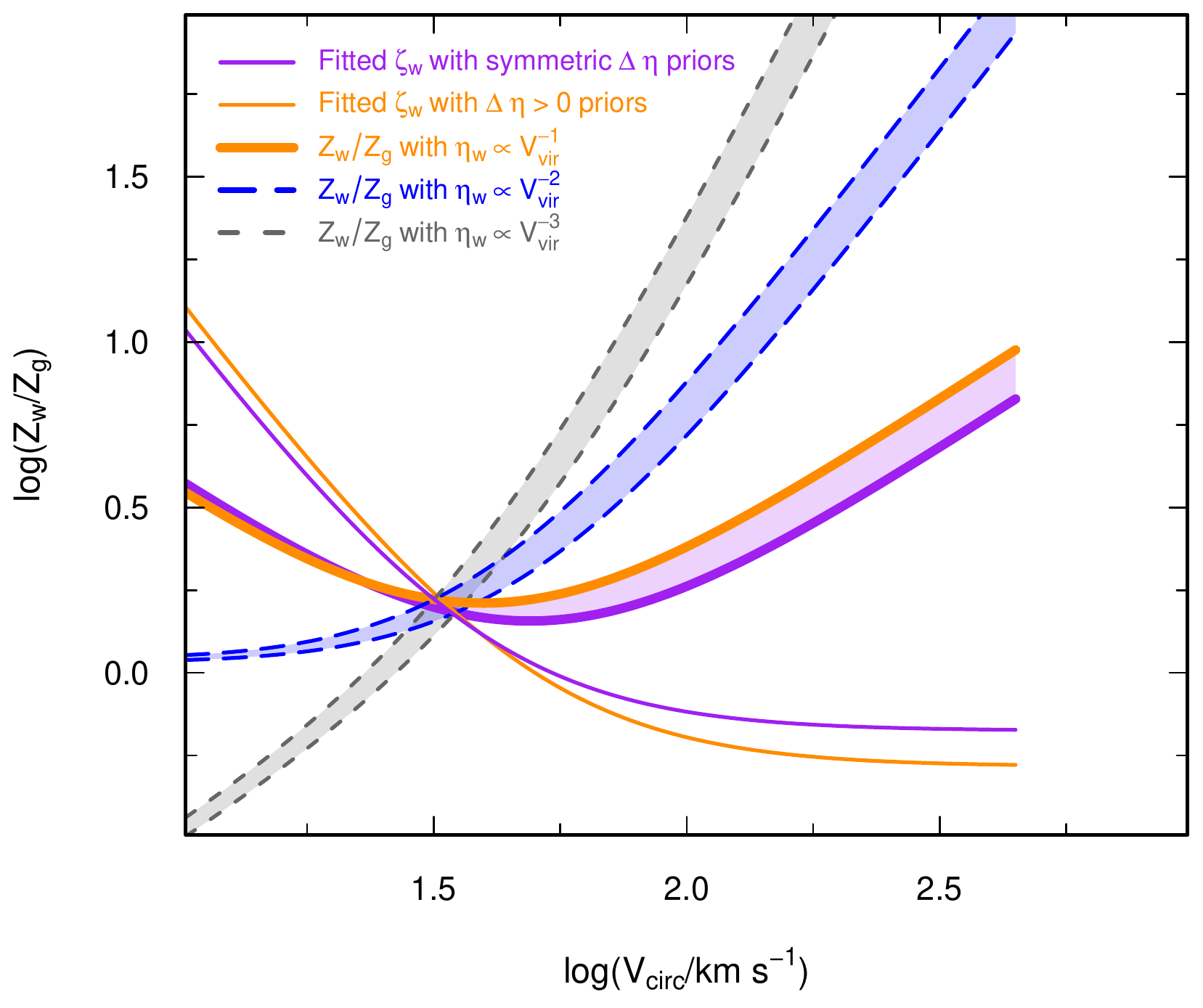}
\hspace{0.02\linewidth}
\includegraphics[width=0.48\linewidth]{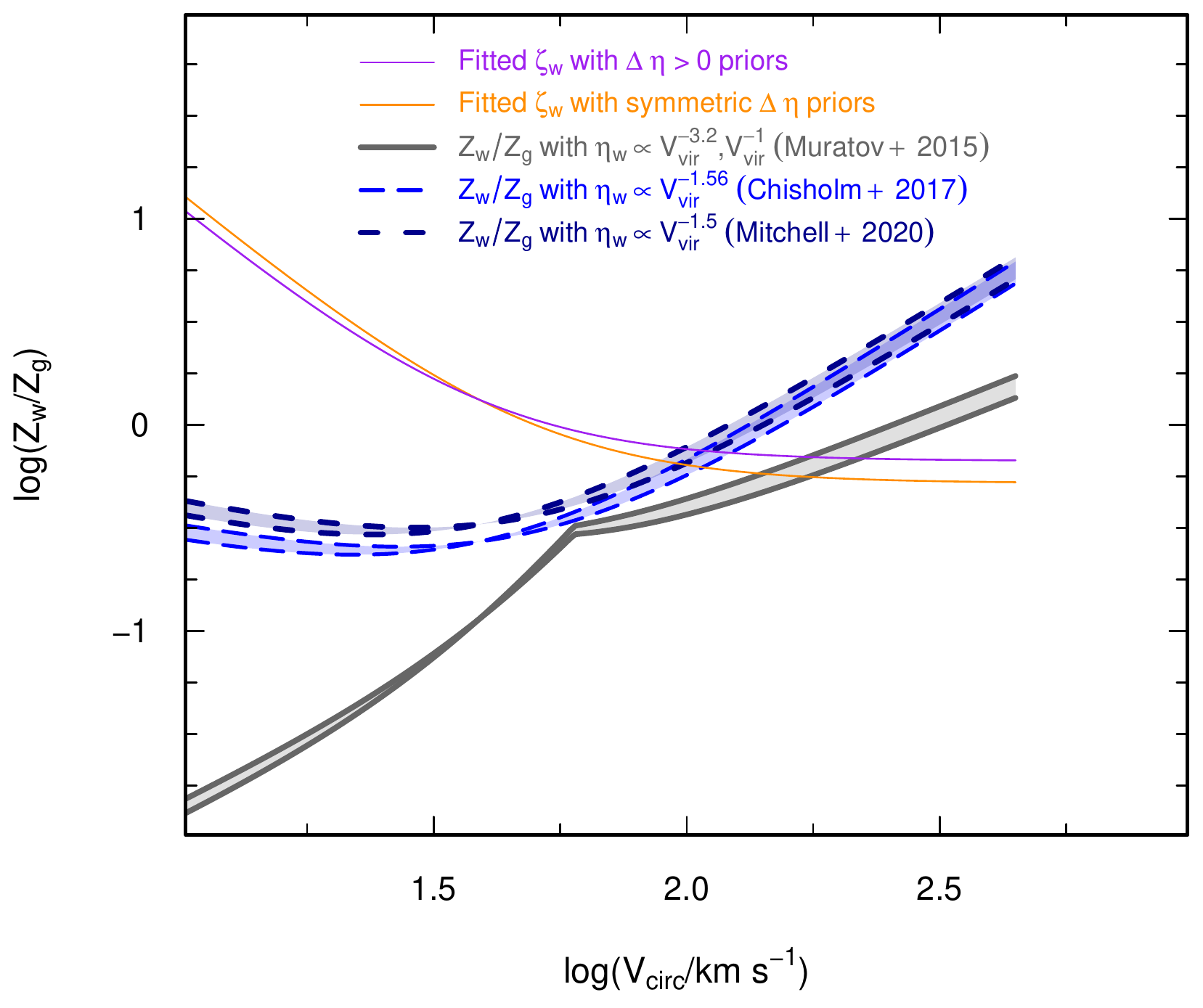}
\caption{Ratio of inferred wind and ISM metallicities \zw/\zg\
plotted against virial velocity \vcirc. Also shown (with arbitrary
scaling) is the shape of Bayesian \zetaw\ as also plotted in Fig.
\ref{fig:loadings_vcirc}; it is clearly not a power law, thus the
different power-law forms of \etaw\ introduce inflections in
\zw/\zg. \textit{Left panel:} As indicated in the legend,
different trends of \zw/\zg\ are inferred by dividing the fitted
form of \zetaw\ [see Eq. \eqref{eqn:zetawmodel}] by various
scalings of \etaw\ (see text for more details). The shaded regions
correspond to the trends allowed by the two different Bayesian
estimates of \zetaw\ according to the different \deta\ priors.
\textit{Right panel:} The broken-power law trend for \etaw\ found
by \citet{Muratov+15} is shown as a solid heavy (gray) curve,
where the inflection at \vcirc\,=\,60\,\kms\ corresponds to a
sharp decrease of \zw/\zg\ (see text for more details). The
short-dashed (dark-blue) curve shows the relation found by
\citet{Mitchell+20_EAGLE}, in the low-mass regime where stellar
feedback dominates, and the long-dashed (blue) curve shows the
observed relation derived by \citet{chisholm18}. }\label{fig:zwzg}
\end{figure*}

In low-mass galaxies, \fent\ is found to be low,
both observationally \citep[e.g.,][]{martin02} and from simulations
\citep[e.g.,][]{maclow99}.
The entrainment fraction is closely related to the mass-loading factor of the wind \etaw,
and probably varies with galaxy mass \citep[gravitational potential, see e.g.,][]{hopkins12}.
The maximum oxygen abundance of SNe ejecta \zej\ is known to within a factor of 2-3, \zej$\sim 0.04-0.1$
\citep[e.g.,][]{woosley95}, and can constrain \fent\ and \etaw\ \citep[e.g.,][]{martin02}.
The increase of \zetaw\ with decreasing virial velocity and stellar mass (see Fig. \ref{fig:loadings_vcirc})
is possibly due to diminishing \fent\ to raise \zw\ closer to \zej.
To verify this, we need to isolate \etaw, the mass-loading factor in the product \zetaw\,=\,\etaw\,\zw/\zg.

\subsection{The driving mechanisms behind stellar outflows} \label{sec:outflows}

One way to do this is to compare the \vcirc\ scaling of \zetaw\ with expected \vcirc\ scalings of \etaw\
(see  Eq.\,\ref{eqn:zetawmodel}).
Galactic outflows powered by stellar feedback are thought to be driven either by momentum,
imparted to the ISM by massive stellar winds and SNe through radiation pressure
\citep[e.g.,][]{murray05,murray10,hopkins11,zhang12},
or by energy, injected into the ISM by massive stars and core-collapse SNe
\citep[e.g.,][]{dekelsilk86,heckman90,dallavecchia12,christensen16,naab17}.

For outflows that conserve momentum, we would expect a power-law
scaling \etaw\,$\propto\,V^{-1}_\mathrm{circ}$ while for
energy-conserving winds, \etaw\,$\propto\,V^{-2}_\mathrm{circ}$
\citep[e.g.,][]{dekelsilk86,murray05,hopkins12}. The only way such
scalings can be reconciled with the (modified power-law) form of
Eq.\,\eqref{eqn:zetawmodel} is by a complementary scaling of \zw.
Examples of such complementary scaling are shown in Fig.
\ref{fig:zwzg} where the ratio of wind and ISM metallicities \zw\
and \zg\ is plotted against virial velocity \vcirc. In the left
panel of Fig. \ref{fig:zwzg}, we have taken the normalization of
\zw/\zg\ to be the same as that as the best-fit of Eq.
\eqref{eqn:zetawmodel} using our Bayesian estimates, simply
\vcirc/V$_0$, where V$_0$\,=\,32\,\kms\ for symmetric \deta\
priors, and V$_0$\,=\,35\,\kms\ for priors with \deta\,$\geq$\,0.
Both fits have power-law index $b\,\sim\,2$. The assumptions of
momentum-conserving or energy-conserving outflows lead to
different estimates for the trend of \zw/\zg.

Because dwarf galaxies are found to have a low entrainment
fraction, \citep[e.g.,][]{maclow99,martin02}, we would expect
\zw/\zg\ to increase toward low \vcirc\ because low \fent\ would
enrich the outflows toward the higher super-Solar abundance of the
SNe ejecta, \zej\ [see Eq. \eqref{eqn:fent}]. Energy-conserving
winds result in a flattening of \zw/\zg\ toward low \vcirc\ (blue
curves), and a cubic trend of \etaw\ with \vcirc\ gives a strongly
declining \zw/\zg\ (gray curves). The trend of
\etaw\,$\propto\,V_\mathrm{circ}^{-1}$ is the only one that gives
a rising trend of \zw/\zg\ at low \vcirc. Thus,
momentum-conserving winds would be favored by our results.

Quantifying this conclusion is, however, dependent on the
metallicity calibration. For the PP04N2 calibration used here, the
roughly quadratic low-mass \vcirc\ dependence is shallower than
the cubic (or quartic) dependence found by
\citet[][]{Peeples_Shankar11} for different O/H calibrations.
Other calibrations such as those by \citet{Tremonti+04} or
\citet{Denicolo+02} give differing trends (with power-law slopes
$b>$3 at low \vcirc), that are inconclusive for determining the
nature of the mass-loading factor \etaw. The PP04N2 calibration
adopted here results in the shallowest increase of \zetaw\ toward
low \vcirc\ (power-law slope $b\approx2$), relative to other O/H
calibrations, and this is reflected in a more incisive low-mass
(low \vcirc) increase in \zw/\zg.

The right panel of Fig. \ref{fig:zwzg} shows the predictions for
\etaw\ from the sub-grid FIRE (Feedback in Realistic Environments)
simulations \citep{Muratov+15}, and the EAGLE simulations
\citep{Mitchell+20_EAGLE}. At low \vcirc\ ($< 60$\,\kms),
\citet{Muratov+15} find the power-law dependence on \vcirc\ to be
steep ($\sim -3.2$), while at higher masses (\vcirc) the winds are
momentum-driven with \etaw\,$\propto$\,\vcirc$^{-1}$. A similar
conclusion was reached by \citet{dave13}, namely that at low
masses, winds are energy-driven (or even more steeply with
\vcirc), while momentum-driven winds dominate at higher masses.
Judging from Fig. \ref{fig:zwzg}, these trends would be
incompatible with our results. In contrast, EAGLE simulations
\citep[e.g.,][]{Mitchell+20_EAGLE} indicate a slope midway
between, namely $\sim -1.5$, leading to a very gradual increase in
metallicities toward lower \vcirc. Comparing the left- and
right-hand sides of Fig. \ref{fig:zwzg} shows that such a
dependence better reflects our results at low \vcirc. At higher
\vcirc\ (masses), the situation is more complex because the
flattening of \zetaw\ is difficult to interpret in terms of a
single power-law dependence of \etaw\ on \vcirc. In any case, we
conclude that at low masses, winds in stellar outflows cannot be
energy driven, but rather are more consistent with momentum-driven
outflows (\vcirc$^{-1}$), or a slightly steeper ``combined'' slope
as found by the EAGLE models with \vcirc$^{-1.5}$.

There is also the question, not touched upon here, of whether or
not outflows are able to expel material from the galaxy halo, thus
enriching the IGM. The entrainment fraction \fent\ would be key to
relating the strength of the outflows with metallicity in the
winds \citep[e.g.,][]{martin02,chisholm18} and deserves further
study. In addition, entrainment is also affected by the gas phases
involved in the outflow. Our analysis traces the metallicity in
the ionized gas (considered the ``cool'' phase by most
simulations, $T\,<\,2 \times 10^4$\,K), thought to carry most of
the mass, whereas physically realistic winds also contain a ``hot
phase'' ($T\,>\,5 \times 10^5$\,K) that harbors most of the energy
\citep{kim20}. Consequently, with MAGMA, we are missing the
diagnostic of the hot X-ray phase.

\section{Summary and future perspectives}\label{sec:conclusions}

In this paper, we explore the impact of ejective feedback on galaxy
evolution using data from the MAGMA sample of 392 galaxies
with \hi\ and CO detections, which collects a homogeneous
sample of local field star-forming galaxies, covering the widest
range of stellar and gas masses, metalliticies and star formation
compiled up to now.
After presenting the sample, and the correlations
among metallicity, SFR, \mstar\ and (cool) gas-phase components in
\citetalias{Ginolfi+20} and \citetalias{hunt20}, in this third paper of the
series, we investigate outflow metal loading factors and effective
yields in terms of gas fraction, mass and circular velocity, using
the formalism of \citet{pagel09}. The main results of the paper
follow.

\begin{itemize}
{\parskip 0.5\baselineskip
\item
We model the chemical enrichment of MAGMA galaxies using a
formalism based on \cite{pagel09}, without any constraint on the
\mgas\ time evolution and including the assumption that wind
metallicity \zw\ is different from the ISM metallicity \zg. Wind
metallicity \zw\ can be greater than the ambient \zg\ if it is
primarily comprised of supernova ejecta or depressed if a
sufficient amount of metal-poor gas is entrained as the wind
propagates out of the galaxy
\citep{Chisholm+15,Creasey2015,Muratov+17,pillepich18}. Using this
formalism we have observationally constrained two parameters
important for stellar feedback: (a)~the metal-loading factor of
the winds \zetaw; and (b)~the difference of the mass loading of
the accretion and winds, \deta\,=\,\etaa\,$-$\,\etaw. This is done
through a Bayesian approach applied to subsamples of galaxies in
different \mstar\ and \mug\ bins.
\item In terms of trends with \mstar, we find that \deta\ is degenerate,
and exceedingly difficult to constrain. However, our Bayesian
estimates based on \mug\ bins show that \deta\ increases
significantly with increasing gas fraction, implying that \deta\
is a strong function of gas content and possibly evolutionary
state (see Appendix \ref{app:toymodel}).
\item
The wind metal-loading factors \zetaw\ are found to increase with decreasing
stellar mass and circular velocity, with a flattening toward larger \mstar. This indicates that the
efficiency of metal-loading in the winds depends on the
potential depth and is strongly enhanced in low-mass systems, in
agreement with previous work based on different methodologies for estimating \zetaw\
\citep[e.g.,][]{Peeples_Shankar11,chisholm18}.
\item
We have also constrained effective yields \ye\ with MAGMA data,
and find that \ye\ increases with stellar mass and \vcirc\ up to a certain threshold,
followed by a flattening toward larger \mstar, and possibly a mild
inversion at the highest masses.
\item Finally, we have inferred possible trends with \vcirc\ for the
wind mass-loading factors \etaw\ and find that our results for low
\vcirc\ (\mstar) favor momentum-driven winds, rather than
energy-driven ones. }
\end{itemize}

Various quantities in our analysis, \zetaw, \ye, and \zg, present
inflections at \mstar\,$ \sim 3 \times 10^{9}$\,\msun,
corresponding to virial velocities $\sim 100$\,\kms. This
corresponds to a ``gas-richness'' mass scale, associated with
transitions of physical processes, separating two different mass
regimes \citep[\papii; see also][]{kannappan04}. It delineates an
``accretion-dominated'' regime at low masses, where stellar
outflows are relevant, and an intermediate mass,
``gas-equilibrium'' regime, where outflows are less effective, and
galaxies are able to consume gas through SF as fast as it is
accreted. There would also be an additional, larger ``bimodality''
mass threshold at \mstar\,$ \sim 3 \times 10^{10}$\,\msun\
\citep[\papii,][]{Kannappan2013} but this is only weakly evident
in MAGMA, mainly because of the exclusion of passive or
``quenched'' gas-poor galaxies and AGN.

These results can also be interpreted within a broader context,
since the mass thresholds defining metallicity and gas properties
are associated with changes of correlations involving stellar
populations and stellar and dark matter distributions. In
particular, inflections with mass of metal-loading and metal
yields at the ``gas-richness'' threshold seem to define changes of
the slopes of total mass density profiles including dark matter
\citep[e.g.,][]{Tortora+19_LTGs_DM_and_slopes}; this threshold
approximatively separates galaxies with mass density slopes
shallower and steeper than $\sim -1$ (see their Figure 1). More
investigations on this mass scale will follow.

As outlined in \citetalias{hunt20}, further improvement of the
MAGMA coverage, with more detections of total gas content in dwarf
galaxies, will strengthen the conclusions of our series of papers.
In addition, it is important to reassess the impact on our
conclusions of assumptions made here, namely that the accretion
rate $\dot{M}_a$ is proportional to the SFR and the
chemically-unenriched nature of the accreted gas. In this paper,
we have also pursued a first step to combine information from
completely independent datasets and observables. Only by including
gas, stellar, kinematical and dynamical observables, will we be
able to determine the efficiency of the physical phenomena which
shape and regulate galaxy evolution.

\section*{Acknowledgements}
We thank the anonymous referee and the editor for profound insight that helped to greatly improve the paper.
We are grateful for funding from the INAF PRIN-SKA 2017 program 1.05.01.88.04.

\bibliographystyle{aa}


\bibliography{Tortora_MAGMA-III.bib}

\appendix

\section{Additional cornerplots for Bayesian best fits}\label{app:bayesian}

\begin{figure*}[!h]
\begin{center}
\hbox{
\includegraphics[angle=0,width=0.33\linewidth]{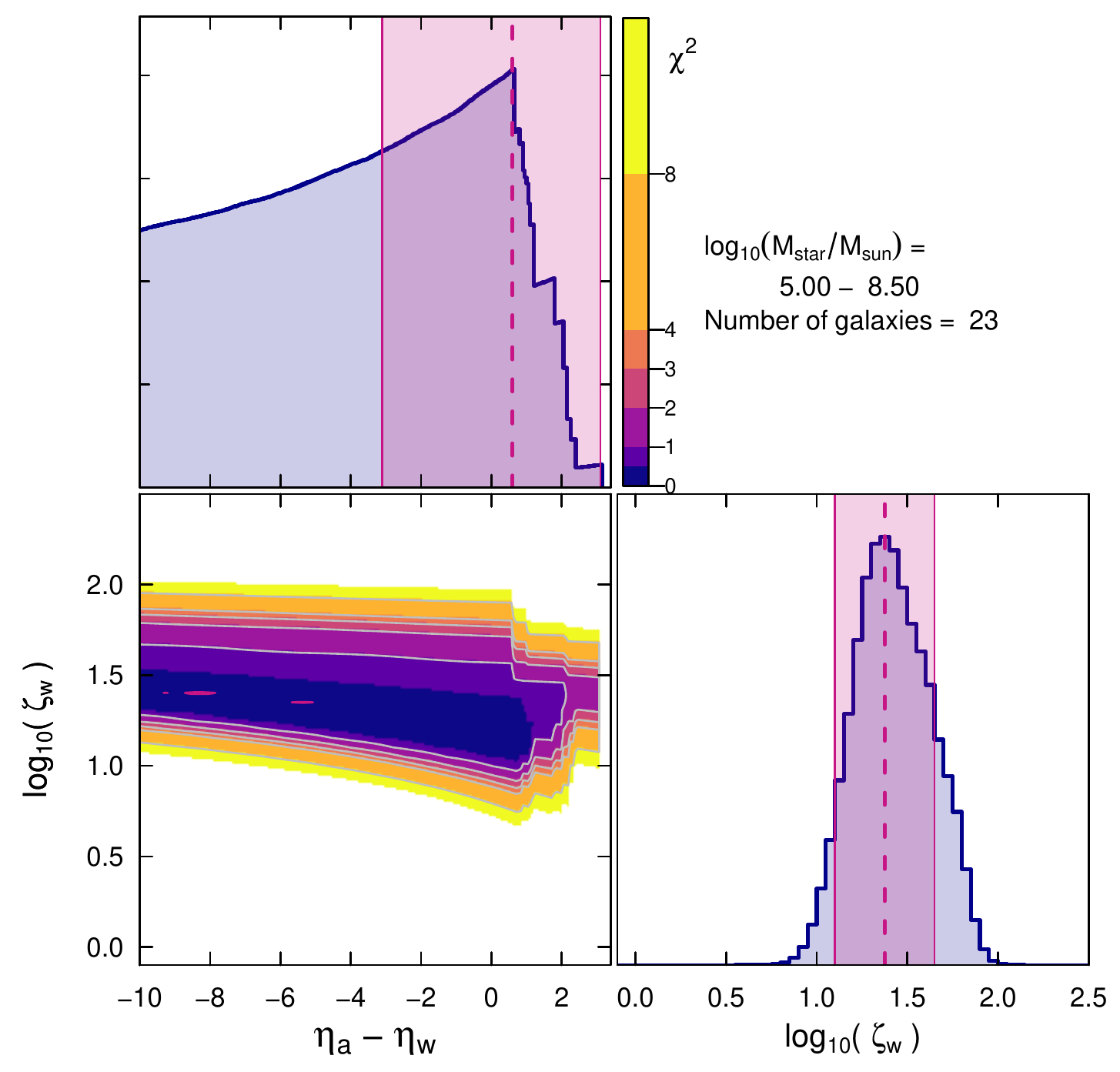}
\includegraphics[angle=0,width=0.33\linewidth]{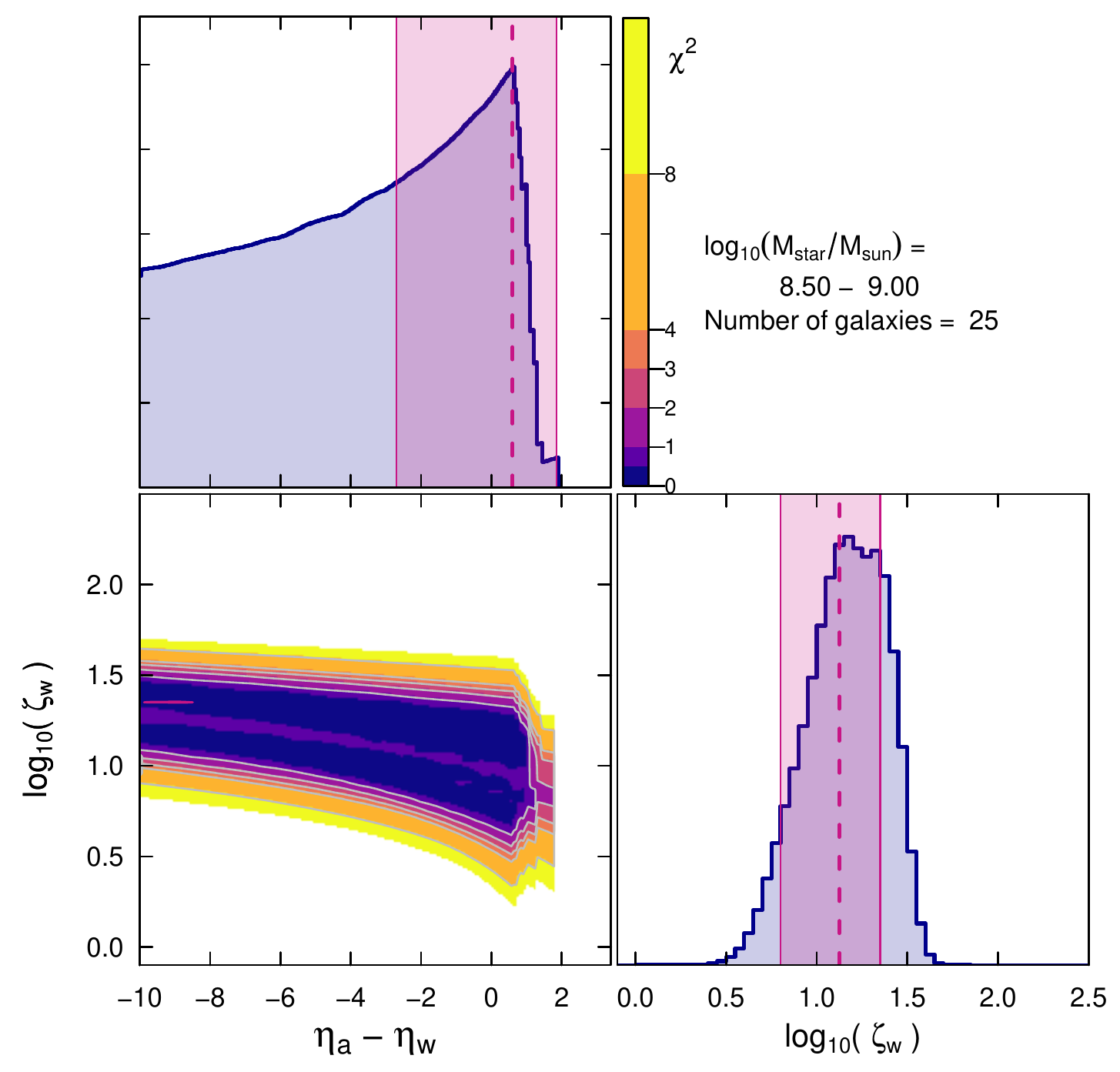}
\includegraphics[angle=0,width=0.33\linewidth]{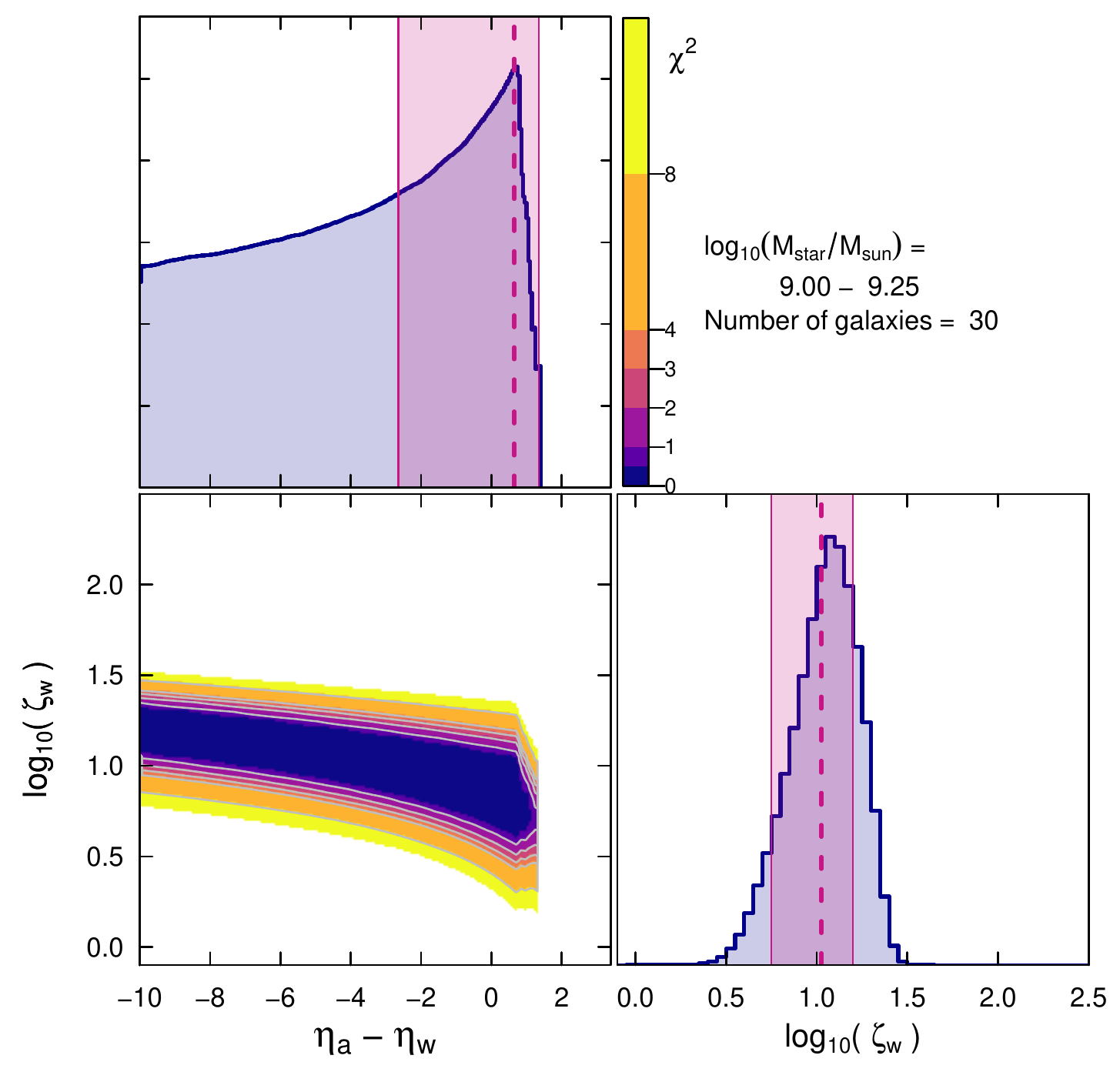}
} \vspace{\baselineskip} \hbox{
\includegraphics[angle=0,width=0.33\linewidth]{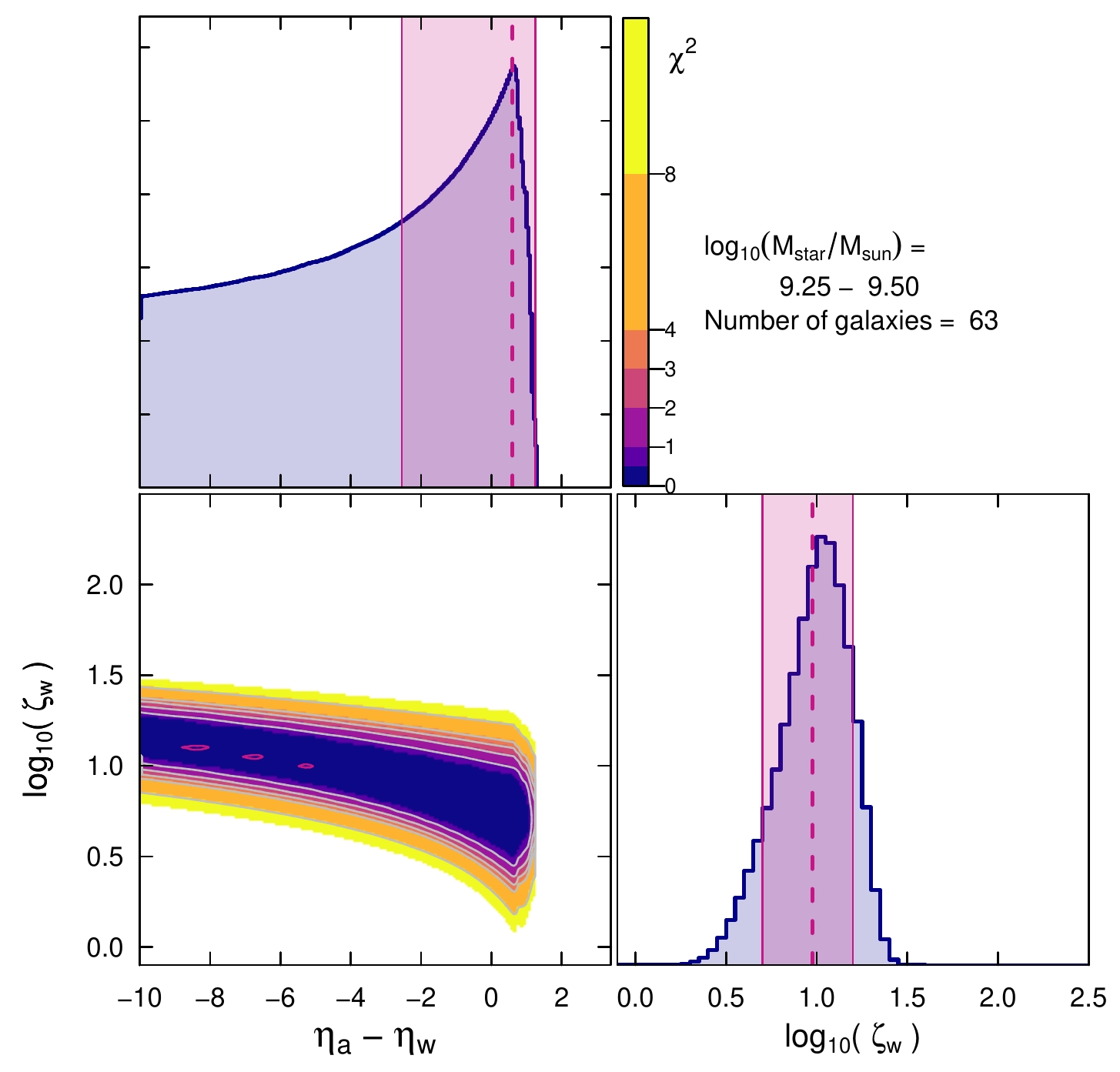}
\includegraphics[angle=0,width=0.33\linewidth]{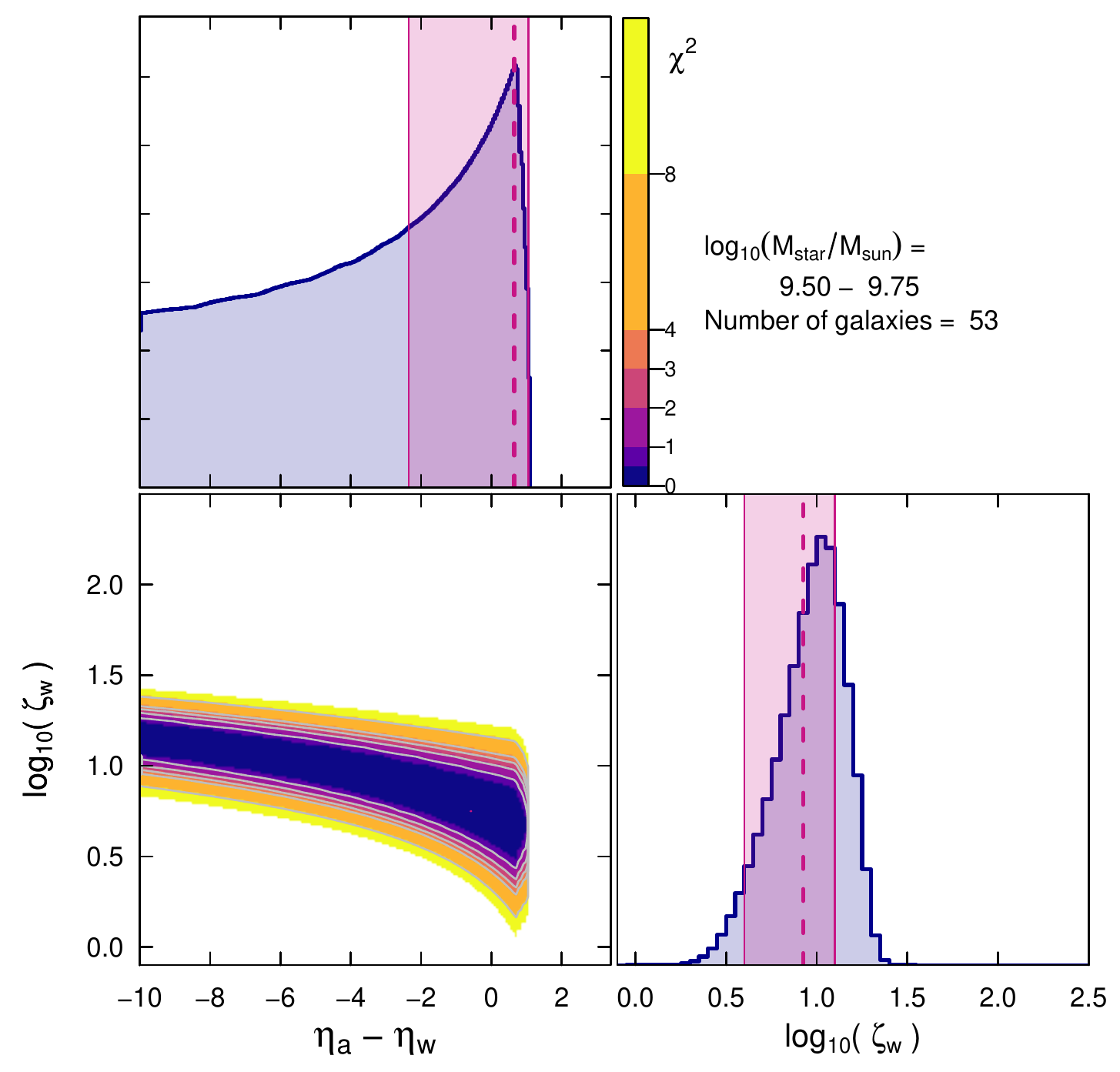}
\includegraphics[angle=0,width=0.33\linewidth]{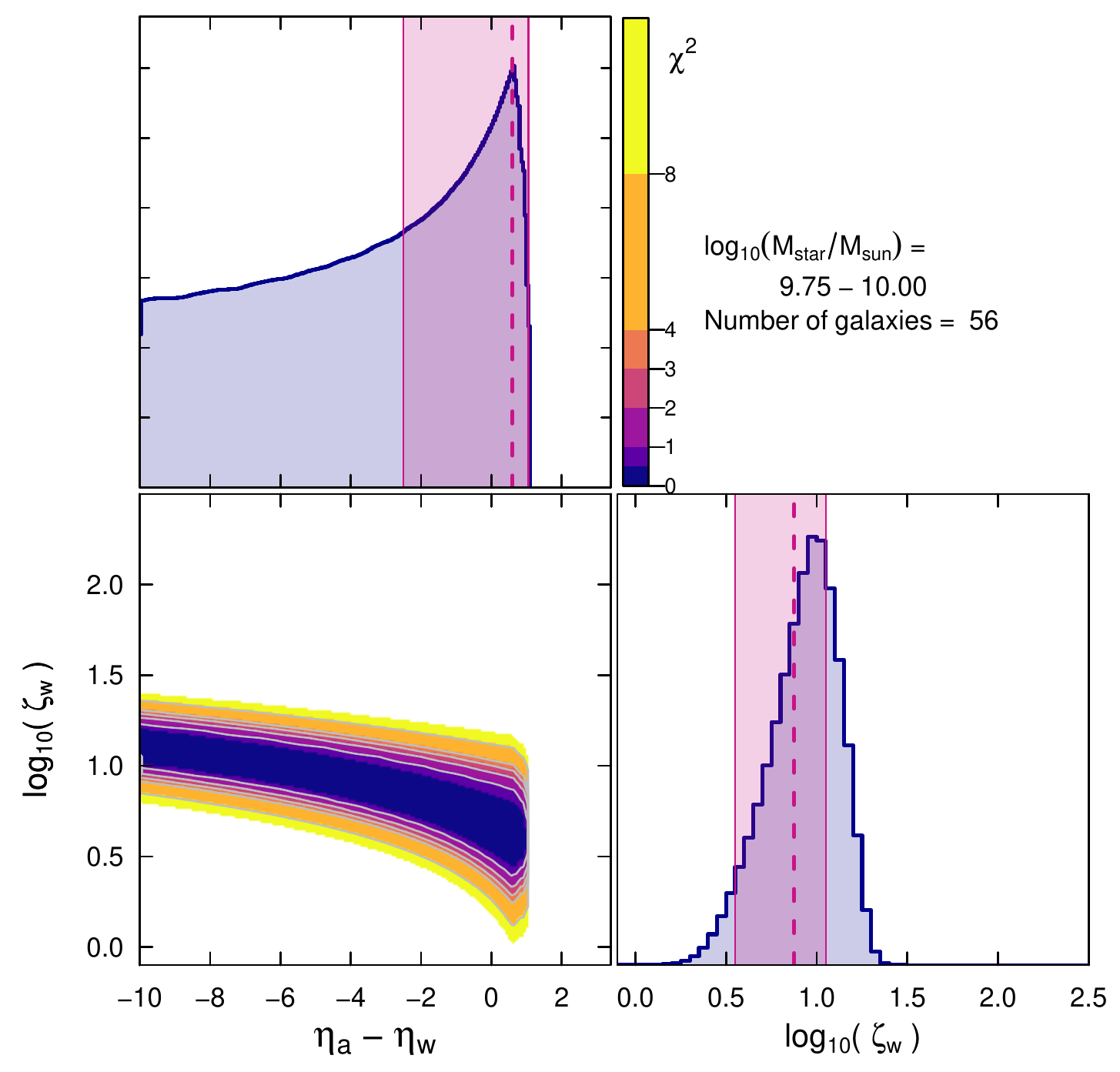}
} \vspace{\baselineskip} \hbox{
\includegraphics[angle=0,width=0.33\linewidth]{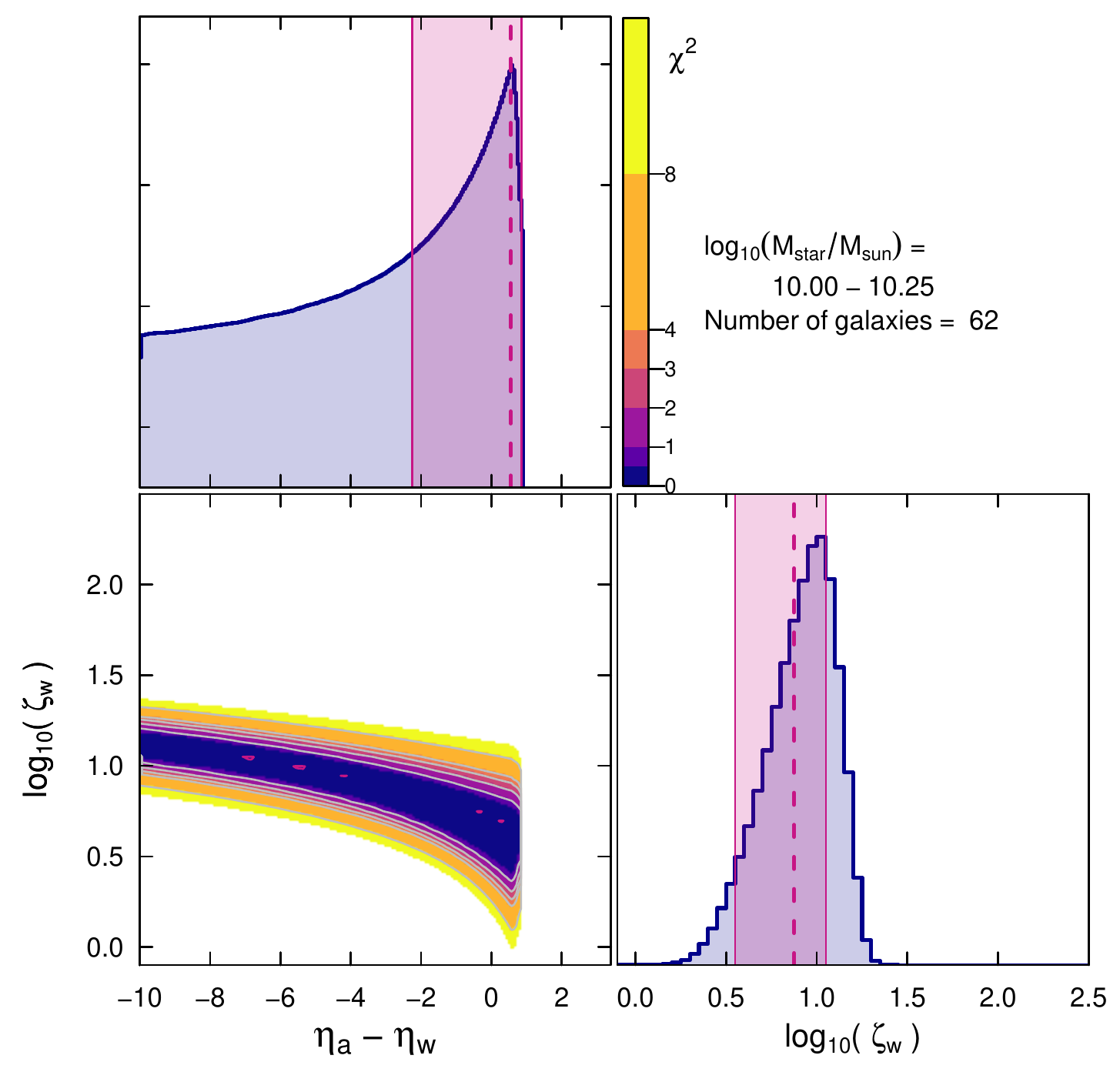}
\includegraphics[angle=0,width=0.33\linewidth]{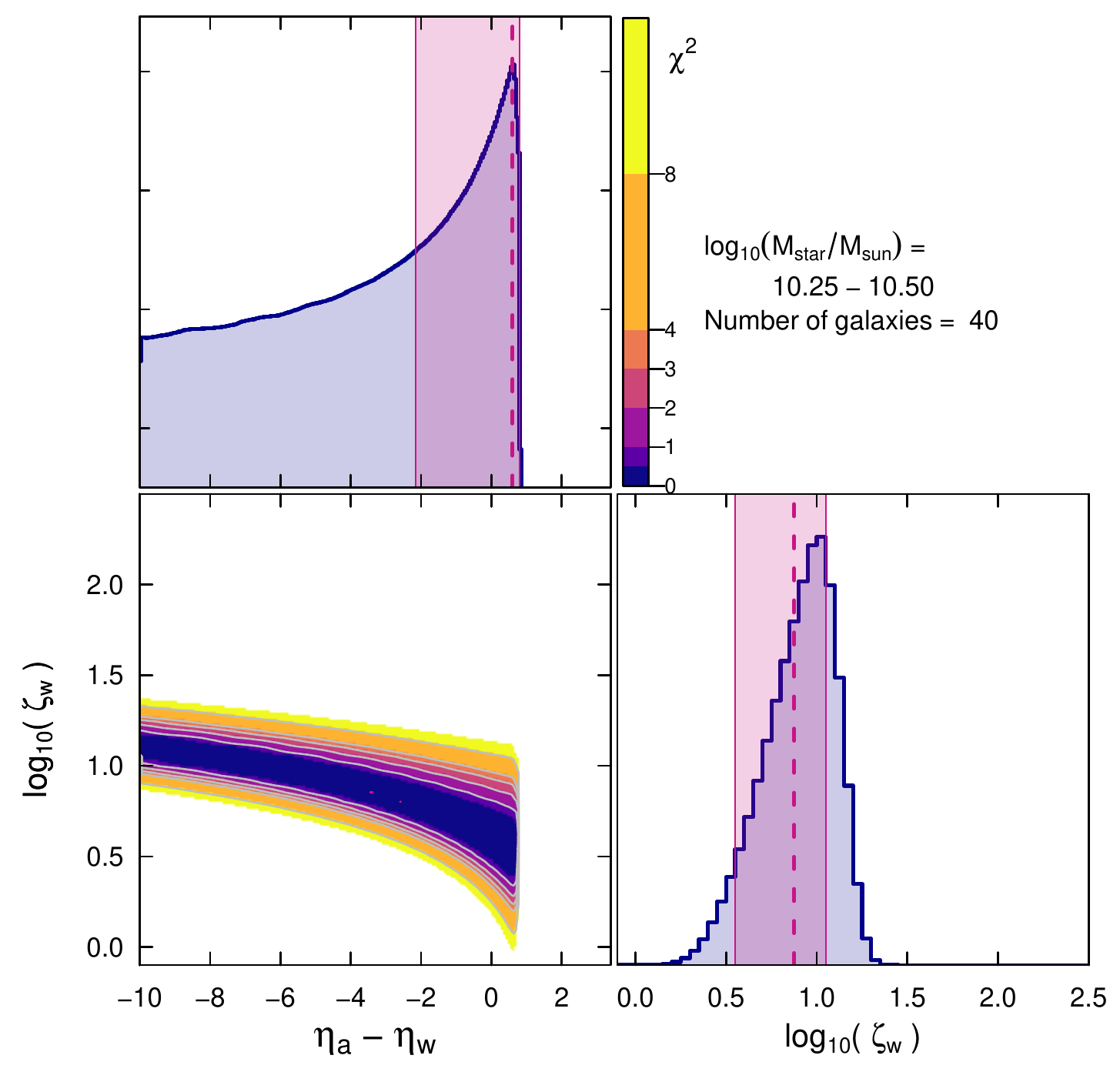}
\includegraphics[angle=0,width=0.33\linewidth]{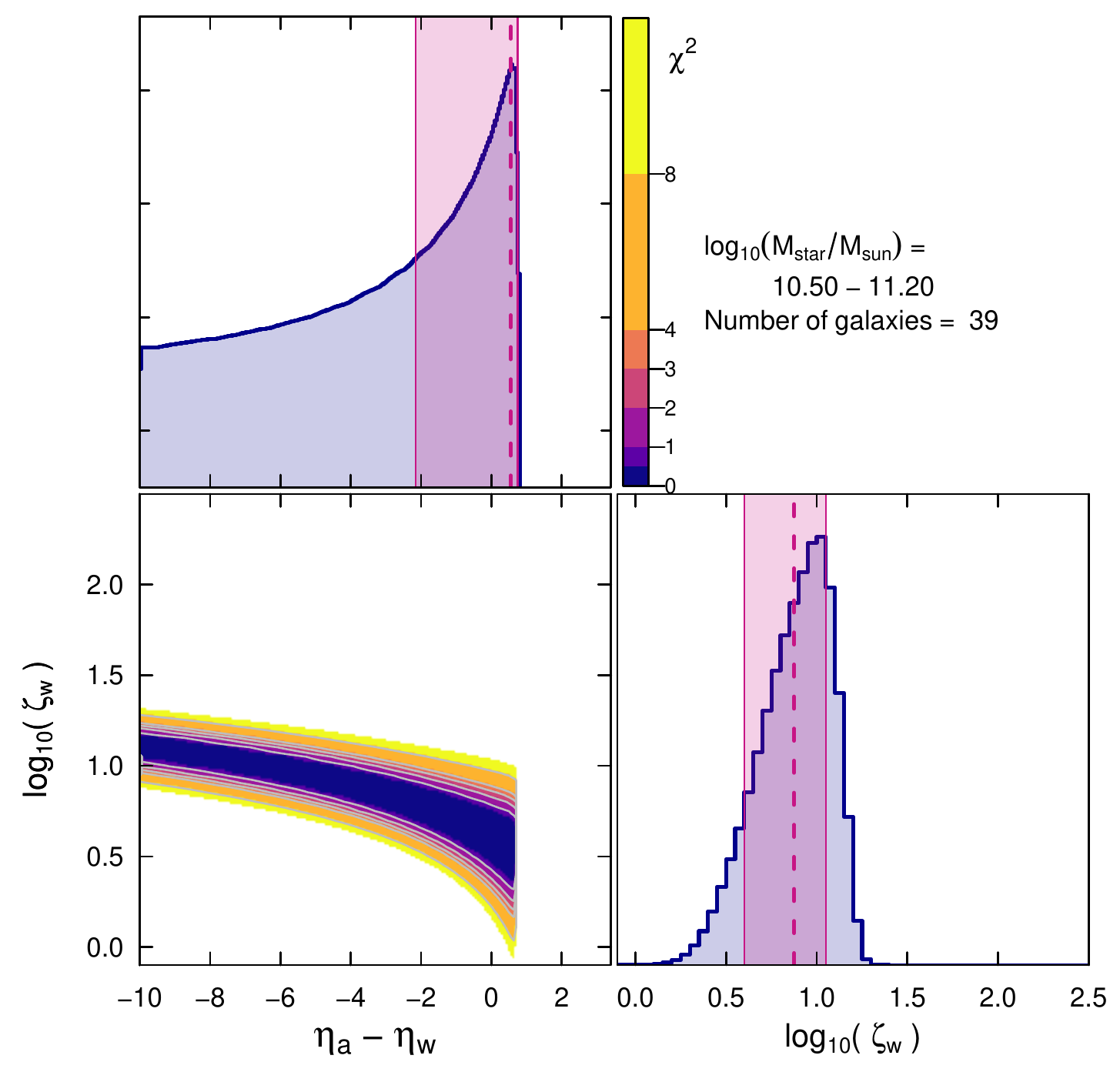}
}
\end{center}
\vspace{-1.2\baselineskip} \caption{Corner plots of \chisq\
surface as a function of the model parameters (\deta, \zetaw) for
MAGMA. The violet contours correspond to the minimum \chisq\
value. The top and right panels of each corner plot report the
probability density distributions for the marginalized parameters;
confidence intervals ($\pm 1\sigma$) are shown as violet-tinted
shaded rectangular regions, and the MLE (PDF mode for
\deta, median for \zetaw) is shown by a vertical dashed line.
Here, \deta\,$\in$\,[-10,3.4] (see Fig. \ref{fig:bayes} for the
``symmetric'' \deta\ intervals). \label{fig:bayes_10}
}
\end{figure*}

\begin{figure*}[!t]
\begin{center}
\hbox{
\includegraphics[angle=0,width=0.33\linewidth]{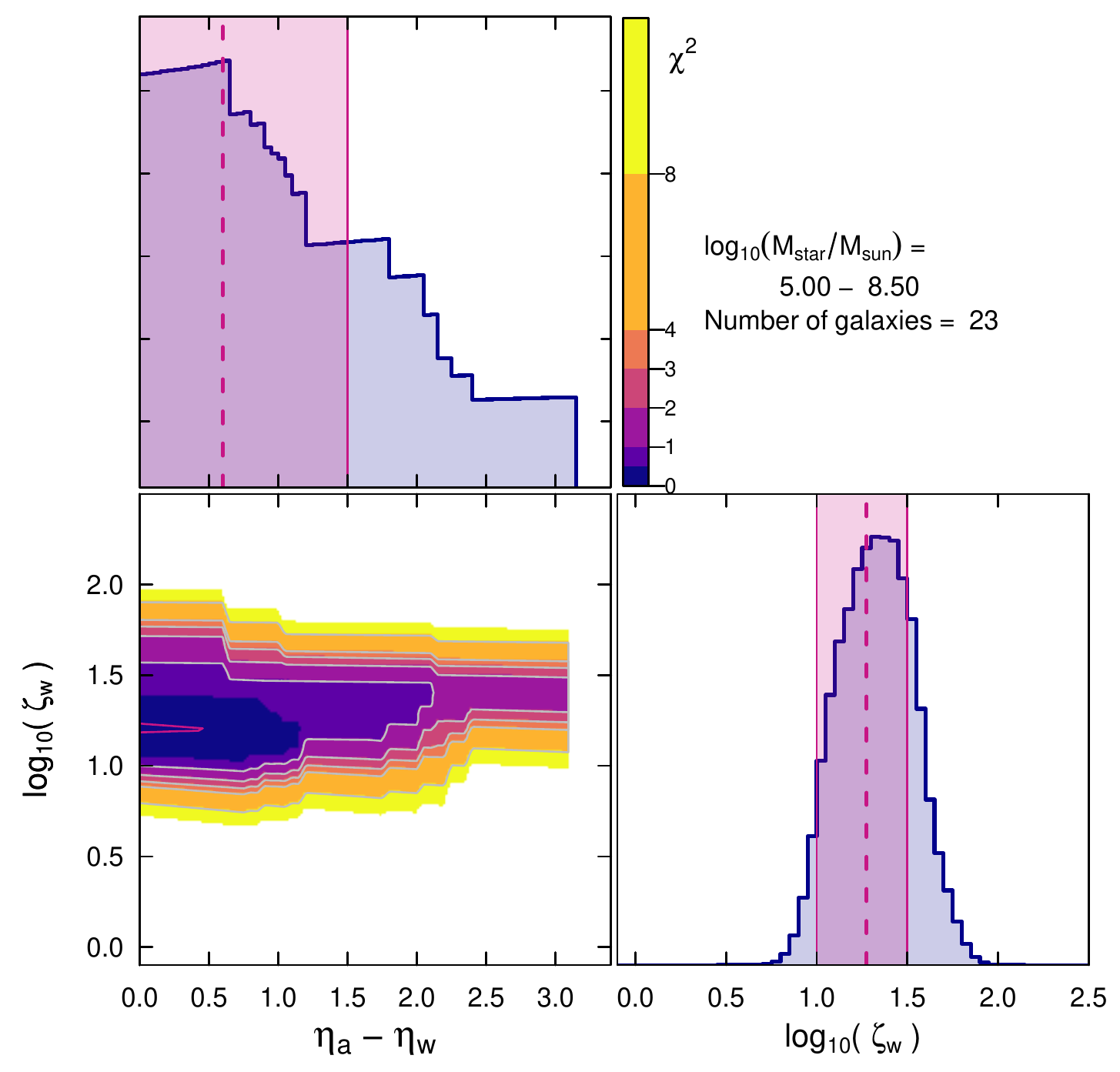}
\includegraphics[angle=0,width=0.33\linewidth]{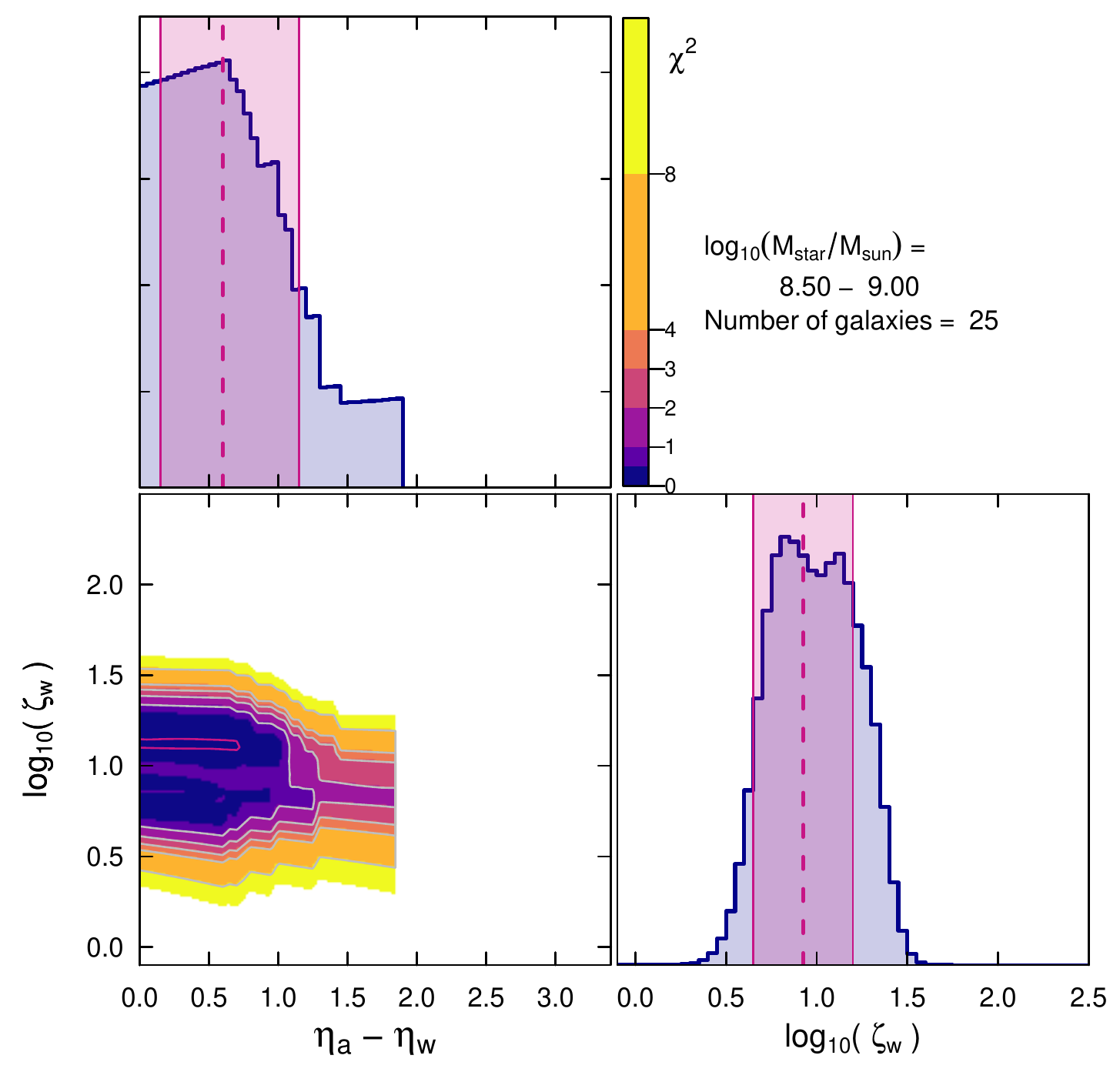}
\includegraphics[angle=0,width=0.33\linewidth]{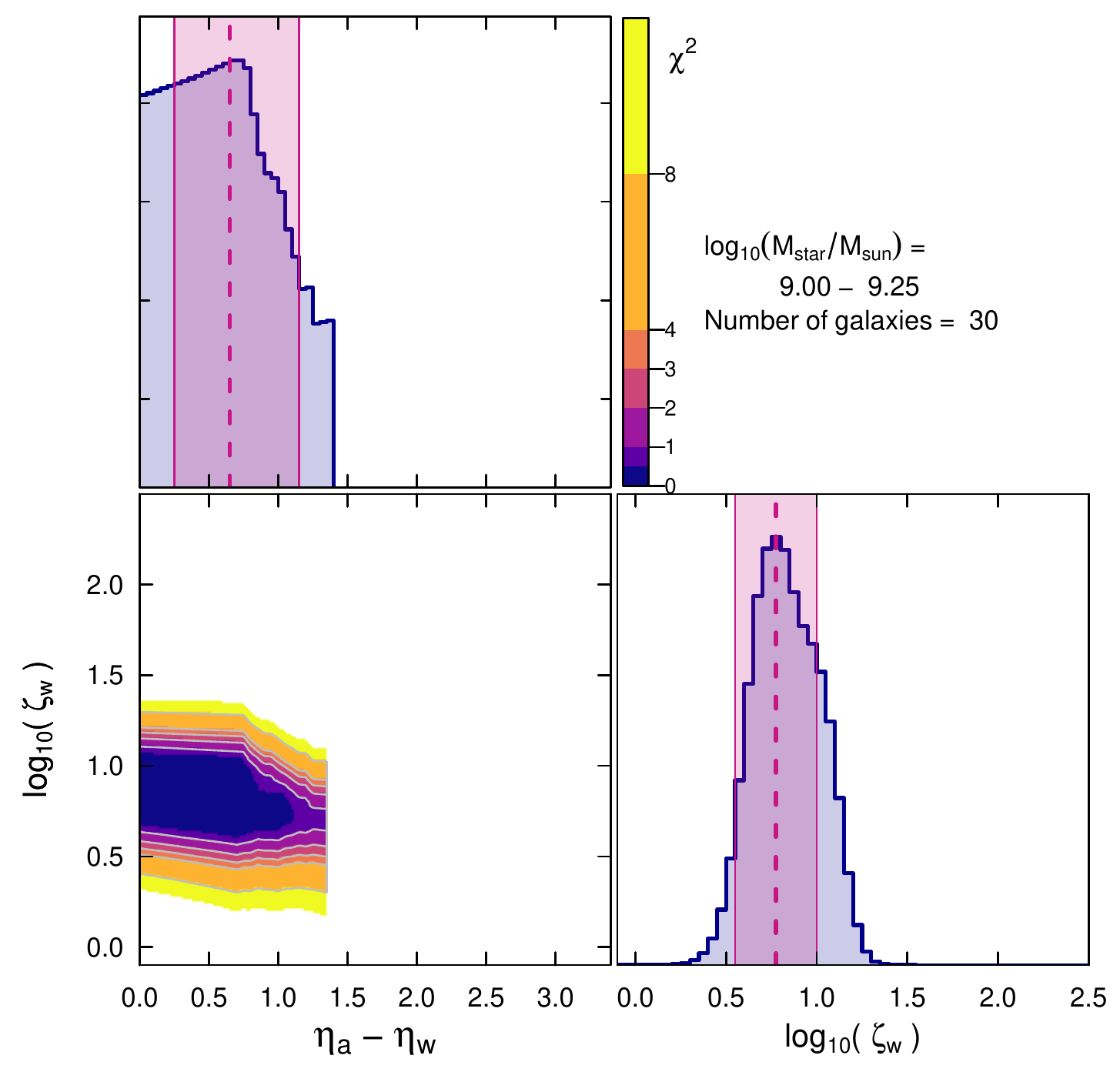}
}
\vspace{\baselineskip}
\hbox{
\includegraphics[angle=0,width=0.33\linewidth]{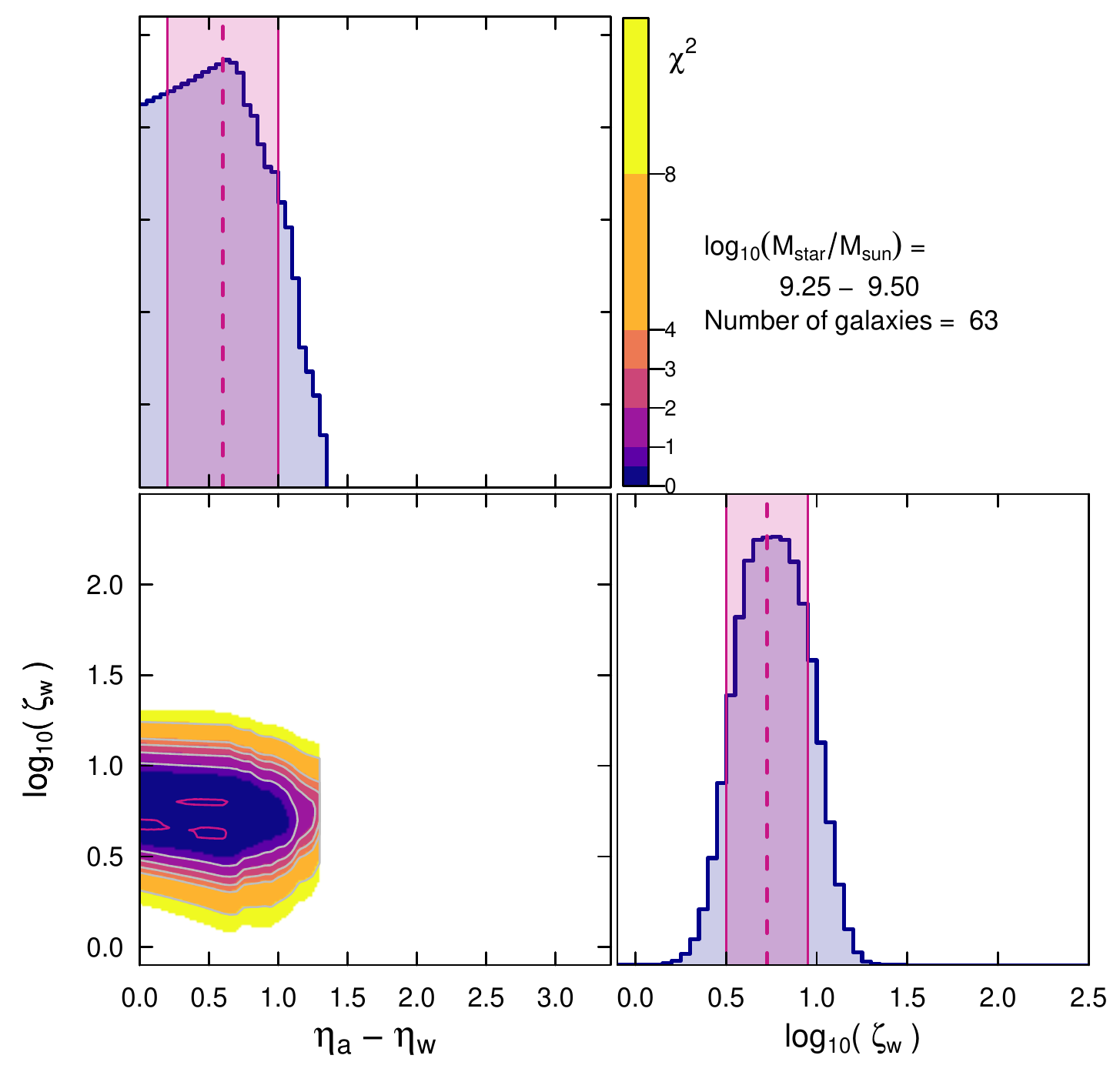}
\includegraphics[angle=0,width=0.33\linewidth]{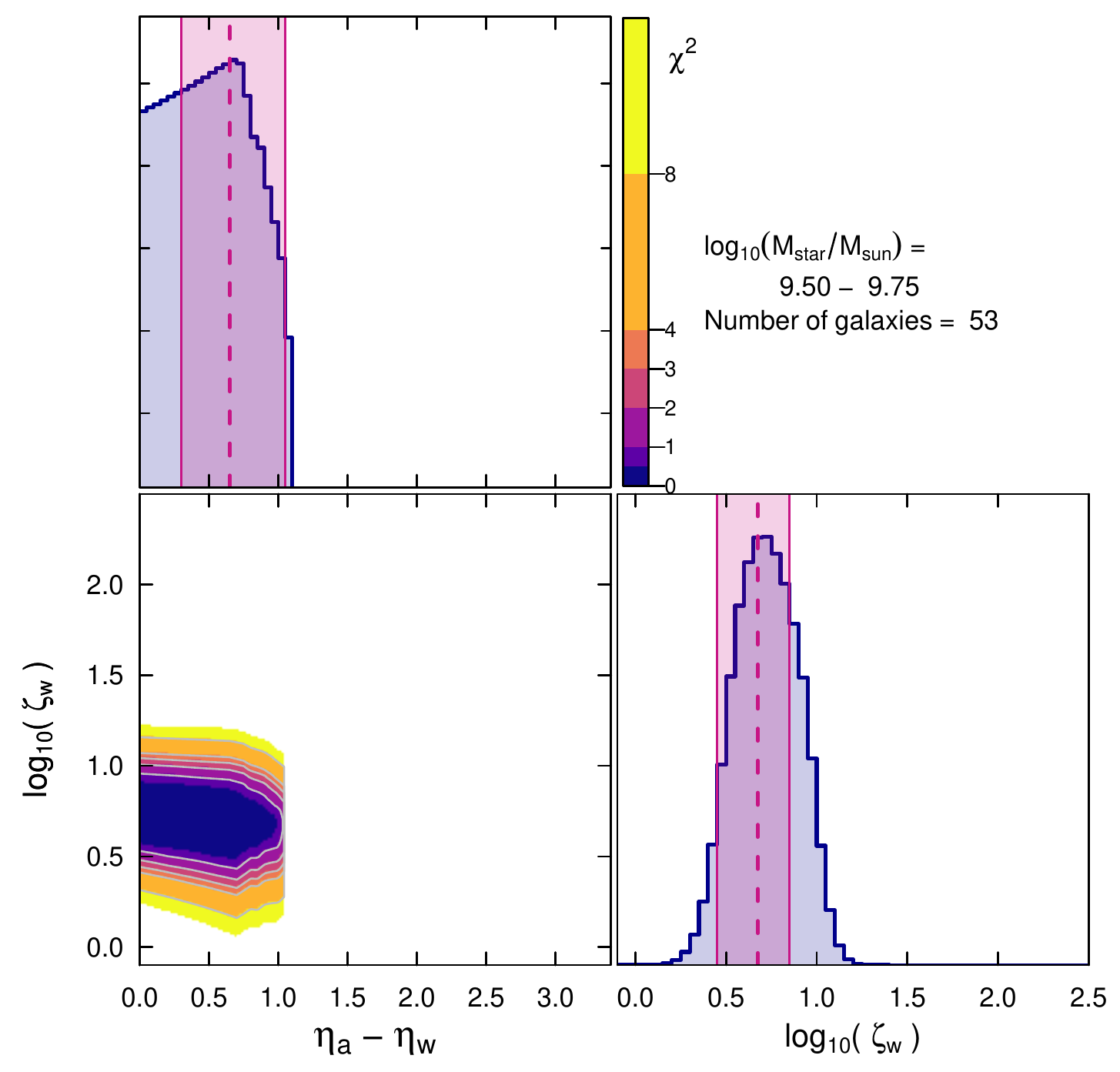}
\includegraphics[angle=0,width=0.33\linewidth]{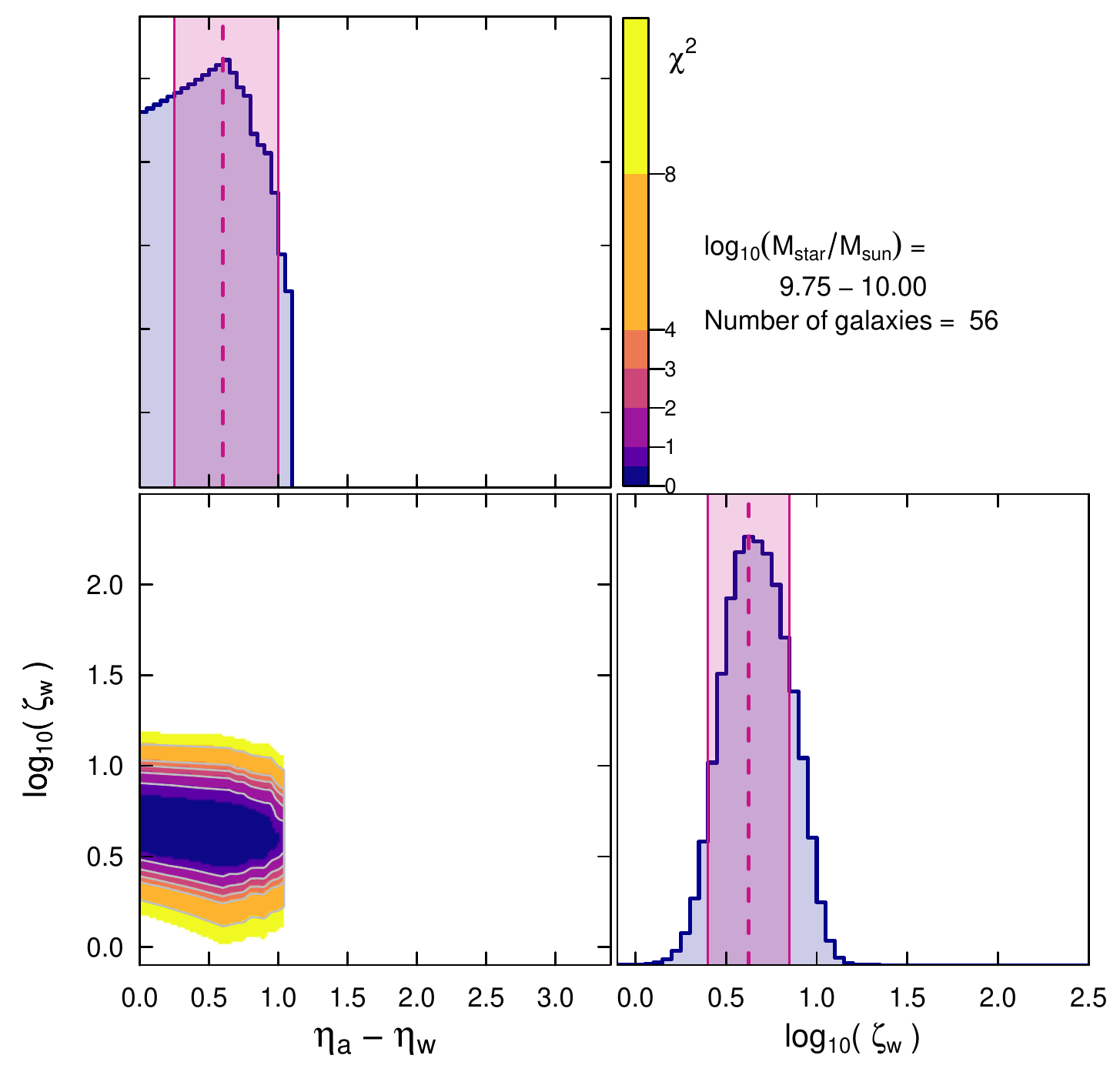}
}
\vspace{\baselineskip}
\hbox{
\includegraphics[angle=0,width=0.33\linewidth]{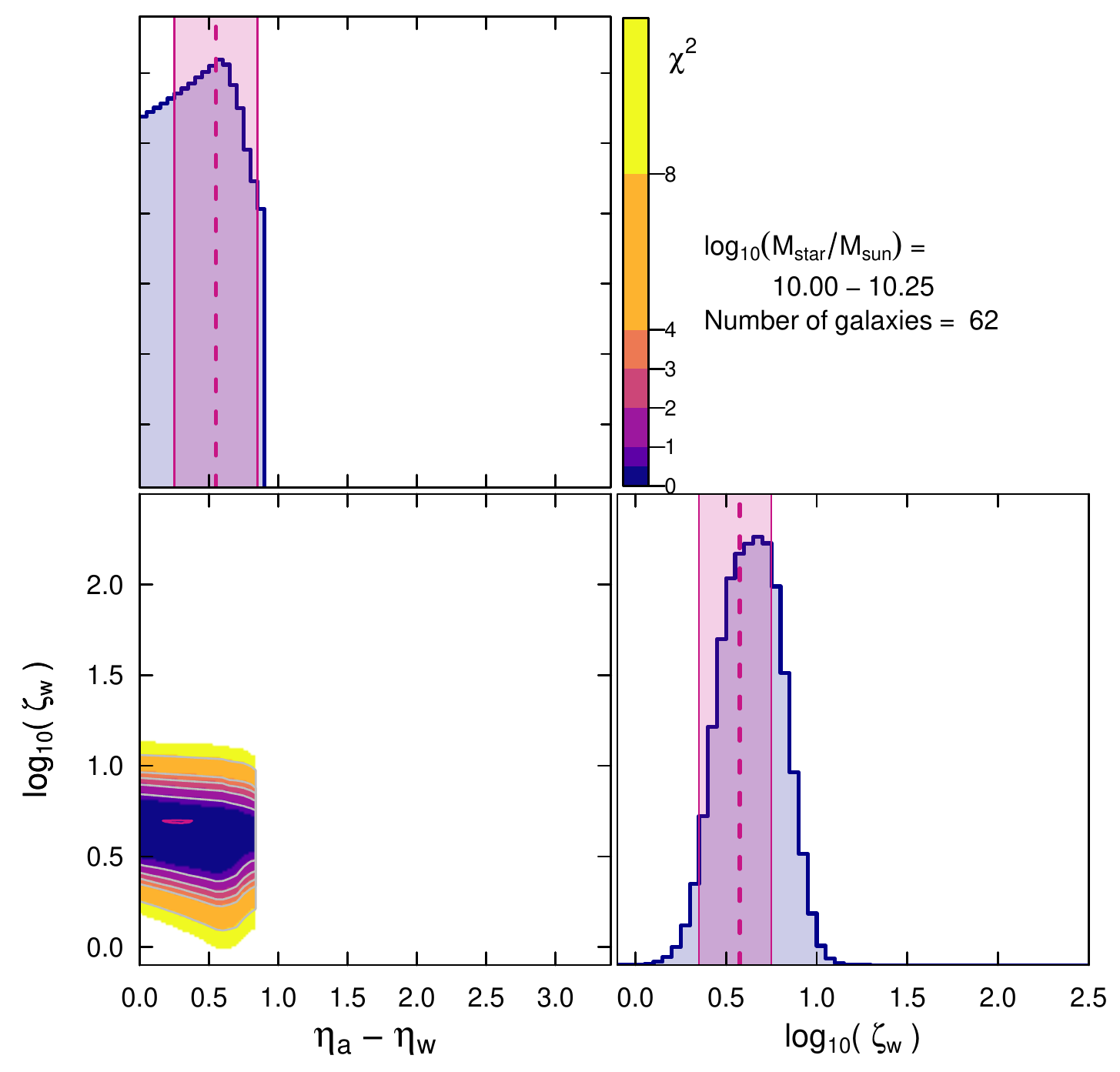}
\includegraphics[angle=0,width=0.33\linewidth]{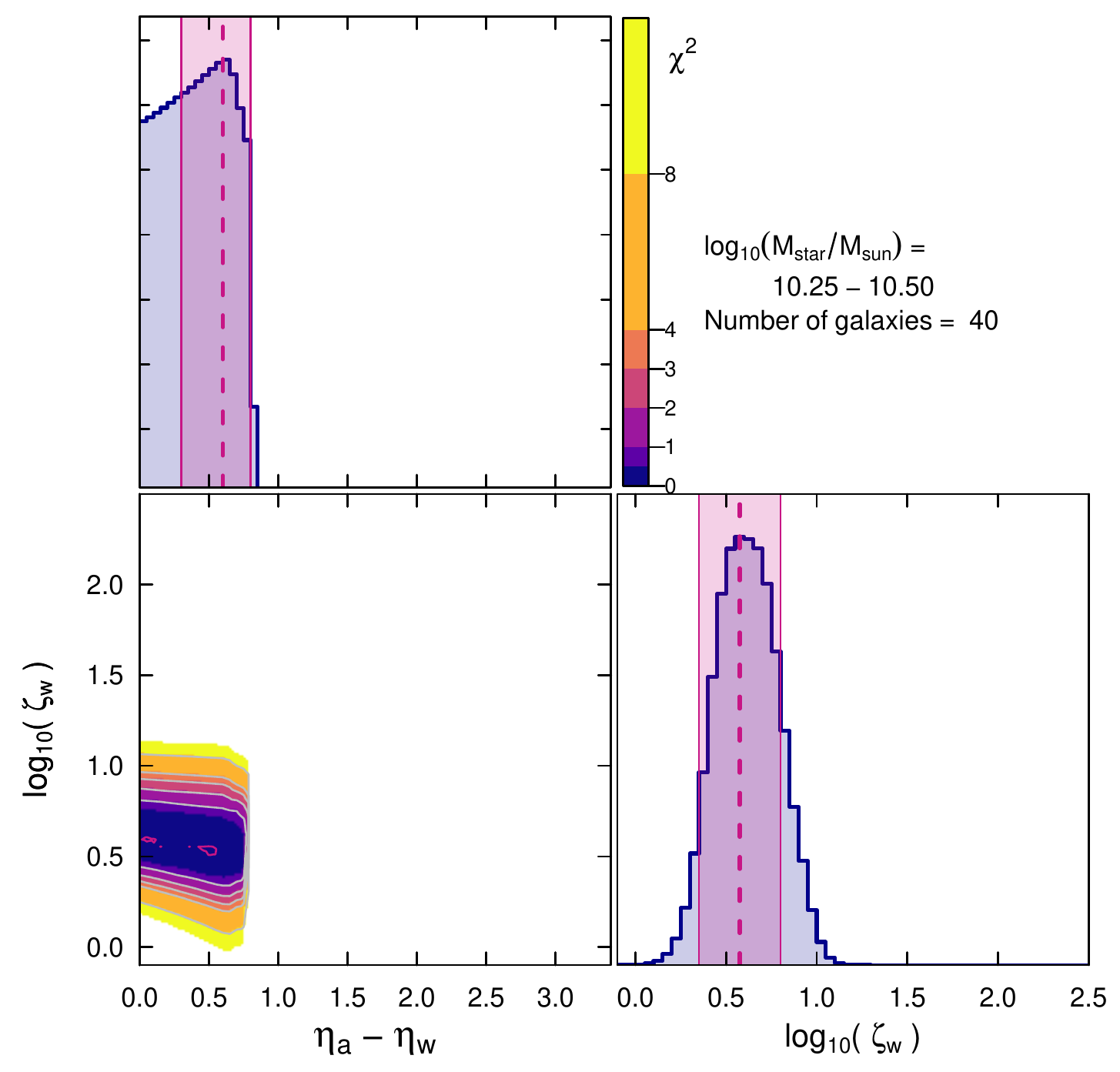}
\includegraphics[angle=0,width=0.33\linewidth]{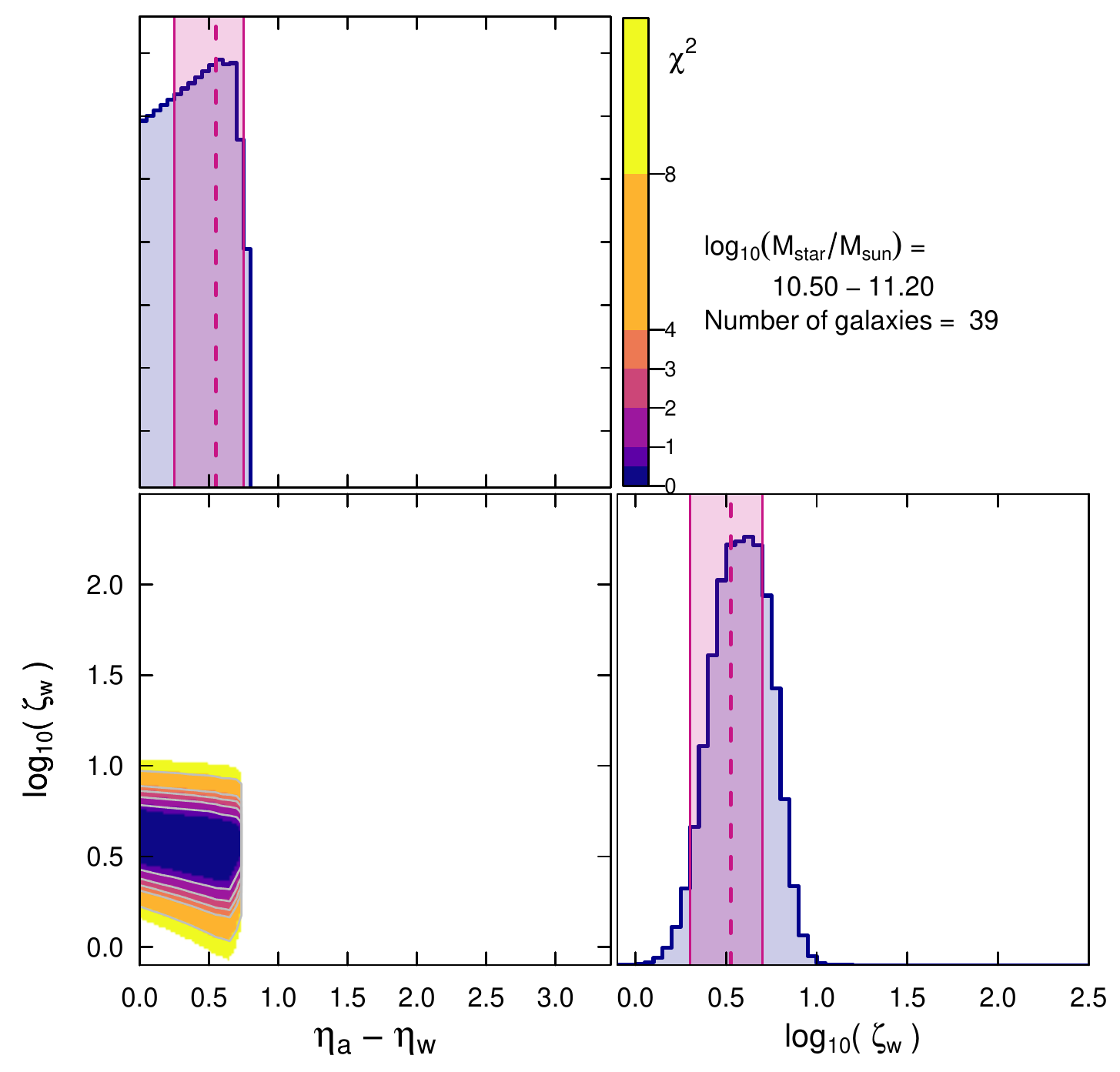}
}
\end{center}
\vspace{-1.2\baselineskip} \caption{Corner plots of \chisq\
surface as a function of the model parameters (\deta, \zetaw) for
MAGMA. The violet contours correspond to the minimum \chisq\
value. The top and right panels of each corner plot report the
probability density distributions for the marginalized parameters;
confidence intervals ($\pm 1\sigma$) are shown as violet-tinted
shaded rectangular regions, and the MLE (PDF median) is shown by a
vertical dashed line. Here, \deta\,$\in$\,[0,3.4] (see Fig.
\ref{fig:bayes} for the ``symmetric'' \deta\ intervals).
\label{fig:bayes_0} }
\end{figure*}

Here we show the additional corner plots of the \chisq\ surface as a function of the model
priors (\deta, \zetaw) for MAGMA, for \deta\,$\in$\,[-10,3.4]
(Fig. \ref{fig:bayes_10}) and \deta\,$\in$\,[0,3.4] (Fig. \ref{fig:bayes_0}).

\null ~~~


\section{SFHs and toy-model predictions for gas and stellar growth}\label{app:toymodel}

In order to investigate the time derivatives of \mgas\ and \mstar\
and their ratio, we introduce an arbitrary SFH. As a simple
example, we have taken an exponential with arbitrary scaling
defined by $\beta$ and the star-formation time scale \tausf:
\begin{equation}
\psi(t)\,=\,\psi_0 \exp \left( \frac{\beta t}{\tau_\mathrm{SF}} \right) \quad .
\label{eqn:sfrtime}
\end{equation}
Equation \eqref{eqn:sfrtime} can be integrated to obtain \mstar\ [see Eq. \eqref{eqn:mstar}],
at the current age of the galaxy \tgal:
\begin{equation}
M_{\Large\star}(t=t_\mathrm{gal})\,=\,\frac{\psi_0\ \alpha\ \tau_\mathrm{SF}}{\beta}\
\left[ \exp \left( \frac{\beta t_\mathrm{gal}}{\tau_\mathrm{SF}}\right) - 1 \right] \quad .
\label{eqn:mstar_tgal}
\end{equation}

If we take the time-scale of star formation \tausf\ to be equal to the gas depletion time \taugas\
[the inverse of \epss, see Eq. \eqref{eqn:sfr}],
there are two unknowns in Eq. \eqref{eqn:mstar_tgal}: \tgal\ and $\psi_0$
($\beta$ is fixed arbitrarily for rising and declining SFHs).
However, $\psi$(\tgal)\,=\,$\psi_0\,\exp(\beta t_\mathrm{gal}/\tau_\mathrm{SF})$\,=\,SFR
and $M_{\Large\star}$(\tgal)\,=\,\mstar, are the observables that we measure, so that
Eq. \eqref{eqn:mstar_tgal} can be solved for $\psi_0$:
\begin{equation}
\psi_0\,=\,\frac{\beta\,M_\mathrm{star}}{\alpha\ \tau_\mathrm{SF}}
+ \mathrm{SFR} \quad . \label{eqn:psi0}
\end{equation}
This toy model also gives the initial gas mass \mi\,=$\psi_{0}$\,\tausf.

The remaining variable needed to define the galaxy's evolutionary state is \tgal,
which can be established with $\psi_0$ through Eq. \eqref{eqn:sfrtime}:
\begin{equation}
t_\mathrm{gal}\,=\,\frac{\tau_\mathrm{SF}}{\beta}\ \ln \left( \frac{\mathrm{SFR}}{\psi_0} \right) \quad (\beta \neq 0).
\label{eqn:tgal}
\end{equation}

In case of a constant SFH ($\beta\,=\,0)$, with $\psi(t)\,=\,\psi_0\,=\,$SFR,
the solutions are quite simple: $M_{\Large\star}(t)\,=\,$\mstar,
$M_\mathrm{gas}\,=\,$\mi\,=\,SFR\, \tausf,
and \deta\,=\,$\alpha$.

\begin{figure}[!h]
\includegraphics[angle=0,width=0.98\linewidth]{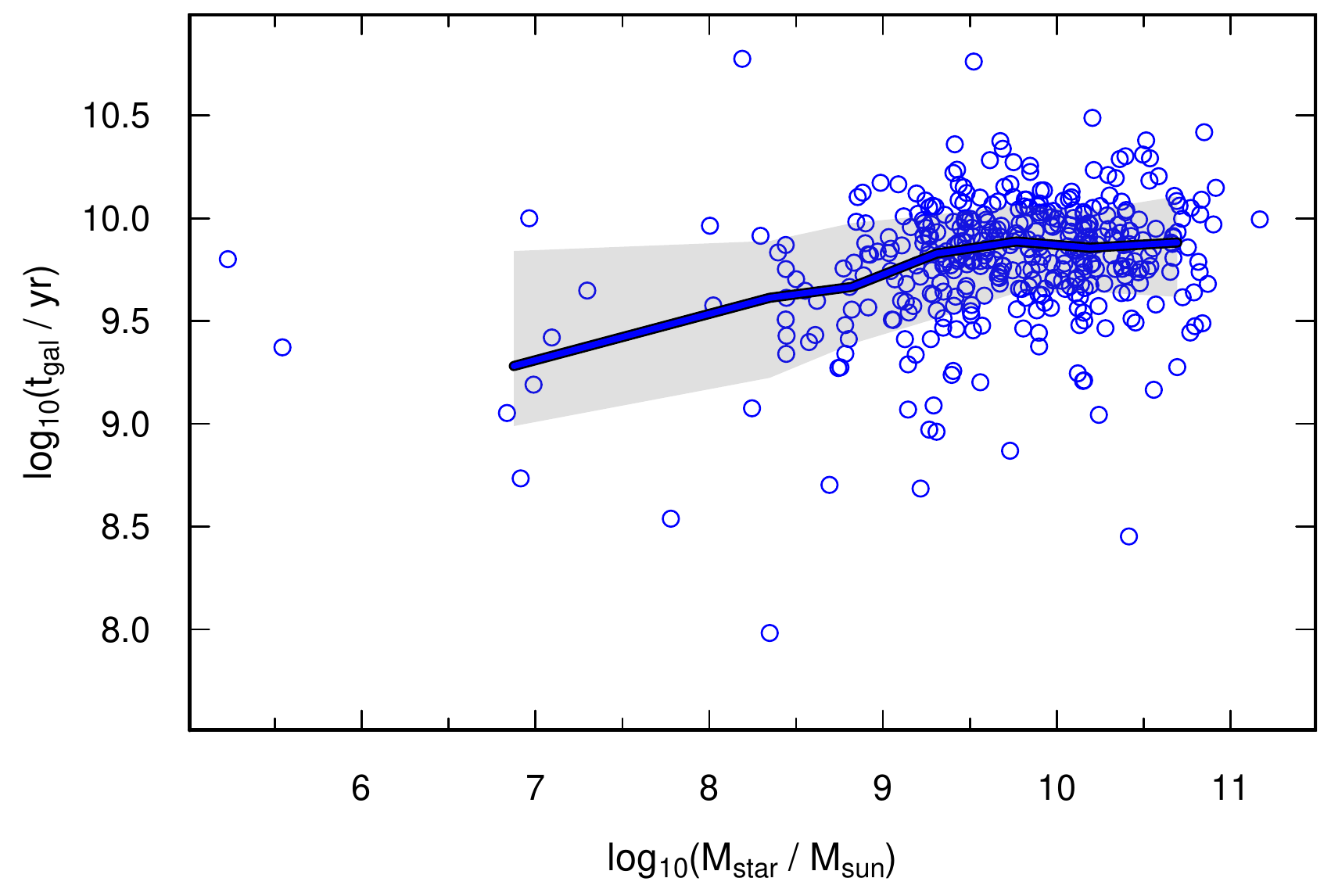} \\
\vspace{0.5\baselineskip}
\includegraphics[angle=0,width=0.98\linewidth]{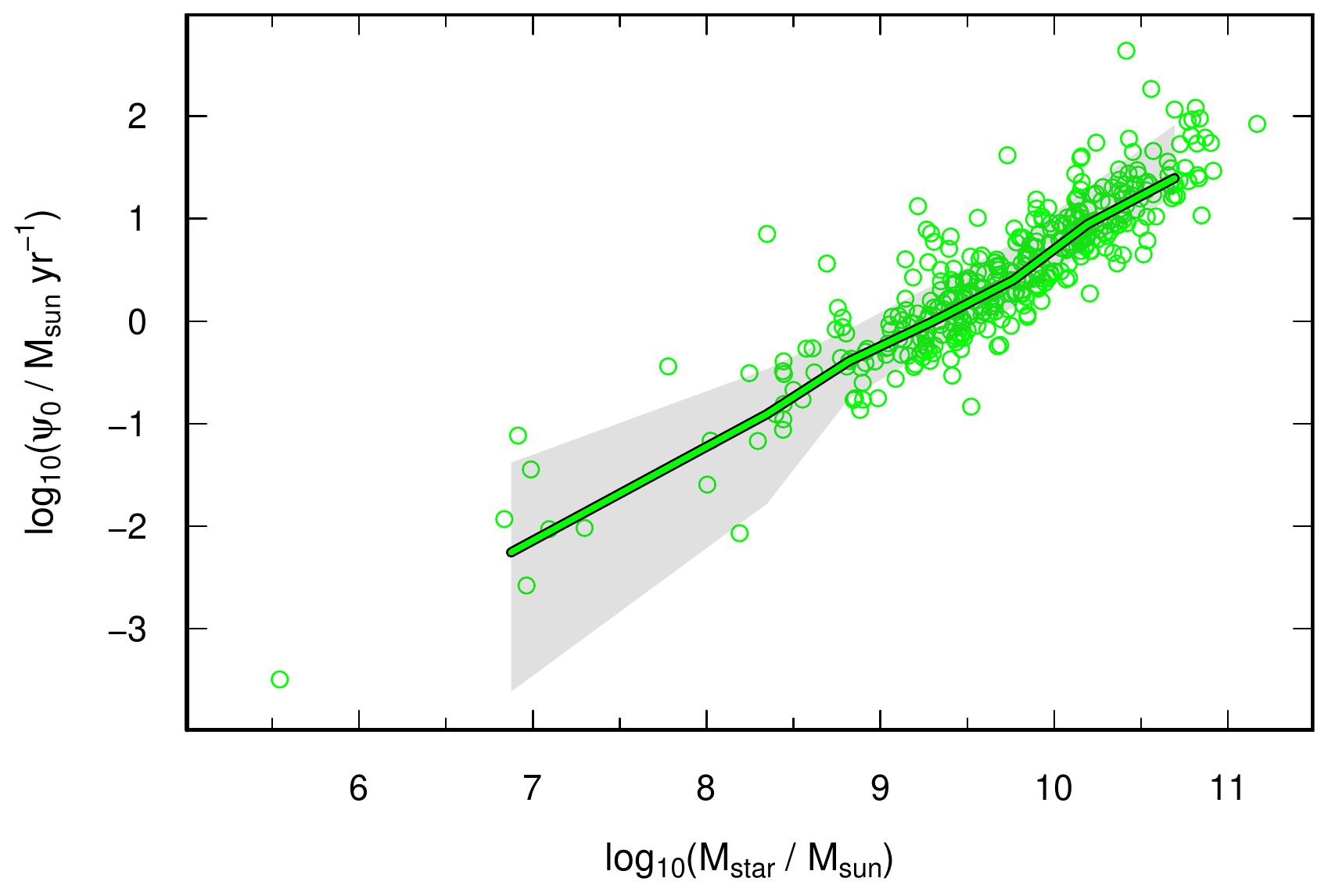} \\
\vspace{0.5\baselineskip}
\includegraphics[angle=0,width=0.98\linewidth]{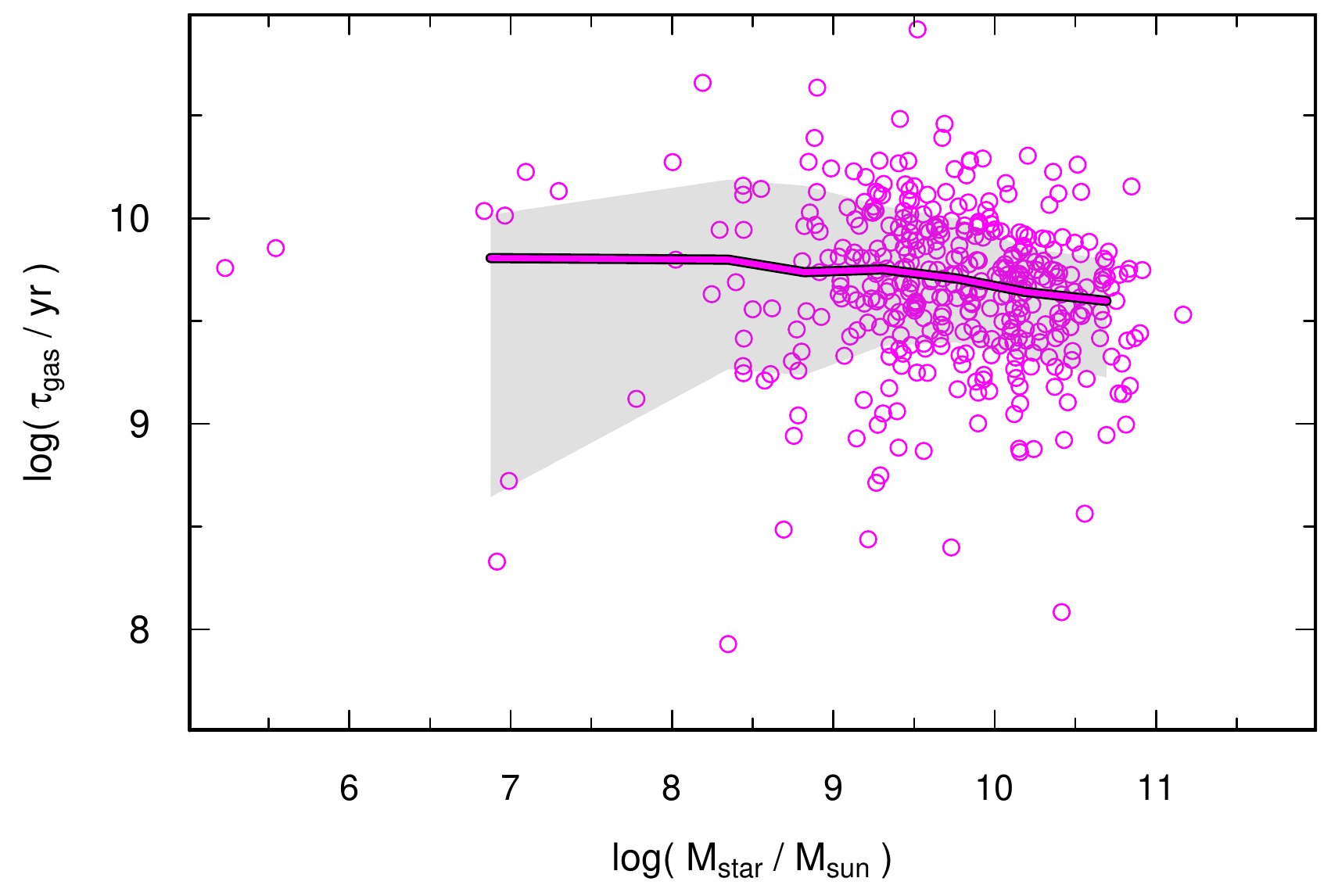}
\caption{Inferred galaxy age \tgal\ plotted against \mstar\ for MAGMA galaxies (top panel);
initial SFR at $t=0$, $\psi_0$ plotted against \mstar\ (middle);
\taugas\ (assumed equal to \tausf) plotted against \mstar\ (bottom).
In the two top panels, we assumed a simple declining exponential SFH with $\beta\,=\,-1$ [Eq. \eqref{eqn:sfrtime}],
and that \tausf\,=\,\taugas\,=\,\epss$^{-1}$.
}
\label{fig:correlations}
\end{figure}

Figure \ref{fig:correlations} shows the values inferred for \tgal\ with
the simple SFH of Eq. \eqref{eqn:sfrtime} with $\beta\,=\,-1$ (declining exponential)
plotted against \mstar\ (top panel) and $\psi_0$ plotted against \mstar\ (middle).
Although there is significant scatter, galaxies with \mstar\,$\la 10^9$\,\msun\
are apparently younger, with ages as low as $10^9$\,yr, rather than the overall
inferred age of more massive galaxies, $10^{10}$\,yr.
The initial SFR $\psi_0$ is very tightly linked with \mstar, over a dynamic range
of $\sim 10^3$.
The bottom panel of Fig. \ref{fig:correlations} illustrates that the total gas
depletion times (i.e., including both molecular and atomic components) are relatively
constant with \mstar.

\begin{figure}[!h]
\begin{center}
\includegraphics[angle=0,width=1.04\linewidth]{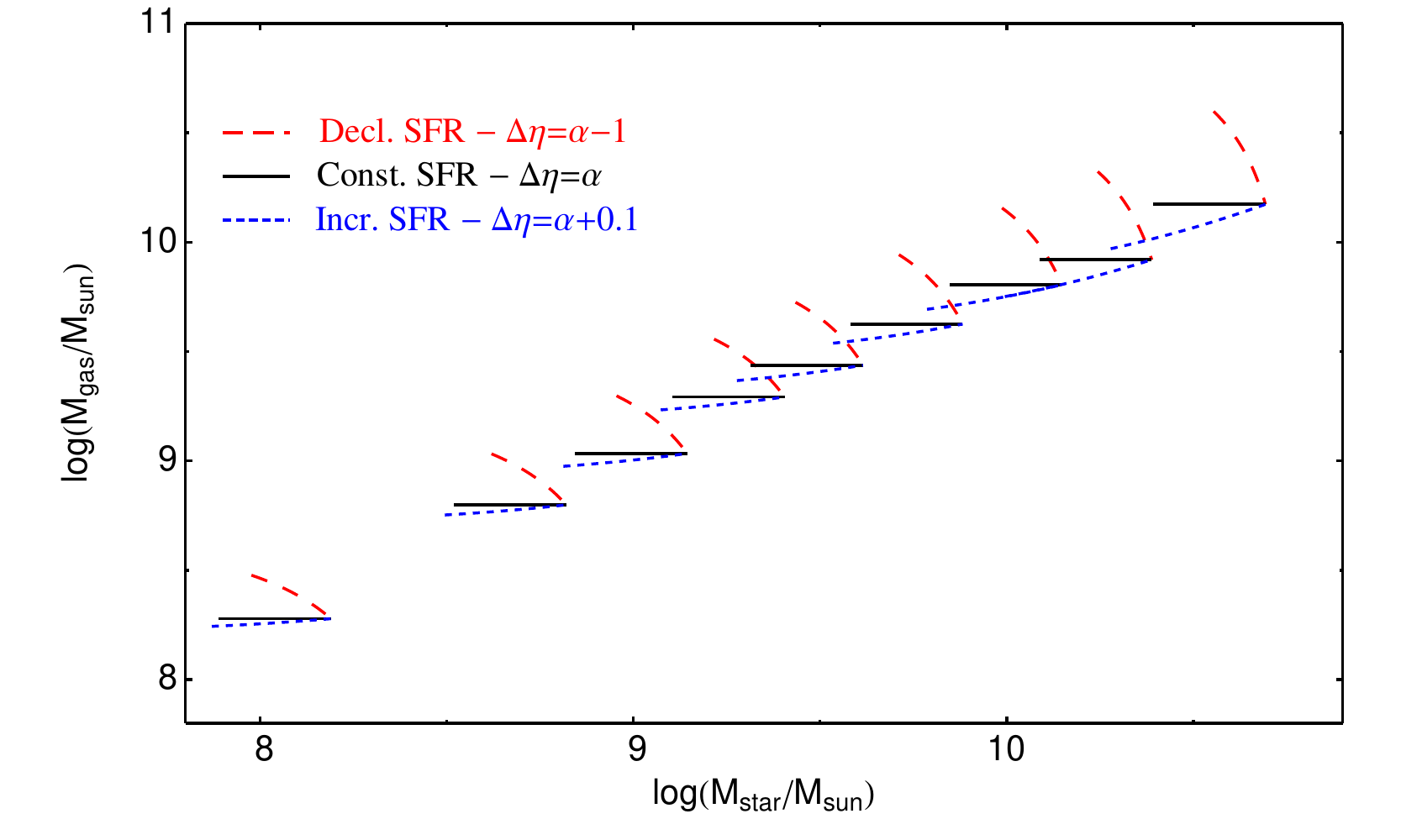}
\end{center}
\vspace{-1.2\baselineskip} \caption{Time evolution of \mgas\ vs
\mstar\ over the last 50\% of the galaxies' lifetime, according to
the toy-model based on three exponential SFHs: $\beta\,=\,(-1,\
0,\ 1)$, and also assuming that the SFR timescale \tausf\ is the
same as the gas depletion time, \taugas\,=\,\epss$^{-1}$. Galaxy
age \tgal\ is given by Eq. \eqref{eqn:tgal}, $\psi_0$ by Eq.
\eqref{eqn:psi0}, and the \mgas\ evolution by Eq.
\eqref{eqn:mg_sol}. The median curves in different \mstar\ bins
are plotted, where the (median observed) gas mass \mgas\ is taken
to be \mgas(\tgal). For \deta\,=\,$\alpha$, the equilibrium
solution, \mgas\,=\,\mi\ throughout the lifetime of the galaxy. }
\label{fig:toymodel}
\end{figure}

Now the only missing ingredient to describe \mgas\ is \deta\ [see
Eq. \eqref{eqn:mg_sol}].
Ultimately, the toy model presented here requires through Eq. \ref{eqn:sfr} that the
SFH given by Eq. \eqref{eqn:sfrtime} is related to the time
evolution of \mgas\ [Eq. \eqref{eqn:mg_sol}].
If \tausf$^{-1}$ is the same as
\epss, as we have assumed, the formalism requires:
\begin{equation}
\Delta \eta\,=\,\alpha + \beta\quad.
\end{equation}
Thus, fixing $\beta$ allows only certain values of \deta. Figure
\ref{fig:toymodel} shows the time evolution of \mgas\ and \mstar\
over the last 50\% of the galaxies' lifetime, assuming that \tgal\
is given by Eq. \eqref{eqn:tgal}, and $\psi_0$ by Eq.
\eqref{eqn:psi0}. The different sets of curves correspond to
different \mstar\ medians and the corresponding medians of \tausf,
\tgal, and $\psi_0$, while the vertical separation is achieved
through the assignment of different \deta\ as required by
$\beta\,=\,(-1,\ 0,\ 0.1)$.
The rising exponential SFH cannot have $\beta$ arbitrarily large
because otherwise Eq. \eqref{eqn:psi0} has negative values;
$\beta\,=\,0.1$ is close to the largest value allowed by the data.
Figure \ref{fig:toymodel} clearly
shows that according to the SFH, and the value of $\beta$ and consequently \deta,
$d\log\,M_g/d\log$\,\mstar\ can be positive or negative. Thus, it
is difficult to assign \deta\ through the application of Eq.
\eqref{eqn:deltaeta}.

\section{The formalism for the ``homogeneous wind''}\label{app:hom_wind}

In this paper we have determined solutions for \zg\ in the most
general case of \zw\,$\neq$\,\zg.
We demonstrate \citep[as already shown by][]{Peeples_Shankar11} that in general this is
a valid assumption since the outflowing gas \zw\ is not the same as
the ISM gas, \zg, especially at low masses
\citep[e.g.][]{Chisholm+15,Creasey2015,Muratov+17}.

However, much earlier work assumed, instead, that \zw\,=\,\zg,
adopting the so-called ``homogeneous wind'' model
\citep[e.g.,][]{pagel09,Erb2008}. This assumption leads to
\zetaw\,=\,\etaw, and modifies Eq.\,\eqref{eqn:zg} as follows:
\begin{equation}
Z_g\,=\,\left( \frac{q}{\eta_a} \right)\, \left[ 1 - \left(
\frac{M_g}{M_i} \right)^{\frac{\eta_a}{\alpha - \eta_a + \eta_w}}
\right] \quad . \label{eqn:zg_hom_wind}
\end{equation}
This solution, already reported by \citet{Erb2008} in her Eq.\,11,
places the dependence of \etaw\ only in the exponent of \mgas/\mi,
while \etaa\ also appears in the denominator in the \zg\ expression.

The very interesting outcome of this exercise is that the \etaw\
derived in this way reproduces the values of \zetaw\ shown in Fig.
\ref{fig:loadings} and obtained using the more general formalism
(using \zw\,$\neq$\,\zg) of Eq.\,\eqref{eqn:zg}. Thus, metal
loading \zetaw\ and mass loading \etaw\ in the winds can be
confused, depending on the assumptions made in the application of
the formalism.

\end{document}